\documentclass[12pt]{article}

\usepackage[
top    = 2.50cm,
bottom = 2.50cm,
left   = 2.50cm,
right  = 2.50cm]{geometry}
\usepackage{jheppub}
\usepackage[sort&compress]{natbib}
\usepackage[utf8]{inputenc}

\usepackage{amssymb}
\usepackage{graphicx}
\usepackage{slashed}
\usepackage{amsmath}
\usepackage{verbatim}
\usepackage{caption}
\usepackage{subcaption}
\usepackage{graphicx}
\usepackage[usenames]{xcolor}

\newcommand{\eg}{{\it e.g.,}\ }
\newcommand{\ie}{{\it i.e.,}\ }
\newcommand{\reef}[1]{(\ref{#1})}

\newcommand{\mt}[1]{\textrm{\tiny #1}}

\newcommand{\be}{\begin{equation}}
\newcommand{\ee}{\end{equation}}
\newcommand{\bea}{\begin{eqnarray}}
\newcommand{\eea}{\end{eqnarray}}
\newcommand{\beq}{\begin{equation}}
\newcommand{\eeq}{\end{equation}}
\newcommand{\beqa}{\begin{eqnarray}}
\newcommand{\eeqa}{\end{eqnarray}}
\newcommand{\beqar}{\begin{eqnarray*}}
\newcommand{\eeqar}{\end{eqnarray*}}

\newcommand{\tfun}{T_\mt{fun}}
\newcommand{\wfun}{\omega_\mt{fun}}
\newcommand{\dt}{\delta t}
\newcommand{\cO}{{\cal O}}

\begin{document}

\title{Holographic Quenches in a Confined Phase}

\author[a]{Robert C. Myers,}
\author[b]{Moshe Rozali,}
\author[b]{Benson Way}

\affiliation[\,a]{
\it{Perimeter Institute for Theoretical Physics, 
Waterloo, Ontario N2L 2Y5, Canada}}

\affiliation[\,b]{
\it{
Department of Physics and Astronomy, University of British Columbia,\\
Vancouver, BC V6T 1Z1, Canada}}

\emailAdd{rmyers@perimeterinstitute.ca}
\emailAdd{rozali@phas.ubc.ca}
\emailAdd{benson@phas.ubc.ca}

\abstract{
We investigate quenches of holographic theories in a confined phase, where the energy injected is insufficient to reach the deconfined phase. In such quenches, thermalization is not associated with gravitational collapse and the formation of a black hole.   
Nevertheless, we attempt to characterize the late-time state of this scenario.  
We check a number of notions of thermalization that do not require horizon formation, and find no evidence for thermalization in our chosen parameters and initial states, even in the weakest sense.

We find that the post-quench behaviour of both local and nonlocal observables exhibit oscillatory behaviour rather than decaying towards equilibrium. We generally find that the response of the nonlocal observables is smoother than that of the local ones. We discuss mechanisms which generate such smoothing, as well as ``beats" which appear in the time-dependence of the nonlocal operators for certain classes of quenches. When tuning the quench parameters such that the smoothing is ineffective, we are able to perform ``entanglement spectroscopy", recovering the spectrum of the confined phase of the theory from the time dependence of the entanglement entropy, as well as other nonlocal observables.

}

\maketitle

\section{Introduction}

The study of quantum many-body systems in far-from-equilibrium settings is a challenging and fascinating subject. In particular, recent advances in cold atom experiments have stimulated a vigorous research
programme into quantum quenches, \eg \cite{more1,more2,more3,more4}. Although such quenches are well understood in the context of quantum mechanics \cite{LL}, much less is known about such processes for many-body systems, as in quantum field theories. Theoretical progress has been made in more specialized
systems, including two-dimensional conformal field theories, \eg \cite{Cala06,Cala07,cala08,sotir08}, (nearly) free field theories, \eg \cite{Cala05,kollath07,cramer08,sotir08,roux2009,sotir09q}, and integrable models, \eg \cite{Cala05,rigol2007,man07,rigol2008,cala11q}. However, it seems that we are still far from the goal of understanding general organizing principles governing the far-from-equilibrium behaviour of many-body systems. Further, there is an absence of theoretical techniques that provide an efficient description of these quenches in a broad variety of systems.

Gauge/gravity duality provides a promising framework where questions about certain strongly coupled field theories are recast into questions about classical gravity, \eg  \cite{joh}. Such holographic models are especially well-suited for the study of quantum quenches since, with relatively modest efforts, one is able to study a broad range of quench protocols, beginning either in the ground state or at finite temperature, for a wide variety of strongly coupled quantum field theories in general spacetime dimensions. Indeed, holographic techniques have been used in many previous investigations of quantum quenches, \eg \cite{Das:2010yw,Rangamani:2015agy,Caputa:2017ixa,
Basu:2013soa,Basu:2012gg,Basu:2011ft,Buchel:2012gw,Buchel:2013lla,
Buchel:2013gba} and the related issue of
``thermalization" \cite{Craps:2015upq,Balasubramanian:2014cja,Buchel:2014xwa,
AbajoArrastia:2010yt,Albash:2010mv,Balasubramanian:2010ce,
Balasubramanian:2011ur,Aparicio:2011zy,Allais:2011ys,Galante:2012pv,Baron:2012fv,
Balasubramanian:2012tu,Keranen:2011xs,Caceres:2012em}. In fact, holographic studies \cite{Buchel:2012gw,Buchel:2013lla,Buchel:2013gba} have lead to the discovery of new universal behaviour in rapid quenches that apply for general quantum field theories \cite{Das:2014jna,Das:2014hqa,Das:2015jka,Das:2016lla}.
A common feature of most of these previous holographic studies is that thermalization in the boundary theory is described by gravitational collapse and the formation of a black hole in the bulk gravity theory.  However, in this paper, we instead study holographic quenches where an event horizon does not appear in the gravitational solution.

Our study can be motivated by the non-holographic studies presented in \cite{pasquale}, which investigates global quenches of a one-dimensional spin chain. In particular, these quenches were made with the Ising chain in both transverse and longitudinal  magnetic fields, which has the distinguishing feature of making the system confined. That is, the otherwise free spinless Majorana fermions found at the critical point are bound as pairs in meson excitations. We will compare their results to those found with quenches in a simple holographic framework which also exhibits confinement. 

In terms of the boundary theory, the holographic model corresponds to a four-dimensional supersymmetric gauge theory compactified with supersymmetry-breaking boundary conditions to produce an effective three-dimensional theory which is confined in the IR  \cite{edward}. The confining ground state is described in the bulk by a particular gravitational solution, known as the AdS soliton \cite{Horowitz:1998ha}, which does not possess an event horizon. We add here that quenches in this gravitation background were already considered by \cite{Craps:2015upq}.\footnote{For other previous studies of holographic quenches in confining models, see \cite{Craps:2013iaa,daSilva:2016nah,Lopez:2017hjg}.}

Importantly, the holographic theory has a mass gap, so quenches that inject less energy than this gap cannot form black holes and will remain in the confined phase.  Since there is no stationary black hole geometry with this final energy (which is conserved after the quench), characterizing the final state itself is interesting, even before discussing the long time dynamics leading to that state. While black hole collapse is prevented, the bulk geometry undergoes strong nonlinear dynamics on long time scales. Analyzing these dynamics is one motivation of the present work. 

We study quenches within this holographic model by numerically evolving the classical bulk gravity equations of motion, subject to a time-dependent boundary condition representative of a quench.  We begin by studying weak quenches, where our analysis of the response functions recovers the spectrum of normal modes of the AdS soliton geometry.  This behaviour is expected from perturbation theory, and serves as a reassuring check on our methods. 

We then study the time series of these local observables for stronger, nonlinear quenches, and for longer times.  Indeed, we find that the time series of the expectation values of local operators shows erratic and nonlinear behaviour.  Our main motivation is deciding whether, and in what sense, the system eventually thermalizes in the long time limit.  As we noted, the question of thermalization in our context is distinct from black hole formation. Instead, we focus on the changes in the power spectra of the time-series of local observables, as the system evolves. What we find is that in all cases there is an initial phase where the excitation energy drifts towards high frequency modes, which we interpret as an initial trend towards thermalization. However, we also find that the initial progression reverses itself, resulting in an oscillatory behaviour, in frequency space, on long time scales. We discuss the implications of that behaviour for various definitions and measures of thermalization. Our evidence indicates that for the range of parameters we probed, and for the duration of time evolution we have considered, the system does not thermalize even in the weakest sense.

An additional goal in this paper is to examine the time-dependence of nonlocal observables. Our investigation considers holographic observables which are analogous to the nonlocal operators studied by \cite{pasquale} on a one-dimensional spin system. In some respects we find similar results: some qualitative details discovered by \cite{pasquale} are reproduced in our higher dimensional holographic context. This includes the suppression of ``light-cone spreading" of correlations and the appearance instead of oscillatory behaviour in a variety of observables.  In particular, we investigate the entanglement entropy and compare to the discussion in \cite{pasquale}. We find that the entanglement entropy of half space is again oscillatory, just as for the one-dimensional spin chain. Further, at least in a particular class of weak quenches, for which the smoothing we discuss presently is ineffective, this quantity reveals the spectrum of normal mode excitations. This generalizes the idea of {\it entanglement spectroscopy} introduced in \cite{pasquale} to higher-dimensional holographic field theories, and to various nonlocal observables.

However, we also find some interesting differences from \cite{pasquale}. Most importantly, there is a significant smoothing of the response of the nonlocal operators in comparison to the expectation values of local operators. We find a number of different mechanisms contribute to the filtering of the high frequency features. Most importantly, the dual holographic constructions of the various nonlocal observables probe deep into the bulk geometry. For example, the entanglement entropy of half space is calculated with an extremal bulk surface that reaches from the asymptotic boundary down to the minimum radius in the bulk. We demonstrate this smoothing and other features of the time-dependence applies for a number of different nonlocal observables in our holographic model.

Our paper is structured as follows: in section \ref{setup} we describe the setup, the equations of motion, and the method for their solution. We describe the basic features of the geometry and the resulting boundary observables. This consists of an oscillatory behaviour with a fundamental frequency, which we relate to the geometry and the spectrum of excitations in the boundary theory. This sets the stage for a more detailed exploration of the time dependence of observables following the quench.
In section \ref{onepo}, we discuss the basic observables, time-dependent expectation values of local operators. We perform time-frequency analysis of that time dependence, and discuss the question of thermalization in our setup. In section \ref{nonloco}, we perform a similar analysis of the time-dependence of nonlocal observables, such as two-point correlation functions and entanglement entropy. We discuss the various smoothing mechanisms for those nonlocal observables, and perform spectroscopy of those observables. We present a brief discussion of our results and of future directions in section \ref{discuss}.

\vspace{20pt}
\section{Setup and Solutions}
\label{setup}

As described above, we will be studying quenches in a holographic framework which was first considered by \cite{edward} to describe a confined phase of ${\cal N}$=4 super-Yang-Mills theory. One of the spatial directions in this four-dimensional theory is compactified, \ie the boundary theory is placed on $\mathbb R^{(1,2)}\times S^1$, and the fermions are given anti-periodic boundary conditions of the compact direction. These boundary conditions break supersymmetry and the effective three-dimensional theory is confining in the IR.

The bulk theory for our studies will be defined by the five-dimensional action 
\begin{equation}
S=\frac{1}{16\pi G}\int \mathrm d^5x\sqrt{-g}\left(R+\frac{12}{L^2}-2 (\nabla\phi)^2\right)\;,
\end{equation}
where $L$ is the AdS length scale.  This action yields the equations of motion
\begin{equation}
\nabla^2\phi=0\;,\qquad R_{ab}+\frac{4}{L^2}\,g_{ab}-2 \nabla_a\phi\nabla_b\phi=0\;.
\label{volum}
\end{equation}
The gravitational solution corresponding to the confining ground state on $\mathbb R^{(1,2)}\times S^1$ is a static geometry known as the AdS soliton \cite{Horowitz:1998ha}. This background has a vanishing scalar $\phi=0$ and the metric is given by
\begin{equation}\label{soliton}
\mathrm ds^2=\frac{L^2}{z^2}\left(-\mathrm dt^2+\frac{\mathrm dz^2}{1-z^4/z_0^4}+\mathrm dx_1^2+\mathrm dx_2^2+(1-z^4/z_0^4)\frac{\mathrm d\theta^2}{4}\right)\;.
\end{equation}
This geometry can be obtained by a double Wick rotation of a planar AdS black hole. Considering the asymptotic geometry, \ie the limit $z\to0$, we find that the boundary metric is naturally given as
\beq
\mathrm ds^2_\partial=-\mathrm dt^2+\mathrm dx_1^2+\mathrm dx_2^2+\mathrm d\theta^2/4\,.
\label{bndry}
\eeq
As the notation suggests, $\theta$ is the compact direction. Further, examining eq.~\reef{soliton}, we see that the proper distance along this compact $\theta$ direction shrinks to zero at $z=z_0$. Hence, $z=z_0$ specifies the location where the bulk geometry closes off in the IR. Regularity at $z=z_0$ enforces the coordinate $\theta$ to have period $\Delta\theta=2\pi\,z_0$. With the metric \reef{bndry}, the proper distance around the compact dimension is $L_c=\Delta\theta/2$. This ``compactification scale" sets the confinement scale in the boundary theory, \eg the deconfinement temperature is given by $T=1/L_c$. 

Compared to the AdS vacuum, this gravitational solution \reef{soliton} has a negative energy \cite{Horowitz:1998ha}
\beq
E= -\frac{L^3}{2\pi G}\,\frac{V_2}{\Delta \theta^3} = -\frac{\pi^2 N^2}8\,\frac{V_2}{L_c^3}\,,
\label{negen}
\eeq
where $V_2$ denotes the spatial volume of the $x_1$ and $x_2$ directions.\footnote{The full four-dimensional stress tensor is given by $T_{ab}=\frac{\pi^2}8\,\frac{N^2}{L_c^4}\,{\rm diag}(-1,1,1,-3/4)$, using the boundary metric in eq.~\reef{bndry} \cite{stress}. \label{footy24}} In the second expression, we have used the standard AdS/CFT dictionary and $N$ corresponds to the rank of the $SU(N)$ gauge group in the dual SYM theory.  This negative energy provides a mass gap for black holes, which exist only for positive energies.

By a scaling symmetry, we are free to set $z_0=1$ without loss of generality. For simplicity, we make this choice in all of our numerical studies of holographic quenches in the following. Since the scale of the confining gap is determined by the size of the circle, this choice is equivalent to fixing the confinement scale.

\vspace{10pt}
\subsection{Quench Metric Ansatz}
Now let us discuss our numerical methods for quenching this system.  To begin, we emphasize that we are only considering {\it global} quenches. That is, the variation of the couplings is uniform across all of the spatial boundary directions, $(x_1,x_2,\theta)$ in eq.~\reef{soliton}. Of course, this choice greatly simplifies the analysis of the holographic quenches since the corresponding bulk metric only depends on two coordinates: the radial coordinate in the bulk and time.

As described above, the AdS soliton has a negative energy \reef{negen}.  If enough energy is injected into the system so that the total energy becomes positive, there will be a phase transition to a deconfined phase described by a planar black hole \cite{Craps:2015upq}.  For our purposes, we only consider quenches that leave us in the confining phase, which places an upper bound on the energy density injected into the system. While this upper bound prevents black hole formation, we will see that for sufficiently strong quenches, the dynamics is still highly nonlinear. 

Our conventions are similar, but not identical to those in  \cite{Craps:2015upq}. We will work with two different metric ansatze. The first of these is used to describe the actual quench process, as follows 
\begin{equation}
ds^2= \frac{L^2}{s(r)^2}\left(-e^{2 h }dt^2 +\frac {e^{2 f}}{g(r)} dr^2+e^{-b}(dx_1^2+dx_2^2)+r^2 e^{2 b} g(r) d\theta^2\right)\;,
\label{metric}
\end{equation}
where $s(r)=(1-r^2)$ and $g(r)=\frac{1-s(r)^4}{4 r^2}$.  This ansatz is similar to the one found in \cite{Craps:2015upq}, so we do not present the equations of motion here.\footnote{In our previous conventions and those of \cite{Craps:2015upq}, this sets $z_0=1$ and hence $\Delta\theta=2\pi$. \label{footy2}} 
The metric functions $h,f,b$ and the scalar field $\phi$ depend on both the radial coordinate $r$ and time $t$, as befits a global quench. In this coordinate system, the conformal boundary is at $r=1$, and the IR endpoint, where the circle closes off smoothly, is at $r=0$. 

Near the IR cap-off point $r=0$, the metric in the $(r,\theta)$-plane describes a Euclidean plane in polar coordinates. Since $g(r=0)=1$ by construction, regularity requires $f(r=0)+b(r=0)=0$, with $\Delta\theta=2\pi$ --- see footnote \ref{footy2}. Furthermore, as our initial condition, we choose  $f(r)=b(r)$, which satisfies that regularity condition and therefore it is sufficient to require $\frac{\partial f}{\partial t} (t,r=0)+\frac{\partial b}{\partial t}(t,r=0)=0$ for the time derivatives of the fields at $r=0$. In our numerical simulations, we impose this boundary condition at each time step. 

Examining the asymptotic geometry in eq.~\reef{metric}, \ie taking $r\to1$, we can read off the conformal boundary metric 
\beq
\mathrm ds^2_\partial=-\mathrm dt^2+e^{-b(r=1)}(\mathrm dx_1^2+\mathrm dx_2^2)+\frac{e^{2 b(r=1)}}4\,\mathrm d\theta^2\;,
\label{bndry2}
\eeq
where we have used $h(r=1)=0$, which is imposed as a boundary condition in our simulations.  We may choose to vary $b(r=1)$ in time as a means of quenching the system (in addition to or in lieu of a scalar field --- see below). Note, however, that (before or) after the quench, we keep $b(r=1)$ constant, in which case the boundary metric \reef{bndry2} can be rescaled to recover the original boundary metric \reef{bndry}.   Since the size of the compact circle in eq.~\reef{bndry2} is given by $L_c(t)=\pi e^{b(t,r=1)}L$, such quenches vary the compactification scale and hence also the confinement scale. 
As for the scalar field, regularity requires $\partial_r\phi(r=0)=0$, and we are free to choose its boundary value $\phi(t,r=1)$ as part of our quench. 

For numerical purposes, we work with the fields
\begin{eqnarray}
B(t,r)=b'(t,r)~~~~~P(t,r)=\exp(f-h)\,\dot{b}(t,r) \nonumber \\
\Phi(t,r)=\phi'(t,r)~~~~~\Pi(t,r)=\exp(f-h)\,\dot{\phi}(t,r)\;,
\end{eqnarray}
which produce a set of first-order differential equations. We use a constrained evolution scheme in which we propagate the dynamical fields $B,P,\Phi,\Pi$ and $f$, and determine the field $h$ by solving a constraint at each time step. If needed, the original fields $b,\phi$ can be obtained by a simple radial integration. 
We use sixth order finite difference discretization, with a grid of a thousand points to discretize the radial coordinate, and fourth order Runge-Kutta evolution, with an adaptive step size. To stabilize the evolution, we use some of the tricks discussed in \cite{Maliborski:2013via}, as well as in \cite{Craps:2015upq}.

Given our setup above, we quench the system starting with a static solution and after the quench is completed, the final state is no longer driven. As we have mentioned above, there are two independent ways in which we perform the quenches, \ie we can drive the holographic system either with the scalar field $\phi$ or the metric anisotropy function $b$ (or both at the same time). For each of these fields, we choose the non-normalizable mode to follow a quench profile of the form
\begin{equation}
f(t)= \frac{a}{2} \,\left( \tanh \frac{t-t_0}{ \delta t}\ -\ 1\right)\,.
\label{proto}
\end{equation}
This profile rises gradually, on a time scale of order the duration $\delta t$, from the initial value $-a$ to the final value which we fixed to be zero. The total change during the quench is then given by the amplitude $a$. The transition occurs in the vicinity of $t=t_0$ and this parameter is chosen for numerical convenience, to ensure the quench is sufficiently gradual. All results quoted are for times well after the quench time at $t=t_0$.

\vspace{10pt}
\subsection{Post-Quench Coordinates}
The coordinate system in the previous section is useful for quenching the system and is adequate for short or intermediate time evolution. However, we found that for the purpose of obtaining long-time evolution, it is advantageous to transform to a different set of variables where the fields are explicitly unsourced. Therefore, we introduce the second metric ansatz:
\begin{equation}
\mathrm ds^2=\frac{2L^2}{1+x}\left[-\hat A\,\hat\delta^2\mathrm dt^2+\frac{\mathrm dx^2}{2(1-x^2)(3+x)\hat A}+\frac{1}{\hat B}(\mathrm dx_1^2+\mathrm dx_2^2)+\frac{(1-x)(3+x)}{16}\,\hat B^2 \mathrm d\theta^2\right]\;,
\label{post9}
\end{equation}
where the boundary lies at $x=-1$ and the circle caps off at $x=1$. Generally, the metric functions $\hat A,\,\hat B,\,\hat\delta$, as well as the scalar field $\phi$, are taken to depend on both $x$ and $t$. The AdS soliton geometry \reef{soliton} is recovered with $\hat A=\hat B=\hat\delta=1$. We have implicitly set $z_0=1$ here and hence the geometry is regular if $\theta$ has period $\Delta\theta=2\pi$.  

Now redefine the metric functions and scalar field to
\begin{equation}\label{eq:redef}
\hat A=1-x_p^2\,\alpha\;,\qquad \hat B=1+x_p^2\,\beta\;,\qquad \hat\delta=1-x_p^4\,\delta\;,\qquad \phi=x_p^2\,\varphi\;,
\end{equation}
where for convenience, we have defined
\begin{equation}
x_m=\frac{1-x}{2}\,,\qquad x_p=\frac{1+x}{2}\,.
\label{lowball}
\end{equation}
The definitions \eqref{eq:redef} essentially enforce the asymptotic boundary behaviour of the metric functions and scalar field (\ie as $x\to-1$), assuming that they are unsourced. Indeed, the boundary metric is the same as that of the AdS soliton \reef{bndry}.  The boundary conditions are guaranteed to be satisfied, so long as $\alpha$, $\beta$, and $\delta$ remain finite, which will always be true when performing numerics.  

At this point, let us define the first-order functions
\begin{equation}
Q=\frac{1}{2}x_p(3+x) \beta'+\beta\;,\qquad \Phi=\frac{1}{2}x_p(3+x) \varphi'+\varphi\;,\qquad P=\frac{\dot \beta}{4\hat A\hat\delta}\,,\qquad \Pi=\frac{\dot\varphi}{4\hat A\hat\delta}\;.
\end{equation}
The evolution equations for the first-order variables are given by
\begin{subequations}\label{eq:evo}
\begin{align}
\dot Q&=\partial_x\left[2x_p(3+x)(1-x_p^4\delta)(1-x_p^2\alpha)P\right]-2x(1-x_p^2\alpha)(1-x_p^4\delta)P\label{evo1}\\
\dot \Phi&=\partial_x\left[2x_p(3+x)(1-x_p^4\delta)(1-x_p^2\alpha)\Pi\right]-2x(1-x_p^2\alpha)(1-x_p^4\delta)\Pi\\
\dot P&=\partial_x[4x_m(1-x_p^4\delta)(1-x_p^2\alpha)Q]\nonumber\\
&\qquad+\frac{2(1-x_p^4\delta)}{(3+x)(1+x_p^2\beta)}\Bigg[2(1-x_p^2\alpha)\left[2x_m(1-x_p^2\beta)Q-2x_mx_pQ^2+x_p^2(3+x)P^2\right]\nonumber\\
&\qquad\qquad-\beta\left[(2+x+x^2)+x_p^2(3+x)\beta\right]\nonumber\\
&\qquad\qquad+\frac{x_p\alpha\left[8(3+x)+x_p(5-x)(2+x)(1+6x+x^2)\beta+{x_p}^3(3+x)(3+14x-x^2)\beta^2\right]}{11-x(2+x)}\Bigg]\label{evo2}\\
\dot \Pi&=\partial_x[4x_m(1-x_p^4\delta)(1-x_p^2\alpha)\Phi]\nonumber\\
&\qquad+\frac{2(1-x_p^4\delta)}{3+x}\left[4x_m(1-x_p^2\alpha)\Phi-\left(2+x(1+x)+\frac{x_p^2[7+x(2-x)][2-x(5+x)]\alpha}{11-x(2+x)}\right)\varphi\right]\;.
\end{align}
\end{subequations}
The constraints are given by
\begin{subequations}\label{eq:constraints}
\begin{align}
\frac{1}{2}x_p(3+x)\beta^\prime+\beta&=Q\\
\frac{1}{2}x_p(3+x)\varphi^\prime+\varphi&=\Phi\\
\alpha^\prime-x_p\left(x_p^2S+\frac{4}{11-x(2+x)}\right)\alpha&=-x_p S\label{eq:constraintalpha}\\
x_p\delta^\prime +(2+x_p^4S)\delta&=S\;,\label{dconstraint}
\end{align}
\end{subequations}
where the quantity $S$ is given by
\begin{align}
S=\frac{8}{11-x(2+x)}\bigg[\frac{3x_pP^2}{(1+x_p^2\beta)^2}-&\frac{(x_p\beta+Q)(4-x_p(1-5x)\beta-6x_mQ)}{(3+x)(1+x_p^2\beta)^2}&\nonumber\\
&\qquad\qquad\qquad+4x_p\Pi^2+\frac{8x_m(x_p\varphi+\Phi)^2}{3+x}\bigg]\;.
\end{align}
Finally, we have the following evolution equations, which are implied by the previous set of equations:\begin{subequations}\label{eq:evo2}
\begin{align}
\dot\beta&=4(1-x_p^2\alpha)(1-x_p^4\delta)P\\
\dot\varphi&=4(1-x_p^2\alpha)(1-x_p^4\delta)\Pi\\
\dot\alpha&=\frac{32x_p^2(1-x_p^2\alpha)^2(1-x_p^4\delta)}{11-x(2+x)}\left[\frac{2[1-x_p(1-2x)\beta-3x_mQ]P}{(1+x_p^2\beta)^2}-8x_m\Pi(x_p\varphi+\Phi)\right]\label{eq:evoalpha}\;.
\end{align}
\end{subequations}
These equations have a number of advantages over the previous ones used for quenching.  First, there are no singularities present in the equations, which improves numerical stability.  In particular, we are able to use spectral discretization of the radial direction. Second, asymptotic quantities can be read directly from these functions without taking any derivatives.  The energy, for example is read directly from the function $E\propto\alpha|_{x=-1}$, which is manifestly conserved from eq.~\eqref{eq:evoalpha}. Both of those advantages greatly improve both stability and accuracy of our numerics.

We evolve the system by repeatedly solving eqs.~\eqref{eq:evo} and \eqref{eq:constraints} within a Runge-Kutta algorithm. Only one of these equations \eqref{eq:constraintalpha} requires a boundary condition.  We take
\begin{equation}
\alpha(t,x=1)=1-\frac{1}{[1+\beta(t,x=1)]^2}
\label{truck}
\end{equation}
which sets regularity at the origin, assuming that $\theta$ has period $\Delta\theta=2\pi$. All remaining boundary conditions have been set by the definition of our evolution variables. We use a Chebyshev pseudo-spectral method for spatial discretization, solving eq.~\eqref{eq:constraints} by LU decomposition. We monitor eq.~\eqref{eq:evo2}, particularly energy conservation as a numerical check.

In the present coordinates, we are unable to quench the system directly, since they were constructed with the assumption that the non-normalizable modes vanish --- see the comment after eq.~\reef{lowball}. As the initial state for long post-quench evolution, we have used the state created by a quench, in the coordinate system \reef{metric} adapted for quenching, translated to the present coordinates. One could imagine other types of initial conditions, for example those generated as some linear combination of the normal modes in the initial state, but those would not necessarily have a quench interpretation. We leave exploration of the long time evolution resulting from different types of initial conditions for future work.

\vspace{10pt}
\subsection{Bulk Solutions -- Basic Features}

Quenching the AdS soliton is quite different from quenching a black hole geometry, or even a geometry which collapses to black hole at late time (\eg the Vaydia geometry). As we shall see later, those differences are apparent within the boundary observables.  Here, however, we will comment on the basic features of the bulk geometry.  Rather than describing the features of one specific quench, we focus on qualitative features typical of all quenches.

\begin{figure}
\centering
\begin{subfigure}{0.49\textwidth}
\includegraphics[width=1.1 \textwidth]{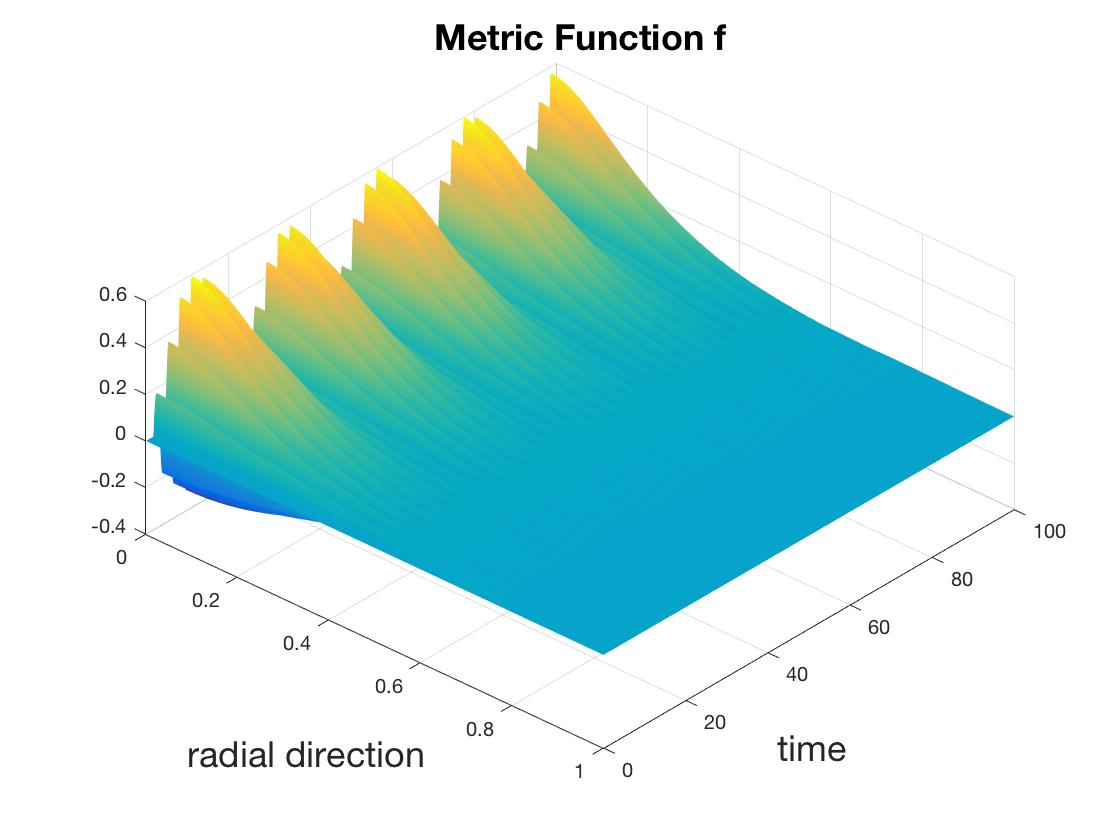}
\end{subfigure}
\begin{subfigure}{0.49\textwidth}
\includegraphics[width=1.1 \textwidth]{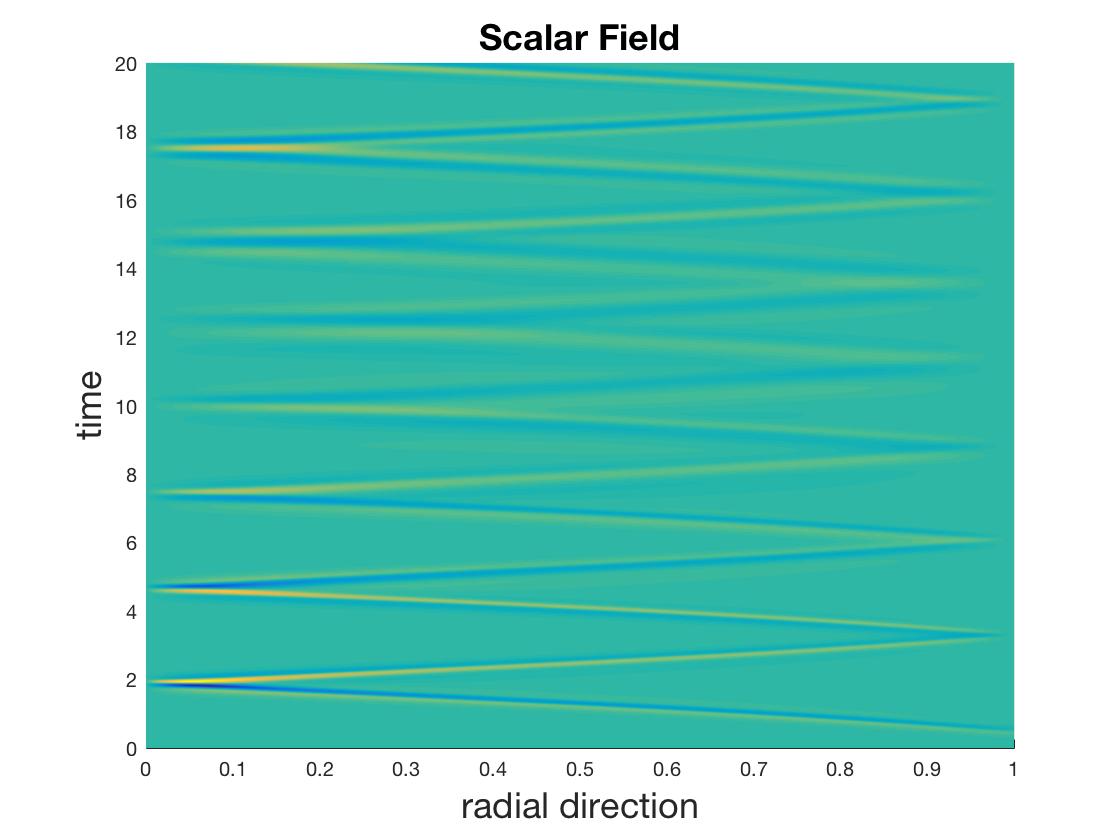}
\end{subfigure}
\caption{Metric and scalar field for times following the quench. The most prominent feature is the approximately periodic behaviour with a period of roughly $\tfun=2.622$. The coordinates are those given in eq.~\reef{metric} and in particular, $r=0$ and $r=1$ correspond to the IR cap-off point and the asymptotic boundary, respectively. }
\label{bulk}
\end{figure}

In figure \ref{bulk}, we display the scalar field and a typical metric function for times following one particular quench.   As we inject energy into the confining geometry, the most prominent characteristic of the subsequent evolution is  the bouncing of the excitation back and forth between the asymptotic boundary and the IR cap-off point. This bouncing frequency has a simple interpretation we now explain.\footnote{For an alternative interpretation of the bounces as  ``revivals", see \cite{daSilva:2014zva,daSilva:2016nah}. Below we will identify a longer time scale, of bouncing in frequency space, which might serve as a more natural time scale for revivals in CFTs \cite{2014PhRvL.112v0401C, 2016JPhA...49O5401C}. We leave that speculative direction for future work.}

For weak, rapid quenches, the geometry remains close to the AdS soliton. Consider then the probe limit where a massless scalar field propagates on a fixed AdS soliton background \eqref{soliton}.
Null geodesics in this geometry approximate the trajectory of a narrow energy pulse, and are given by $\frac{dz}{dt}=\sqrt{1-z^4/z_0^4}$. Thus the period of a bounce from the asymptotic boundary to the cap-off point and back is
\begin{equation}
T_\mt{fun}=2 \int_0^{z_0}\frac{dz}{\sqrt{1-z^4/z_0^4}}=2\,z_0\,\frac{\sqrt{\pi}\,\Gamma(5/4)}{\Gamma(3/4)}\simeq 2.622057554\,\frac{L_c}\pi\,.
\label{fun}
\end{equation}
Recall that $L_c$ is the size of the compact direction in the boundary theory. This bounce period $\tfun$ is closely related to the frequencies of the linearized excitations in the AdS soliton background \reef{soliton}, as we now show.

Consider the normal modes for a free massless scalar field $\phi$ in this geometry. They satisfy
\begin{equation}
{z^3}\,\partial_z \left[\frac{1-z^4/z_0^4}{z^3}  \,\partial_z \phi \right] =-\omega^2 \phi\,.
\end{equation}
To identify  the behaviour of high level modes, we can solve this equation with a WKB ansatz $\phi \sim e^{iS/\hbar}$, where $\hbar$ is a formal small parameter. As usual, the derivatives act only on the exponent to leading order in $\hbar$, which then gives $S'=\frac{\hbar \omega}{\sqrt{1-z^4/z_0^4}}$. The quantization condition for normal modes then becomes
(setting $\hbar=1$)
\begin{equation}
\omega \int_0^{z_0} \frac{dz}{\sqrt{1-z^4/z_0^4}}= \pi n + C\,,
\label{wkb}
\end{equation}
where the constant $C$ depends on the precise boundary conditions which we impose on the normal modes, and $n$ is an arbitrary integer. We expect eq.~\reef{wkb} to be accurate for large values of $n$ and comparing this result to the period \reef{fun} of bouncing of a null geodesic. We see that the level spacing of the higher normal modes by is 
\beq
\Delta\omega=\frac{2 \pi}{\tfun}\simeq2.396280470\,\frac{\pi}{L_c}\,.
\label{wkb2}
\eeq
Hence we can anticipate that this bouncing period \reef{fun} also sets the basic frequency for the response of the field excitations in our holographic quenches (both for the scalar modes analyzed above, and for the metric modes). 

More accurate results for the normal mode frequencies can be determined with a more precise WKB analysis \cite{Minahan98, Constable99}. However, one can also solve numerically for the energies of the lowest lying modes, as shown in table \ref{table33}. To do that, one constructs the operator of linearized fluctuations around the AdS soliton background, and find its eigenvalues (and eigenfunctions, if desired). In the table, we show the lowest lying frequencies for linearized excitations of the massless scalar field and the ``scalar" graviton mode associated with the metric anisotropy, which are the two fields excited in our holographic quenches. In terms of the boundary theory, these both correspond to $0^{++}$ glueballs in nomenclature of \cite{Brower99,Brow00}.  As one can see from the table, when measured in units of the fundamental frequency $\omega_{fun}$ the higher modes differ in near-integer values, confirming the validity of the WKB approximation.
\\

\begin{table}[htbp]
\centering
 \begin{tabular}{|c|| c| c|c| c|c| c|c|c|c|} 
 \hline
 Level & 1  & 2& 3& 4& 5 & 6 & 7 & 8&9 \\ [0.5ex] 
 \hline
 Scalar &  1.4206& 2.4521 & 3.4658 & 4.4734& 5.4783 & 6.4816 & 7.4841&8.4860& 9.4874\\ [0.5ex]
\hline
Metric & 0.9749 & 2.3025 & 3.3677 & 4.3995& 5.4187 & 6.4316 & 7.4410& 8.4480 & 9.4536   \\ [0.5ex]
 \hline
\end{tabular}
\caption{The frequencies (given in units of $\wfun=2\pi/\tfun$) calculated for the linearized modes corresponding to the lowest lying excitations of a massless scalar field and the ``scalar" graviton mode associated with the metric anisotropy. Note that for the splitting of the higher modes satisfies $\Delta\omega=\wfun$ to a good approximation. The frequencies calculated above agree with the results given in table 4 of \cite{Brow00}.}
\label{table33}
\vspace{20pt}
\end{table}

\vspace{10pt}
\subsection{Boundary Observables -- Time Series}

We have identified the time scale $\tfun$ in the system associated geometrically with the bouncing of excitations between the two ends of the geometry. The time scale of the bounces is essentially the shortest time scale associated with the AdS soliton. With $L_c=\pi$ (\ie $z_0=1$), we have roughly $\tfun \simeq 2.622$, as given in eq.~\reef{fun}, and the associated frequency is $\wfun=2\pi/\tfun \simeq 2.396$. This same frequency essentially sets the frequencies of the normal modes in the AdS soliton geometry, \eg see table \ref{table33} and also the asymptotic level spacing in eq.~\reef{wkb2}.  

\begin{figure}
\centering
\begin{subfigure}{0.49\textwidth}
\includegraphics[width=1.3 \textwidth]{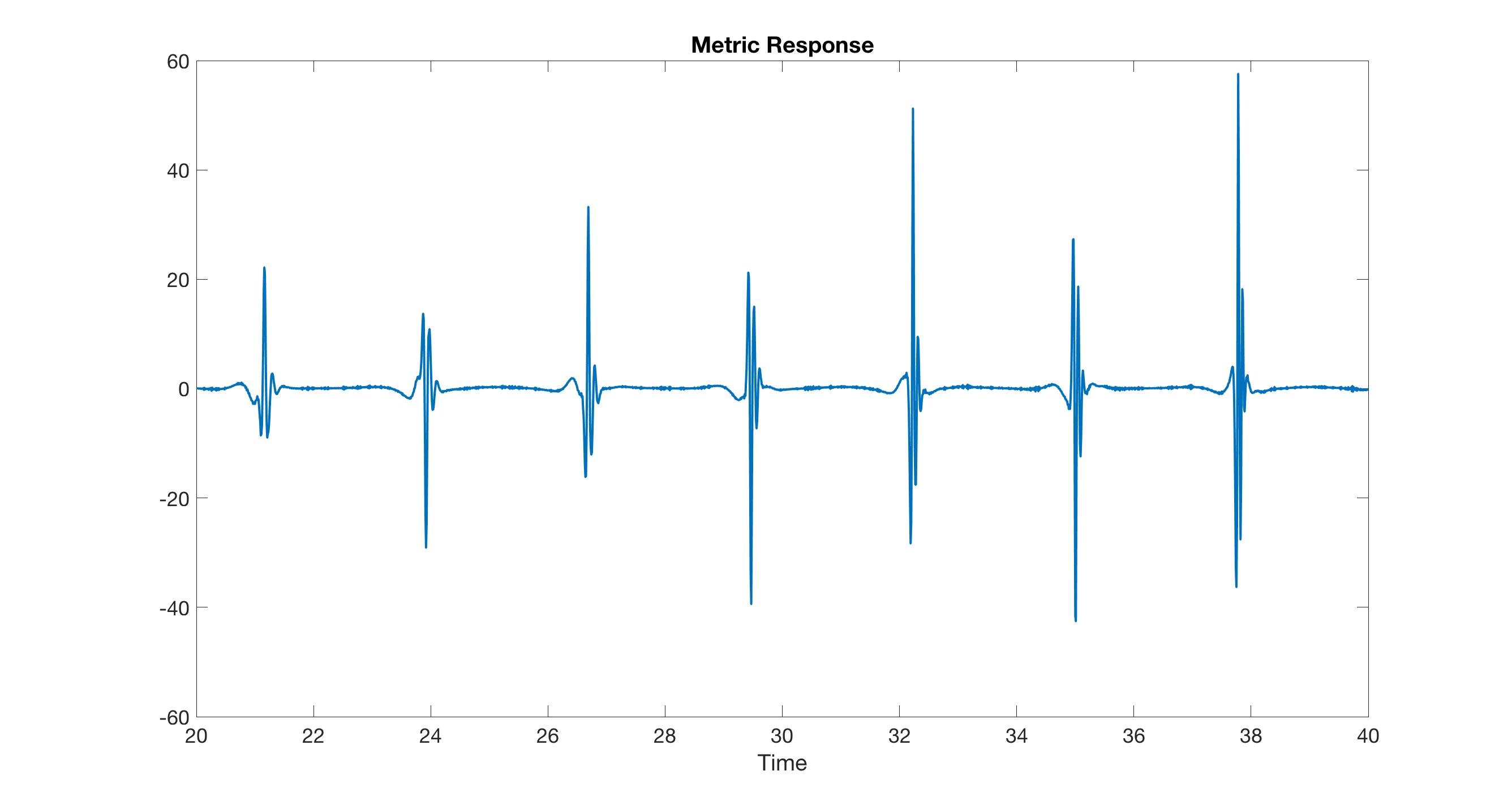}

\end{subfigure}
\begin{subfigure}{0.49\textwidth}

\includegraphics[width=1.0 \textwidth]{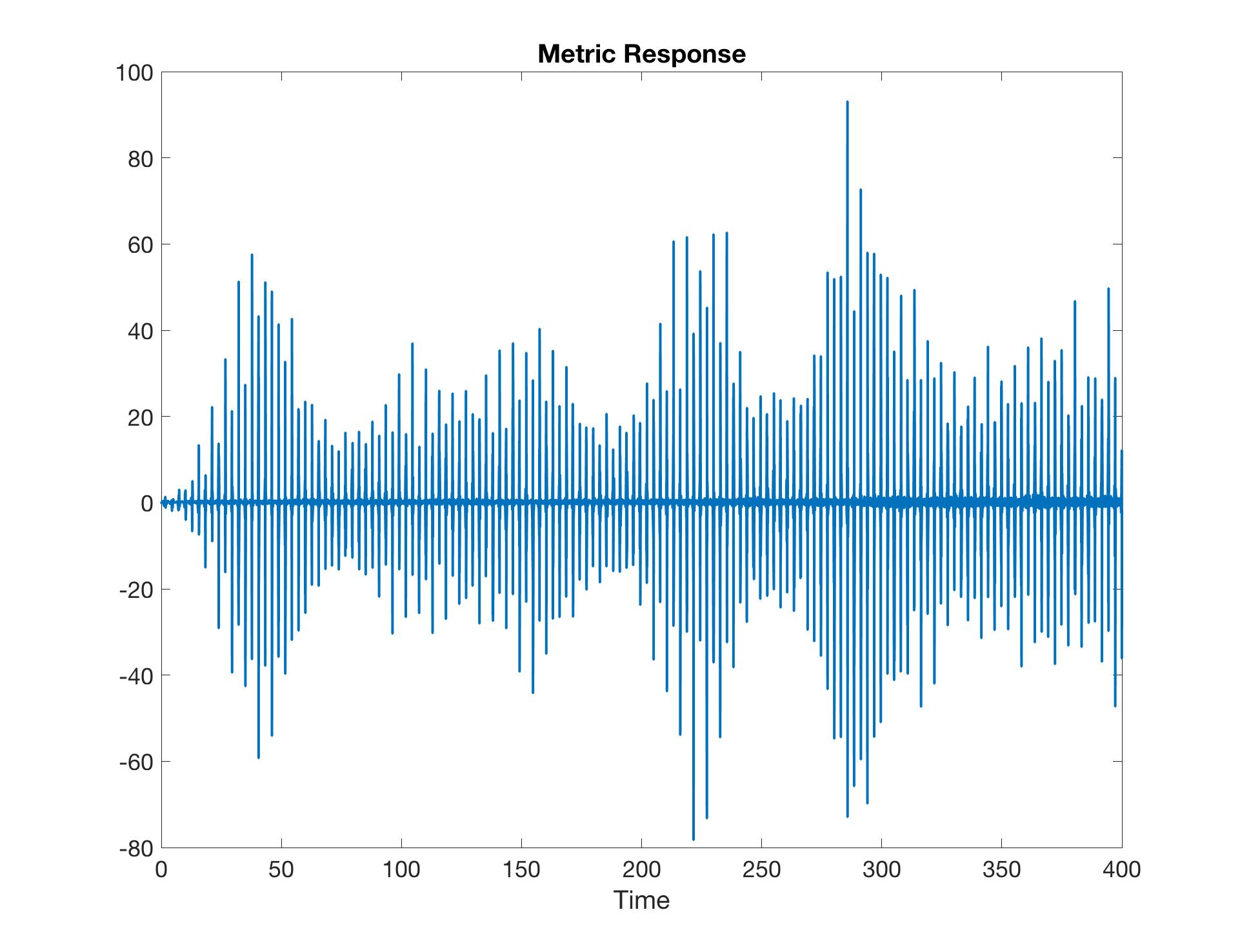}
\end{subfigure}
\caption{Typical time series for the one-point functions in the time-dependent geometry. Here we plot the metric anisotropy response, corresponding to the pressure anisotropy in the boundary theory, for a quench with $a_\phi=0.05$, $a_b=0$ and $\delta t=0.5$. The left panel shows the structure at short time scales with a few bounces in the capped geometry. The right panel shows the structure on long time scales. The time evolution is erratic on both those time scales.}

\label{scalarQ}
\end{figure}

The time series of observables inherits this fundamental frequency, but shows many more features, both on short and long time scales. In figure \ref{scalarQ}, we show a typical example of a time series of boundary observables. On short time scales, we see that as the signal bounces back and forth in the geometry, it is deformed in an erratic way. On the right panel, we see that the same signal has interesting features on long time scales as well. In the rest of the paper, we analyze the time series of various observables and extract from it interesting statements about the dynamics of the theory following the initial quench.

For the sake of an easy interpretation of subsequent dynamics, we will always choose sufficiently short quench durations. That is, we take $\delta t \ll\tfun \simeq 2.622$. For slower quenches, there is no separation between the time scale of the quench and that of the bouncing in the geometry (or as we saw, the lowest normal mode frequencies), which complicates the analysis. Further we expect that for such slow quenches, it would be very difficult to unravel any kind of universal behaviour\footnote{The opposite limit, $\delta t \gg \tfun$ corresponds to driving, rather than quenching, the system. Periodic driving can lead to interesting universal results, \eg see \cite{Auzzi:2013pca, Rangamani:2015sha} for holographic examples.}.

\vspace{20pt}
\section{One-Point Functions} \label{onepo}

Let us now discuss the one-point functions extracted from the bulk scalar and gravitational fields following the quench. These correspond to the time-dependent expectation values of the scalar operator and the boundary stress tensor, respectively.
To be precise, the normalizable mode of the bulk scalar yields $\langle{\cal O}\rangle(t)$ where ${\cal O}$ is the marginal scalar operator in the boundary theory dual to $\phi$. Similarly the anisotropy field (\ie the normalizable mode of $b$ in the quench metric \reef{metric}, or simply $\beta$ in post-quench coordinates) yields a combination of stress tensor components: $\langle(4 T_{\theta\theta}-3T_{tt})\rangle(t)$.\footnote{Of course, we can express this in terms of other linear combinations by making use of the vanishing trace of the stress tensor. In any event, one can easily verify that the expectation value of the above combination vanishes in the (static) confining phase --- see footnote \ref{footy24}.} 
With our symmetries, the stress tensor is fully characterized by the (total) energy (which is conserved after the quench and therefore contains no dynamics) and the anisotropy field. The output of our numerical simulations contains a time series for these boundary observables, which we now analyze using various tools. 

\begin{figure}
\centering
\begin{subfigure}{0.49\textwidth}
\includegraphics[width=1.1\textwidth]{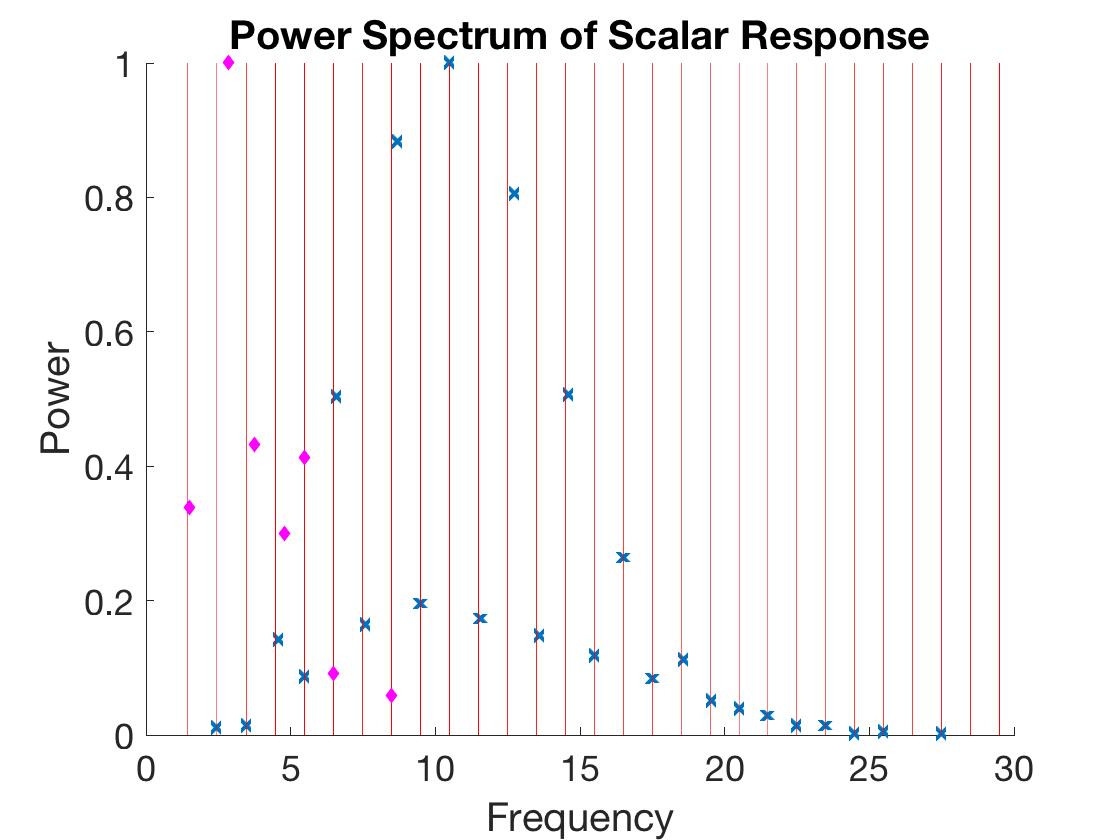}
\end{subfigure}
\begin{subfigure}{0.49\textwidth}
\includegraphics[width=1.1\textwidth]{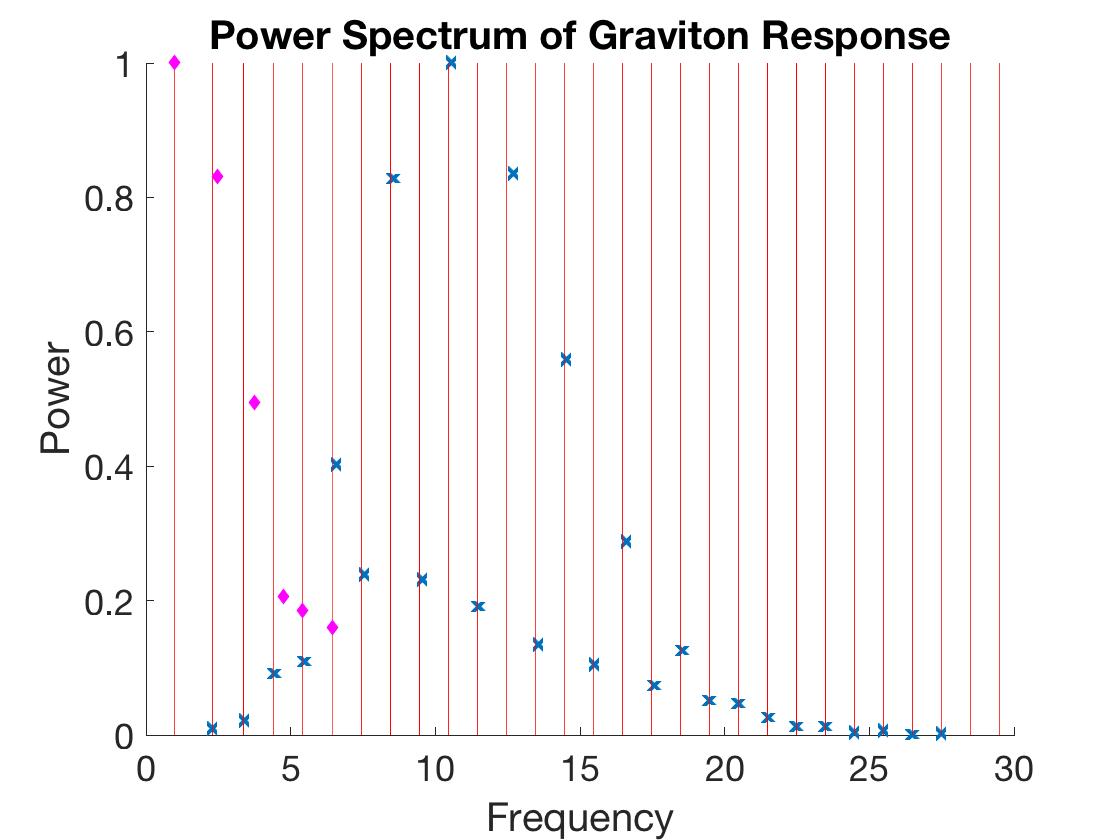}
\end{subfigure}

\caption{Power spectrum of the scalar response (left) or the pressure
anisotropy (right) following two particular weak quenches. The power spectra shown with blue x's follow a quench with amplitudes $a_b=a_\phi=10^{-5}$,
and duration $\delta t =0.1$. The power spectra with magenta diamonds follow a quench with amplitudes $a_b=a_\phi=10^{-5}$,
and duration $\delta t =0.3$. We measure frequencies in units of the fundamental frequency $\omega_\mt{fun}\simeq 2.396$. To focus on the shape of those spectra, we have chosen normalization such that the two spectra have a common maximum. In both cases,   the vertical lines correspond to the normal
modes of the unperturbed background. The data points are obtained
from a Fourier transform of the time evolution, taken at interval $\Delta t=100$ immediately following the quench.  Note that for a wider quench (\ie with larger $\delta t$),
as expected, the power spectrum is concentrated in lower normal modes.}
\label{scalar1pt}
\end{figure}

\vspace{10pt}
\subsection{Weak Quenches} \label{weakone}

We start by quenching the system weakly. When the amplitude $a$ (\ie in eq.~\reef{proto}) is sufficiently small, we inject a small energy density into the bulk, and a linearized analysis should provide a valid description of the dynamics. In this limit, the system is described by the normal modes of the bulk fields, \ie glueballs in the boundary theory, which propagate almost freely. This picture breaks down for sufficiently strong quenches, or for a sufficiently long time evolution.

One way of doing a spectral analysis of our solution is to directly project the spatial data into a spectrum of normal modes.  However, since the normal mode frequencies and the corresponding eigenfunctions are not available analytically, we will instead analyze the spectral properties of our solutions by taking a Fourier transform of the time series for our one-point functions.  This method also has the added advantage that the fundamental time scales can be captured for nonlinear quenches that necessarily differ from the normal modes of the AdS soliton, as we will see below when we analyze these strong quenches. 

To demonstrate that this method adequately captures the spectrum of normal modes,  we show the Fourier transform of the time series for the one-point functions of the scalar field (in the left panel) and of the anisotropy field (in the right panel), in figure \ref{scalar1pt}. We plot the power spectra following two separate weak quenches. We see that indeed, those power spectra have peaks corresponding to the spectrum of normal modes, \eg see table \ref{table33}. Our results here can be compared to the {\it quench spectroscopy} of \cite{pasquale}. There, the one-point function of the order parameter showed oscillatory behaviour after quenching the spin chain and the dominant frequencies in the corresponding power spectra matched the masses of various mesonic bound states in the confining phase. 

\begin{figure}
\centering
\includegraphics[width=.6\textwidth]{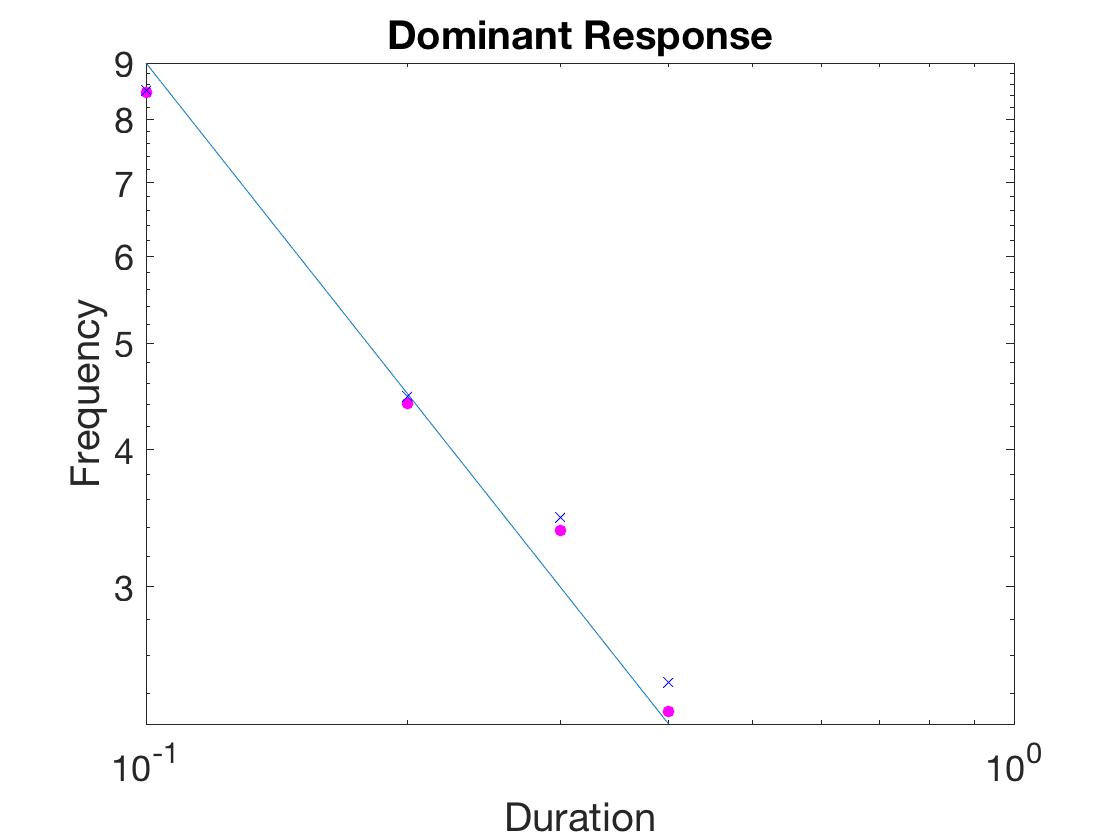}
\caption{The dominant frequency in the response of the system for various weak quenches with $a_b=a_\phi=10^{-5}$, as function of the duration $\delta t$. The magenta dots and black ex's correspond to the response of the scalar and anisotropy fields, respectively. The log-log plot was fit with a line of slope --1, and the fit seems quite reasonable with deviations likely due to quantization of normal mode frequencies. }
\label{quenches}
\end{figure}

We note that the shape of the power spectrum changes little when we vary the amplitude $a$ of the quench, as long it remains weak, \ie $a\ll1$. There is also little variation if we quench the scalar field, the metric, or both.\footnote{Of course, if we do not excite the scalar field, it stays zero at all subsequent times.} Otherwise, the response of the system in the weak regime depends mostly on the quench duration $\dt$.  Hence it is instructive to look at the shape of the spectra for different $\delta t$. In figure \ref{quenches}, we plot the dominant frequency, the location of the maximum of  the frequency distribution, as function of that duration, for a few weak quenches. We see that the dominant frequency varies roughly as $1/\delta t$, for weak quenches (of fixed amplitude). A similar trend can be seen for the width of the frequency distribution, although the data for that quantity contains more noise.

Finally let us note that the state produced in these quenches are not generic, \eg we do not find a Boltzmann distribution among all of the normal modes. Hence we might expect some form of thermalization to take place in the subsequent evolution. However, such thermalization would require the gravitational interactions to play the role of redistributing the energy amongst the various normal modes. Let us re-iterate that the weak quench excites various normal modes in the AdS soliton background and these modes evolve freely, \ie their interactions are negligible. We will, of course, see differences between the real time evolution and the ``simple" oscillations of linearized fluctuations when the quench is sufficiently strong, but the nonlinearities can also become important if the system evolves in time for a very long period.
In the weak quench regime, a simple estimate of the interaction time scale produced by the nonlinearities of the bulk equations of motion would be roughly $\tfun/a$, where $a$ is the (dimensionless) quench amplitude in eq.~\reef{proto}. It is less clear what the subsequent thermalization time scale would be, but this certainly indicates that it is impractical to look for thermalization in the weak quench regime with our numerical simulations. A better setting in which to investigate this issue will the nonlinear quenches discussed in the next section. In fact, we will see some indications that the states produced by those quenches do not thermalize even at very long times after the quench.

\vspace{10pt}
\subsection{Nonlinear Quenches} \label{nonQ}

We have seen that focusing on the power spectrum of the response, as measured by the one-point functions, is a good way to summarize the information contained in the time series. In the case of weak quenches, that power spectrum remains constant in time (at least for sufficiently short time scales --- recall the discussion in the introduction of this section). We now turn to stronger quenches\footnote{With injected energy around 40\% of the mass gap.}, where mode mixing leads to interesting dynamics as can be seen by the time dependence of the power spectrum.

From the perspective of the bulk theory, we are injecting sufficient energy to produce interesting nonlinear gravitational dynamics. From the perspective of the boundary theory, we are injecting an energy of order $N^2$ into the system, but it is still insufficient to reach the deconfined phase of the theory. Hence the number of low energy degrees of freedom (\ie glueballs) remains of order one, and so those glueball states are highly occupied. These large occupation numbers explain how we can have nontrivial dynamics or interactions of the glueballs in the classical large $N$ limit. 

A useful tool for exhibiting the change in the power spectrum as function of time is a windowed Fourier transform.\footnote{We have also used various wavelet transforms that yield similar results.} Roughly speaking, one divides the time series into a series of (possibly overlapping) windows of some fixed duration and then calculates the Fourier power spectrum in each window separately.  More specifically, we use the so-called ``Hamming" windows (\eg see \cite{hamming}), with  $25\%$ overlap between successive intervals. Varying the window size  is a tradeoff between accuracy in time or frequency, we will focus on results that do not depend on the window size, nor on the overlap between successive windows. However, we emphasize that fine grained details of the figures presented below do depend on these choices.

The nonlinear quenches start out by initially exciting the system's normal modes with a profile similar to what is exhibited in figure \ref{scalar1pt}. But due to strong nonlinear dynamics in the bulk, these modes interact and mix. Varying parameters, we find a wide range of behaviours which we only partially explore. We can characterize the behaviour by dividing the subsequent evolution into moderate and long times (compared to the fundamental period  $\tfun \simeq 2.622$, given in eq.~\reef{fun}), and we discuss them separately.

\vspace{10pt}
\subsubsection*{Moderate Times}

\begin{figure}
\centering
\includegraphics[width=.6\textwidth]{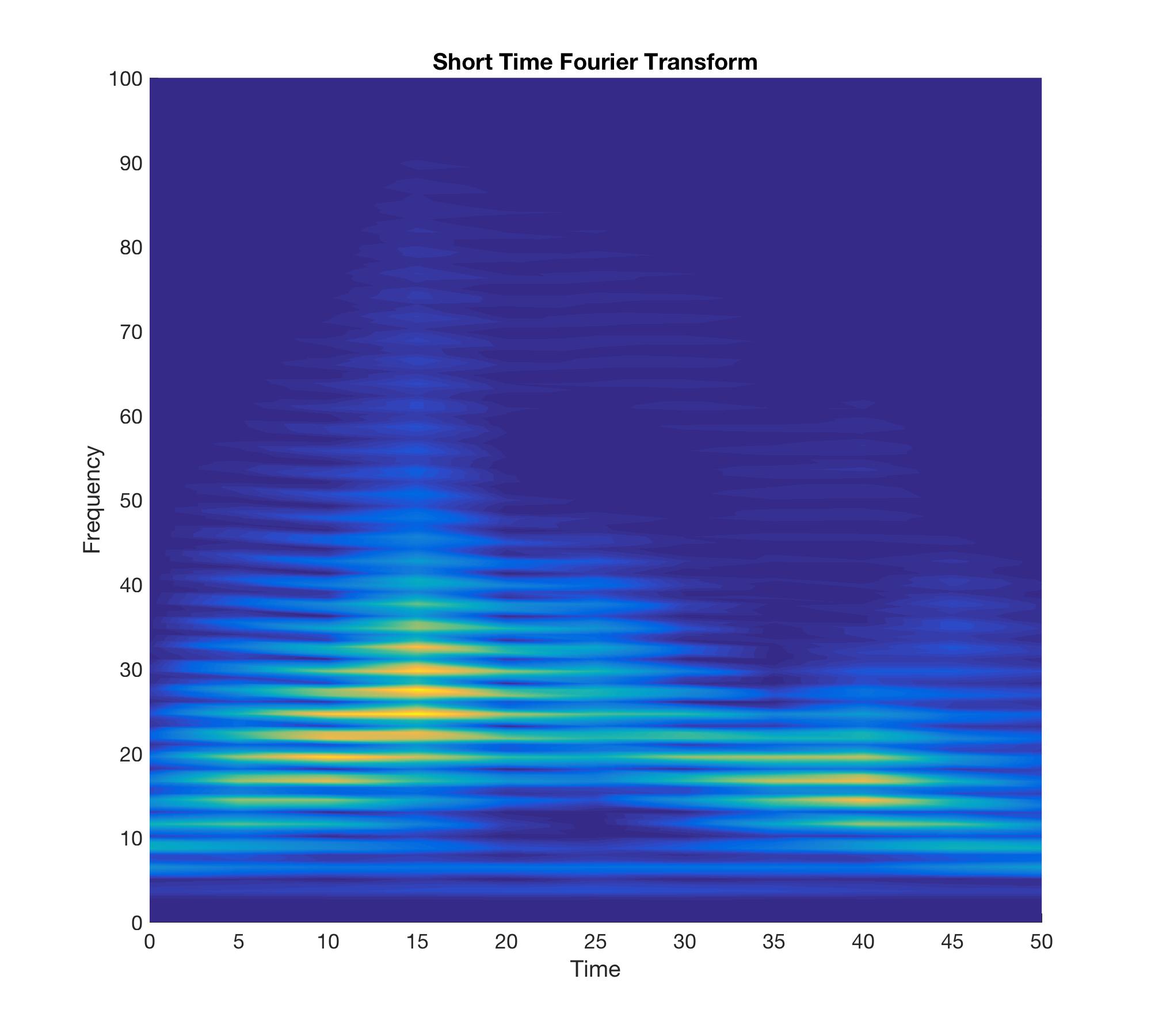}
\caption{Short time Fourier transform (spectrogram)
for the response of the
scalar field, for quench of duration $\delta t=0.5$ and amplitudes $a_b=a_\phi=0.05$.
This shows the power spectrum, colour coded, for each frequency (represented
on the vertical axis), as function of time. Blue represents low amplitudes
whereas warmer colours  represent larger power. We see that the some of the power initially contained in a few low lying resonances
is gradually transferred to higher frequency modes, representing a ``direct cascade". But this process stops and power returns to lower frequencies in later times. }
\label{drift}
\end{figure} 

At sufficiently short times, but still long compared to the fundamental bouncing time $\tfun \simeq 2.622$, we observe that the strong nonlinearities always lead to a ``direct cascade" where some of the initial power moves to higher frequency modes.  We might expect that this behaviour is related to some form of thermalization. As described above the energy is of order $N^2$ while the number of degrees of freedom is only order one, and therefore we would expect that the effective temperature would diverge in the large $N$ limit. Hence we would expect that with thermalization, the power would continue to be transferred to arbitrarily high frequencies. 
This behaviour might be taken as an operational definition of ``thermalization" in a closed system evolving unitarily --- see further discussion of thermalization below.

In figure \ref{drift}, we demonstrate the direct cascade in a particularly clear example, where that initial phase lasts for a relatively long time. We see that in this example, the power is progressively transferred to higher frequencies as function of time.\footnote{Note that the granularity of frequencies in figure \ref{drift} is due to the finite window size we use in the short-time Fourier transform; those fine details depend on the precise way that transform is performed. In particular, this discreteness is {\it unrelated} to the spectrum of normal modes in the geometry.} The details of this cascade process depend on choice the parameters, but the initial spread of power to higher frequencies seems universal in sufficiently strong quenches.

As described above, this initial phase is what we expect from thermalizing systems, which leads to a natural question of whether the system actually thermalizes after a long time evolution. While we cannot give a definite answer using a numerical simulation, which necessarily evolves for a finite duration, we see some interesting phenomena which point to a rich variety of potential long-time behaviours. We leave the complete discussion of the asymptotically long time behaviour (which would necessitate much longer time simulations than what we have attempted here) for future research, and present here a few preliminary comments.

\vspace{10pt}
\subsubsection*{Long Times and (Non)Thermalization}

Let us be more precise about the issues of thermalization. In \cite{2011PhRvL.106e0405B}, a distinction is made between {\it strong} and {\it weak} notions of thermalization. The former means that after a long time, the density matrix of subsystems is approximately thermal, whereas the latter only demands approximately thermal behaviour of the time-averaged observables. Furthermore,  \cite{2011PhRvL.106e0405B} investigates a non-integrable model where both weak and strong thermalization, as well as no thermalization at all, can be achieved depending on parameters.

An even weaker notion of thermalization was developed in \cite{Balasubramanian:2014cja,Buchel:2014xwa}, which requires the long-time averages of observables to become approximately time-independent, regardless of whether or not those averages are described by a thermal ensemble. Such a nonthermal steady state could be indicative of extra conserved quantities existing in the system. In that case, the steady state may be described by a generalized Gibbs ensemble \cite{gge1,gge2}.

\begin{figure}
\centering
\includegraphics[width=.6\textwidth]{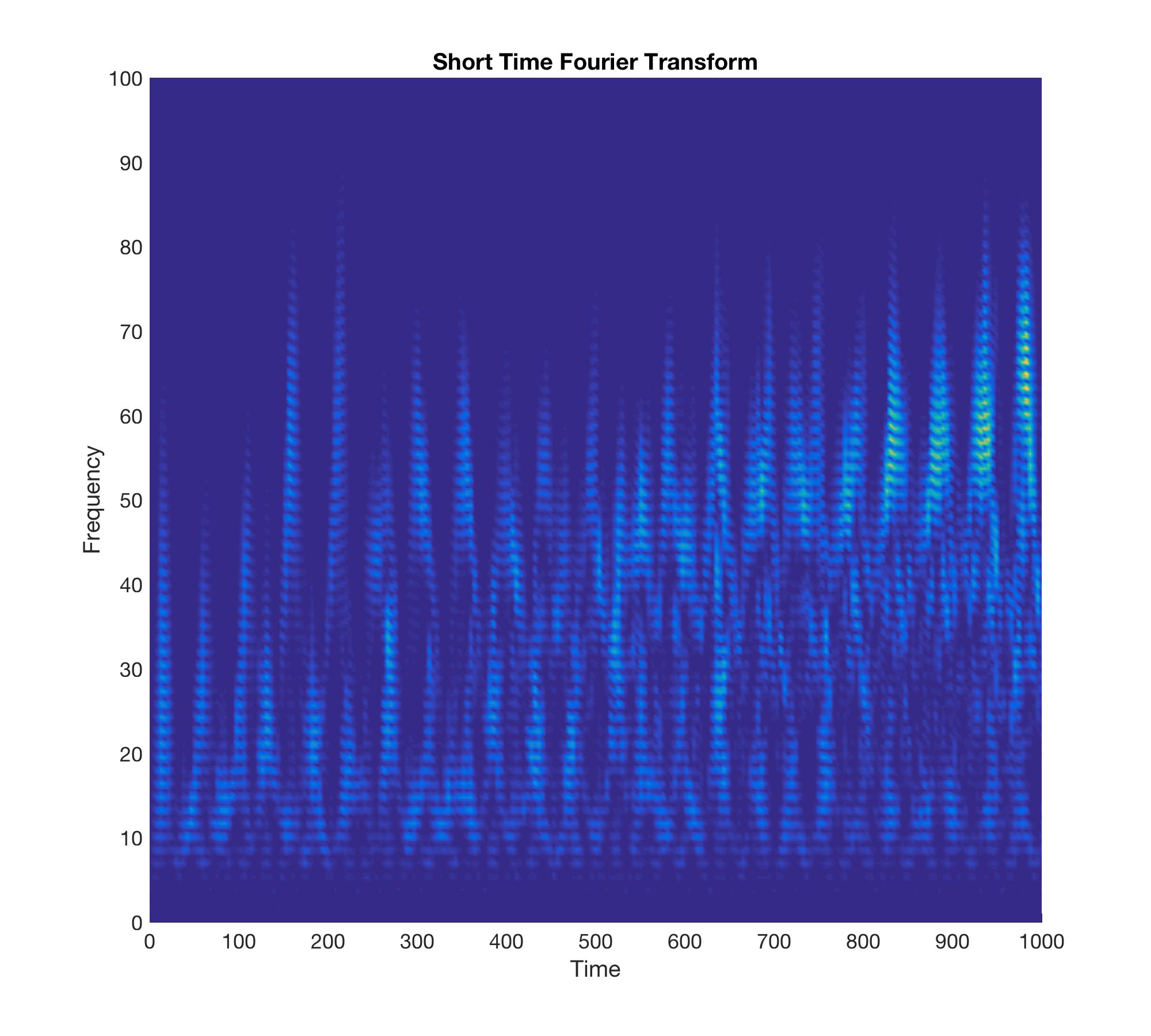}
\caption{Short time Fourier transform (spectrogram)
for the response of the
metric anisotropy, for quench of duration $\delta t=0.3$ and amplitudes $a_b=a_\phi=0.03$.
This shows the power spectrum, colour coded, for each frequency (represented
on the vertical axis), as function of time. Blue represents low amplitudes
whereas warmer colors  represent larger power. We see oscillatory behaviour in the power spectrum on long time scales. }
\label{spectro}
\end{figure}

As we have mentioned, for strong quenches in our system, the power spectrum initially drifts towards higher frequencies, as expected in a thermalizing system.  However, at later times, we inevitably see an ``inverse cascade", where the spectrum returns to lower frequencies.  
A spectrogram demonstrating such long time behaviour is exhibited in figure \ref{spectro}, where we show the subsequent time evolution of a quench similar to the one depicted in figure \ref{drift}. We see  a series of direct cascades alternating with {\it inverse} cascades, where power drifts to higher frequencies and then returns to lower frequencies. The evolution of the power spectrum on these long time scales is erratic and quasi-periodic.

We note in passing that the modulation present in the long-time behaviour found here is different from that observed in \cite{Craps:2015upq}. They also examined the time series of the one-point functions, albeit in weak and purely scalar quenches. The phenomena observed there was explained in terms of  ``beats" in driven harmonic oscillators. We emphasize however that the quasi-periodic behaviour in the power spectrum, which was found in our nonlinear quenches, is distinct from the phenomena of ``beats," where the frequency content for the system would simply be dominated by two nearby frequencies. Further, the modulation which we see is irregular, and thus is not explainable by a simple pattern of superposition of small number of oscillatory modes. We will return to the ``beat" phenomena observed in \cite{Craps:2015upq} in the next section.

The structure of alternating direct and inverse cascades is reminiscent of the holographic results of \cite{Buchel:2012uh,Balasubramanian:2014cja,Buchel:2014xwa} in the case of collapse in global AdS space.   In the so-called ``islands of stability" where black holes do not seem to form in the long time limit, and even in some cases where a black hole eventually forms at late time, the same pattern of direct and inverse cascades appears, resulting in the periodic ``sloshing" of the power spectrum back and forth between high and low frequencies. In our system, the precise period for this behaviour depends on the details of our choice of quench parameters, but it is always  much larger than that fundamental period \reef{fun}: $\tfun \simeq 2.622$.

\begin{figure}
\centering
\includegraphics[width=0.7\textwidth]{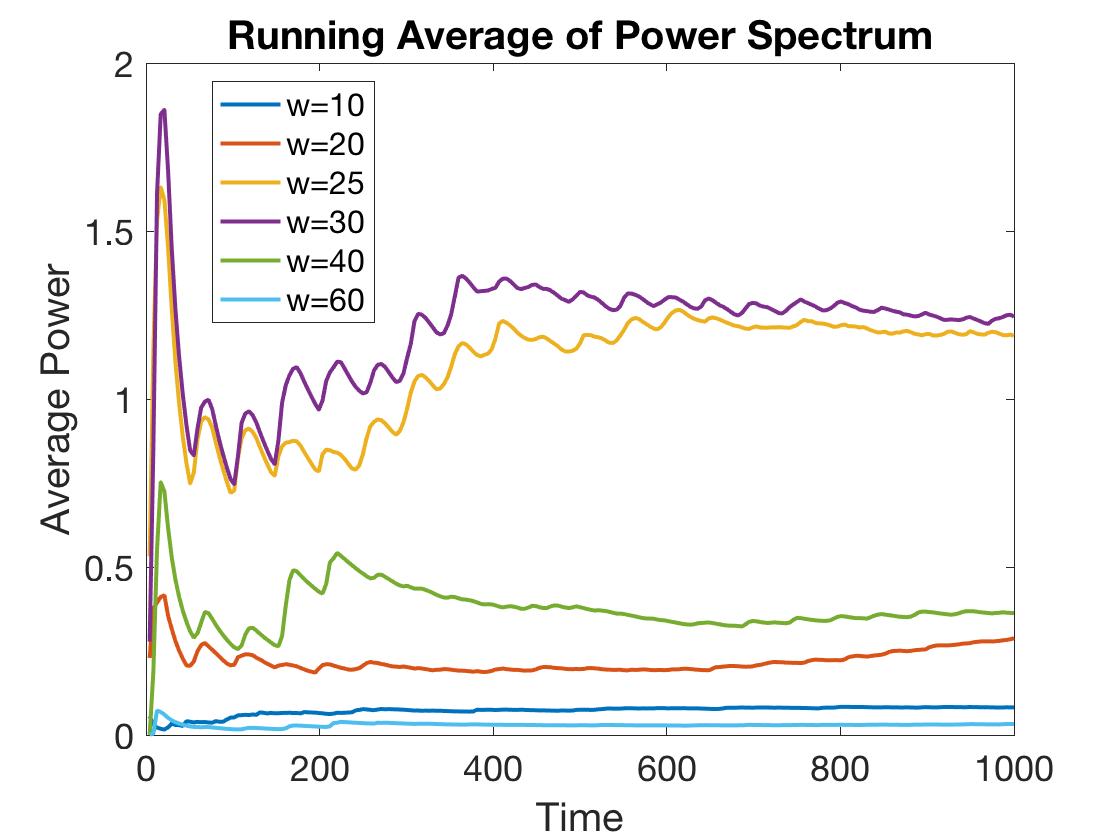}
\caption{The running average of the (unnormalized) power spectrum $\frac{1}{T} \int_0^T\! P(t,\omega)\, dt$ of the scalar response for the quench depicted in figure \ref{spectro}, for some representative frequencies. We see that in the long time limit both low and high frequencies reach a steady state, where the average power in each becomes constant. However some frequencies still show evolution and oscillatory behaviour at the longest time scales we have probed. }
\label{long}
\end{figure}

One way to examine the weak notion of thermalization, which was developed in  \cite{Balasubramanian:2014cja,Buchel:2014xwa}, is to average the power in each mode over time. These averages will saturate after long times for a system that thermalizes  in the weakest sense discussed above, \ie in the sense that long-time averages are time-independent, but they do not necessarily take the naively thermal values. In figure \ref{long}, we depict the long-time averages of the power in a few representative frequencies for the quench whose power spectrum is depicted in figure \ref{spectro}. We find that for longer time scales, the average power spectrum stabilizes for some frequencies, but not all. Note that even for the frequencies that look stabilized in figure \ref{long}, some oscillations remain, even in the long-time limit. The amplitude of these oscillations being relatively small may simply be an artifact of the fact that we are averaging over very long times. 

We can conclude that at least in the time regime that we were able to probe here, \ie roughly $400\,\tfun$ after the quench, the system does not thermalize even in the weakest sense for our nonlinear quenches. While we cannot be sure of the results for asymptotically long times, the quasi-periodic structure we observed, \eg in figure \ref{spectro}, where tendency towards thermalization stops, suggests that the system will not thermalize even on much longer time scales.

\vspace{20pt}
\section{Nonlocal Probes}\label{nonloco}

In \cite{pasquale}, global quenches of a one-dimensional spin chain in a confining phase were also examined with various nonlocal probes, \eg entanglement entropy and two-point correlation functions, and the response of these probes exhibited a rich and interesting structure. Hence we now turn to examining analogous probes in our holographic quenches.  

\vspace{10pt}
\subsection{Entanglement Entropy} \label{HEEr}

One such probe considered in \cite{pasquale} was the entanglement entropy of half space. There the entanglement entropy was found to exhibit an interesting oscillatory behaviour, where the corresponding power spectra showed peaks at frequencies related to the energies of many of the bound states of the system. This was dubbed {\it entanglement spectroscopy}, and is somewhat analogous to discovering the glueball spectrum using the one-point response function, as shown in figure \ref{scalar1pt}. 

Here, we will begin by discussing the entanglement entropy of a strip of width $\ell$, as function of time, but we quickly focus on the simpler case of the entanglement of half space, for reasons described below. To apply the Ryu-Takayanagi formula \cite{RT1,RT2}, or more precisely its covariant extension \cite{HRT}, we find an extremal bulk surface which ends on entangling surface corresponding to the specified boundary strip at fixed time. Below, we start the discussion by describing the calculation and reviewing the results for the static case.

\vspace{10pt}
\subsubsection{Initial State}

\begin{figure}
\centering
\begin{subfigure}{0.45\textwidth}
\includegraphics[width=\textwidth]{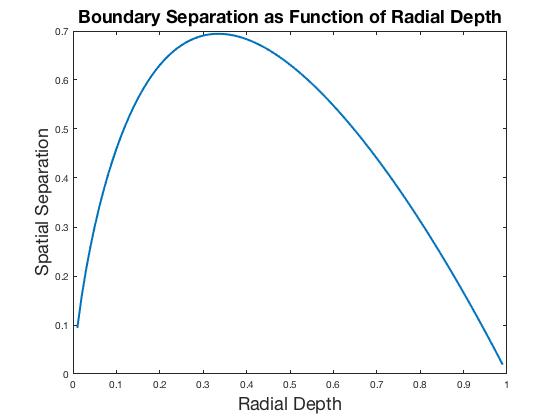}
\end{subfigure}
\begin{subfigure}{0.45\textwidth}
\includegraphics[width=\textwidth]{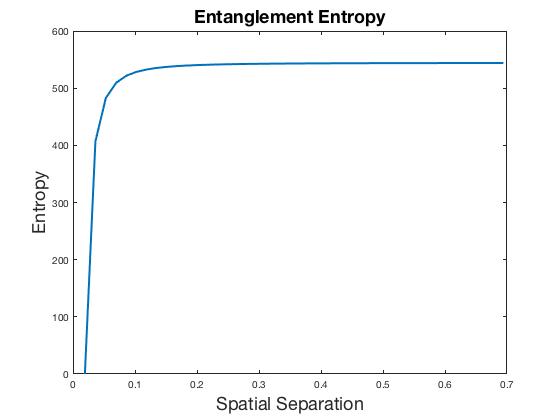}
\end{subfigure}
\caption{Entanglement entropy for the AdS soliton. In the left panel, we exhibit the boundary separation as function of radial depth: each separation can receive contributions from two possible connected minimal surfaces. However, the deeper surfaces have larger action and so do not contribute to the entanglement entropy. In the right panel, we show the entanglement entropy as function of the width of boundary strip, for the connected surfaces. For wider strips, the entanglement entropy comes from disconnected surfaces, and hence it becomes a constant independent of the spatial separation. }
\label{EEloc}
\end{figure}

In the static soliton background \reef{soliton}, narrow strips with small width $\ell$ have an entanglement entropy which is the (regularized) area of the  connected minimal surface staying close to the boundary. As we increase the width of the boundary strip, those surfaces go deeper and deeper into the bulk. In the left panel of figure \ref{EEloc}, we plot the relation between the bulk depth of the connected extremal surfaces and the width of the corresponding boundary strip. In contrast to geodesics which will be discussed in the next section, we see that relation is not monotonic. For each boundary width $\ell$, there are two connected bulk surfaces ending on the appropriate boundary strip. Further, for sufficiently wide strips, no such connected extremal surface exists.

This behaviour led the authors of \cite{Klebanov:2007ws} to identify a phase transition in the holographic entanglement entropy. For narrow strips, the entanglement entropy is determined by the shallower connected surface (since it has the minimal area) in the left panel of figure \ref{EEloc}. The corresponding entanglement entropy for these narrow strips is shown in the right panel in figure \ref{EEloc}, as a function of $\ell$. 

However, for sufficiently wide strips, the holographic entanglement entropy is given by two disconnected bulk surfaces, which simply fall straight into the bulk and close off at the IR cap-off point \cite{Klebanov:2007ws}. Hence, the entanglement entropy is independent of $\ell$ in this regime. For later purposes, we observe that the area of one such extremal surface yields the entanglement entropy of half of a constant time slice in the boundary theory.

Let us note that the connected surfaces, which reach deep into the IR, play no role in determining the entanglement entropy for either the narrow or wide strips. Hence  the connected surfaces, which are relevant for relatively small $\ell$, do not probe the deep IR of the bulk. While some time dependence is seen in the area of such surfaces, the probe with the most interesting time-dependence is that of the disconnected surfaces which go deep into the bulk geometry. For this reason, we restrict our discussion below to the entropy of half-space, given by the regularized area\footnote{In the following, we regulate the areas by subtracting the area of the corresponding extremal surface in the static background (with the same compactification scale $L_c$).} of a single disconnected surface.

\vspace{10pt}
\subsubsection{Time Dependence}

To calculate the entanglement entropy of half space, as a function of time, we begin with the background geometry \reef{metric} and seek extremal surfaces starting at $r=0$ at some $t=t_0$. Then integrating out through this dynamic geometry towards the boundary at $r=1$, we find that the extremal surface typically reaches the boundary at some later time $t_b \geq t_0$. In the various plots when we display the regularized area, it is as function of this boundary time $t_b$.

We can broadly classify the response according to the type of quench that is performed. In particular, there are three distinct behaviours according to whether it is a weak scalar quench (in which only the scalar field is sourced) or a weak gravitational quench (in which the amplitude of the metric source is larger or equal to that of the scalar) or nonlinear quenches (with either scalar or metric sources). Of course, there are no sharp boundaries separating these three regimes and the corresponding responses, but we give an illustrative example of each of the three cases below. Following the description of those cases, we attempt to interpret the differences in terms of different mechanisms of smoothing the nonlocal observables, relative to the one-point function described above, in the next subsection.

\begin{figure}[!h]
\centering
\begin{subfigure}{0.7\textwidth}
\centering
\includegraphics[width=.8 \textwidth]{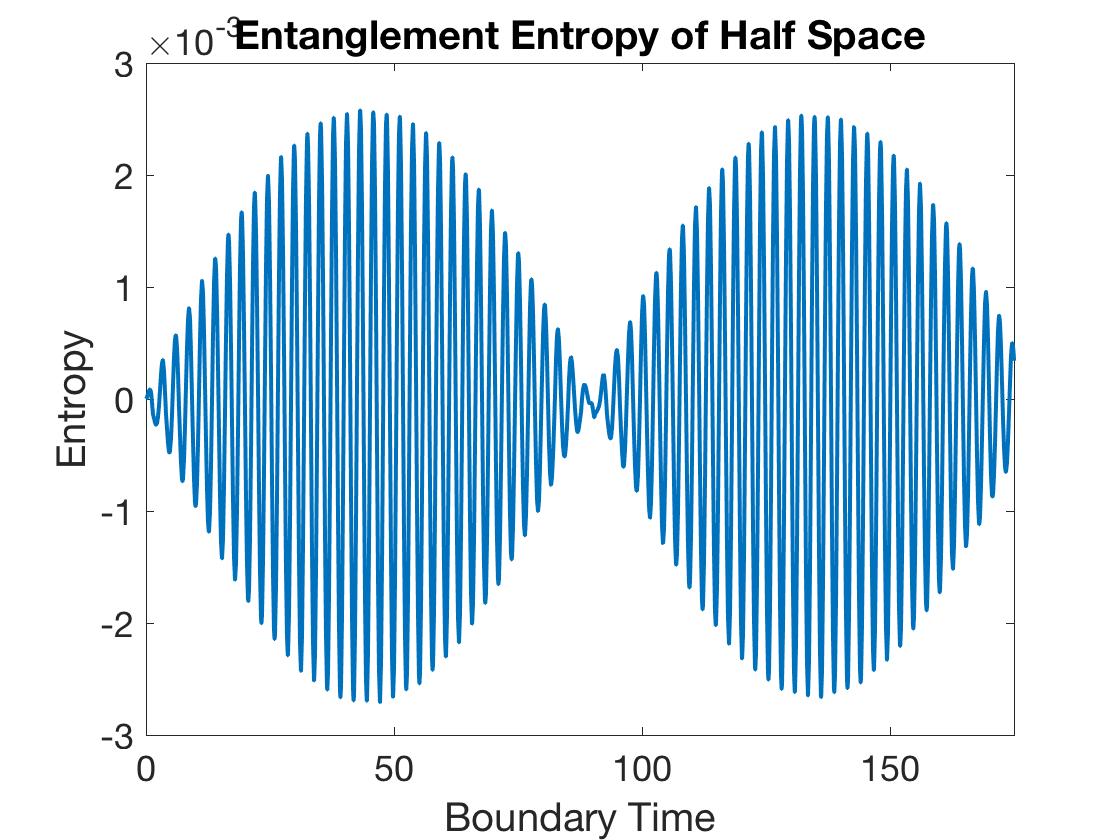}
\end{subfigure}
\centering
\begin{subfigure}{0.42\textwidth}
\includegraphics[width=1.2 \textwidth]{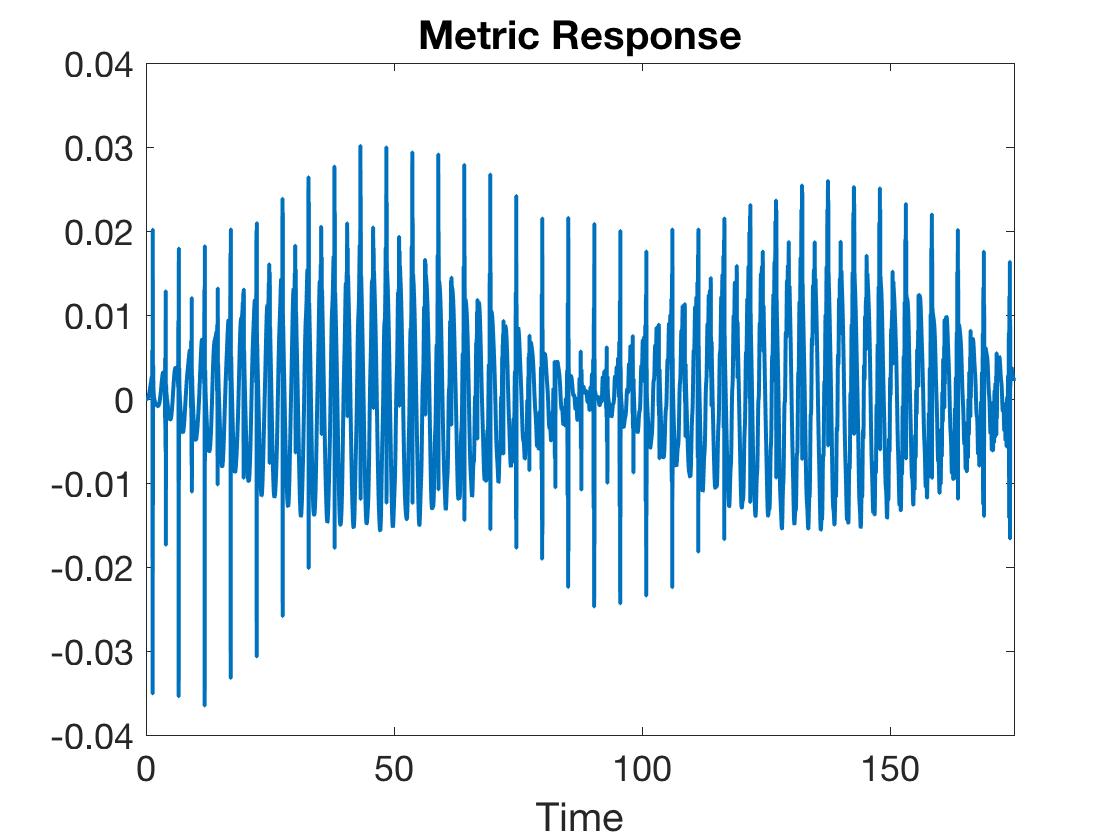}
\end{subfigure}
\begin{subfigure}{0.42\textwidth}
\includegraphics[width=1.2 \textwidth]{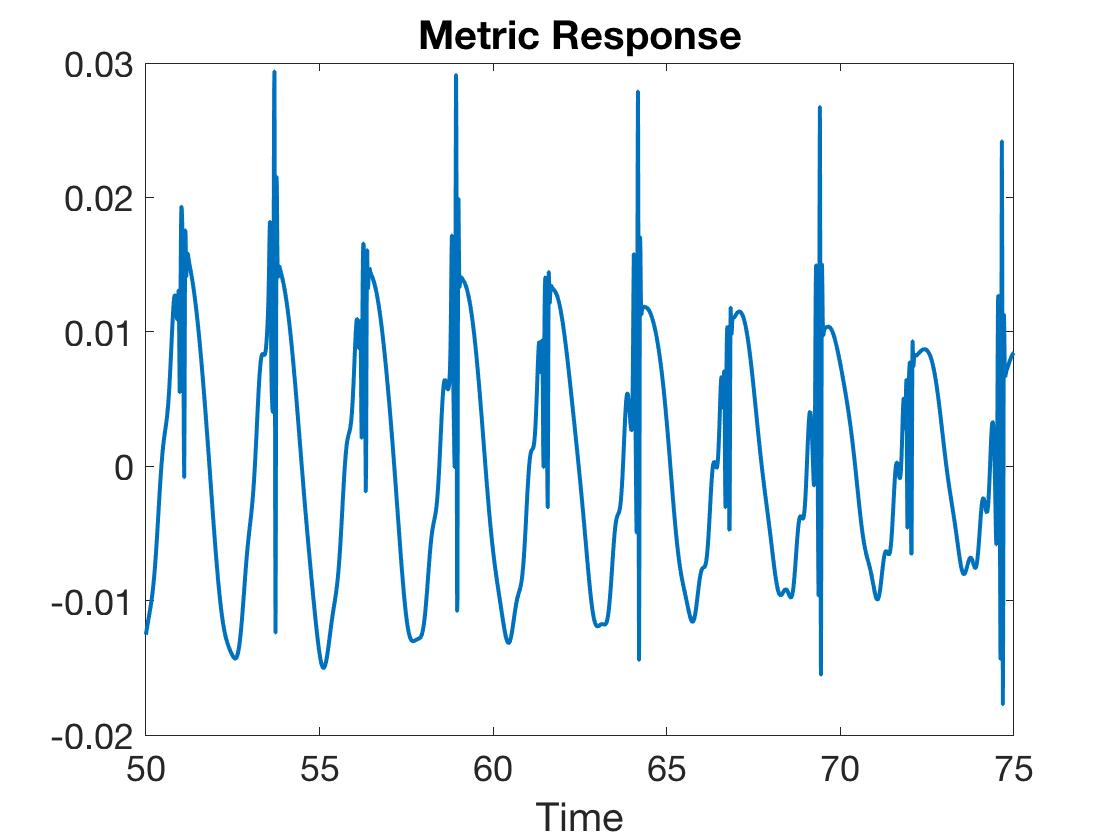}
\end{subfigure}
\caption{In the top panel, we show the entanglement entropy of half-space for a weak scalar quench with $a_\phi=10^{-4}$, $a_b=0$ and $\delta t=0.05$. We see distinct ``beats" in the time series, where the oscillatory behaviour with $\omega \sim \wfun$ is modulated on a longer time scale. In the bottom panels, we see a somewhat more irregular pattern in the metric one-point response for the same quench. The bottom right panel, which zooms in on a short time interval, emphasizes that this response consists of a combination of slow oscillations with $\omega\sim\wfun$ and series of pulses, similar to those shown in figure \ref{scalarQ}. The bottom left panel shows that the amplitude of the one-point response also shows some modulation on longer time scales, similar to the entanglement entropy. }
\label{timeEE}
\end{figure}
First, in the top panel of figure \ref{timeEE}, we show the time-series for the entanglement entropy of half-space for a weak scalar quench.
We see that the response here is relatively smooth and regular. It is entirely dominated by oscillations with $\omega\sim\wfun$, however, these oscillations are in turn modulated on longer time-scales. This ``beat'' phenomena is similar to what was observed by \cite{Craps:2015upq} in the one-point functions and we return to a discussion of this feature below.   We contrast the smooth behaviour seen for the entanglement entropy with the response seen in the one-point function of the anisotropy field for the same quench, which is shown in the bottom two panels. This one-point function exhibits oscillations with $\omega\sim\wfun$ but it also has  structure on shorter time-scales, with the pulses characteristic of the bounces illustrated with figure \ref{scalarQ}. However, the bottom left panel also shows that the amplitude has some modulation on longer time scales, similar to the entanglement entropy. In any event, the short-time structure indicates that the one-point function has a much richer power spectrum than the entanglement entropy, even though both are different measurements of the metric response.
\begin{figure} 
\centering
\hspace{-70pt}
\begin{subfigure}{0.48\textwidth}
\includegraphics[width=1.0 \textwidth]{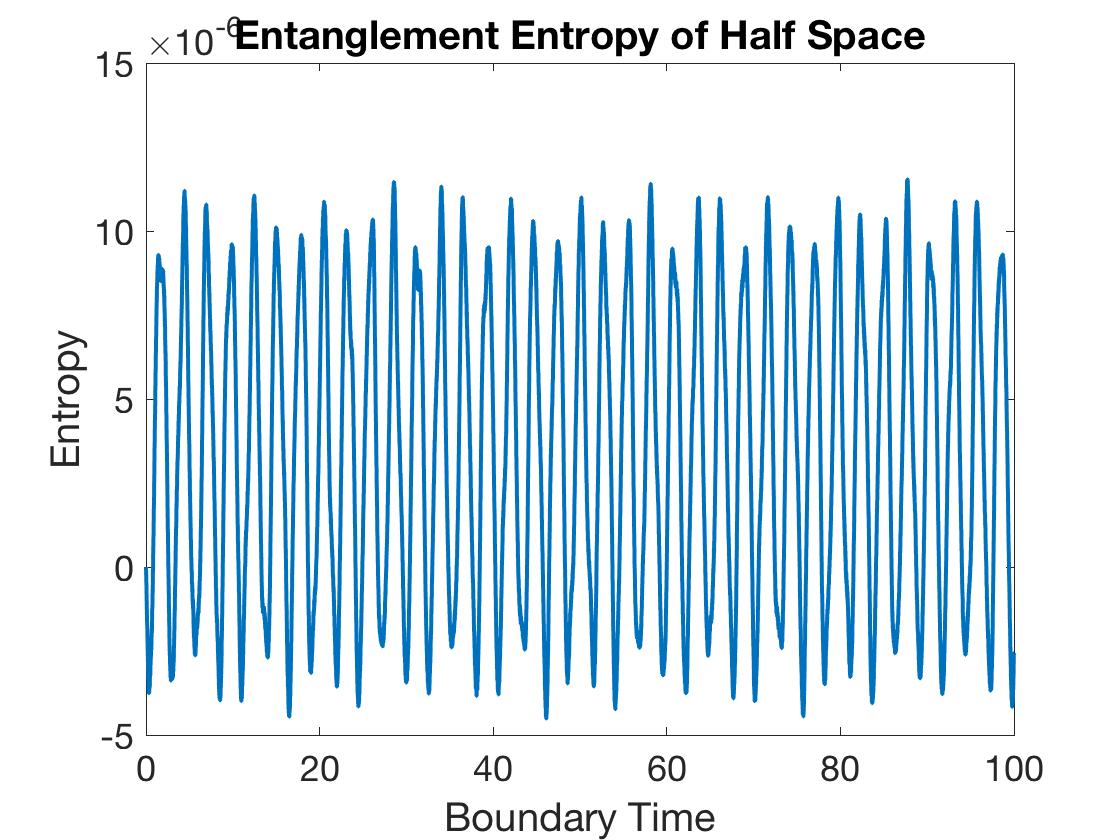}
\end{subfigure}
\begin{subfigure}{0.4\textwidth}
\includegraphics[width=1.4 \textwidth]{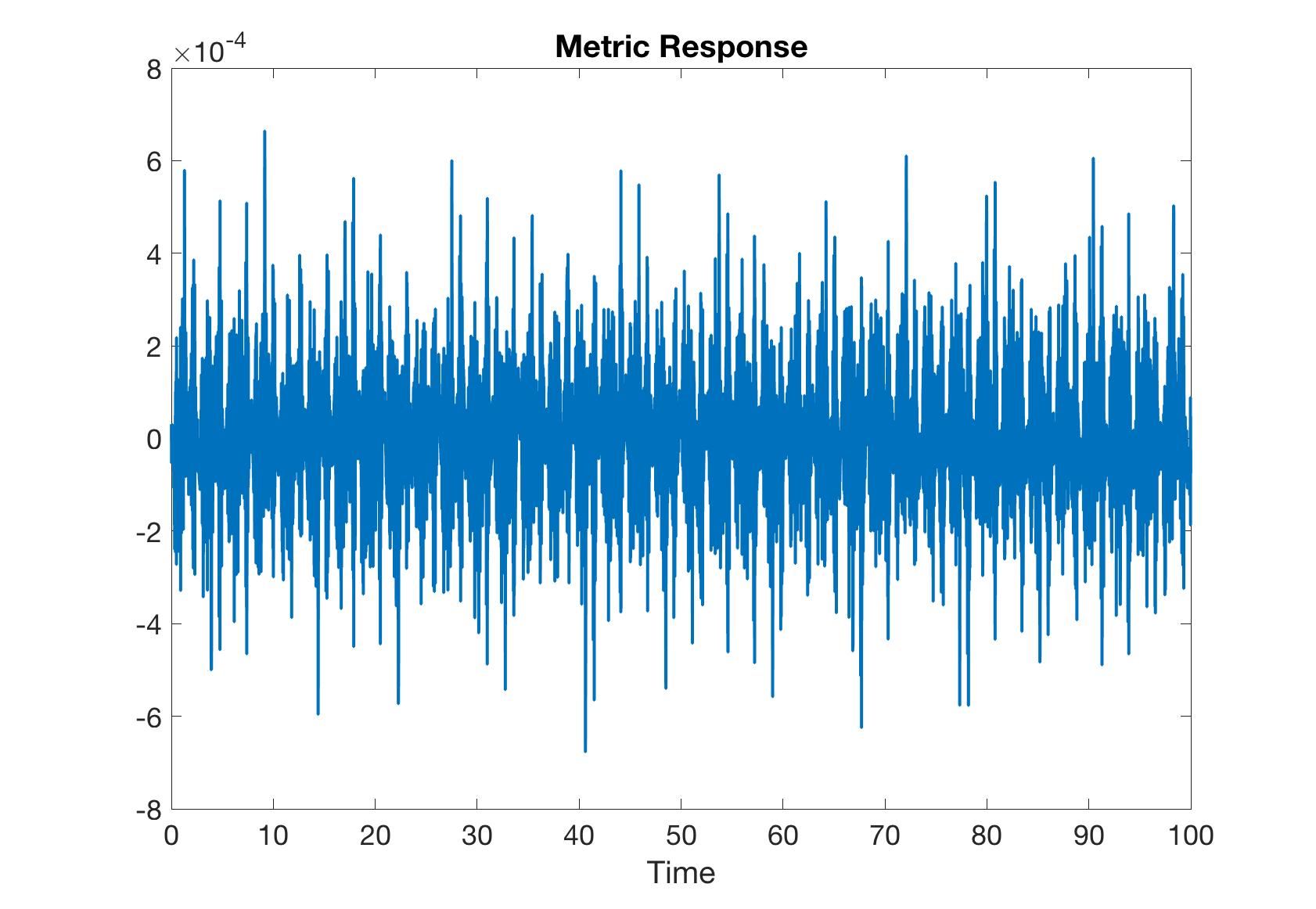}
\end{subfigure}
\caption{In the left panel, we show the entanglement entropy of half-space for a weak gravitational quench with $a_b=a_\phi=10^{-5}$ and $\delta t=0.3$. In the right panel we see the  response in the metric one-point function for the same quench. It is clear that the entanglement entropy is smoothed out compared to the one-point response, but also that it does not exhibit beats. }
\label{timeEEw2}
\end{figure}

In figure \ref{timeEEw2}, we turn to a weak gravitational quench, where  both the scalar and anisotropy field were sourced. The left panel shows the time-series for the entanglement entropy of half-space and we see that this response is somewhat more irregular than found for the weak scalar quench. However, we can also compare this response to the one-point function of the anisotropy field for the same quench, shown in the right panel. The latter response clearly shows a great deal of structure at high frequencies and so we can conclude that the response in the entanglement entropy is still relatively smooth, in comparison. That is, the short-time structure in the one-point function will produce in a much richer power spectrum than for the entanglement entropy. 

\begin{figure}
\centering
\begin{subfigure}{0.48\textwidth}
\includegraphics[width=1.0 \textwidth]{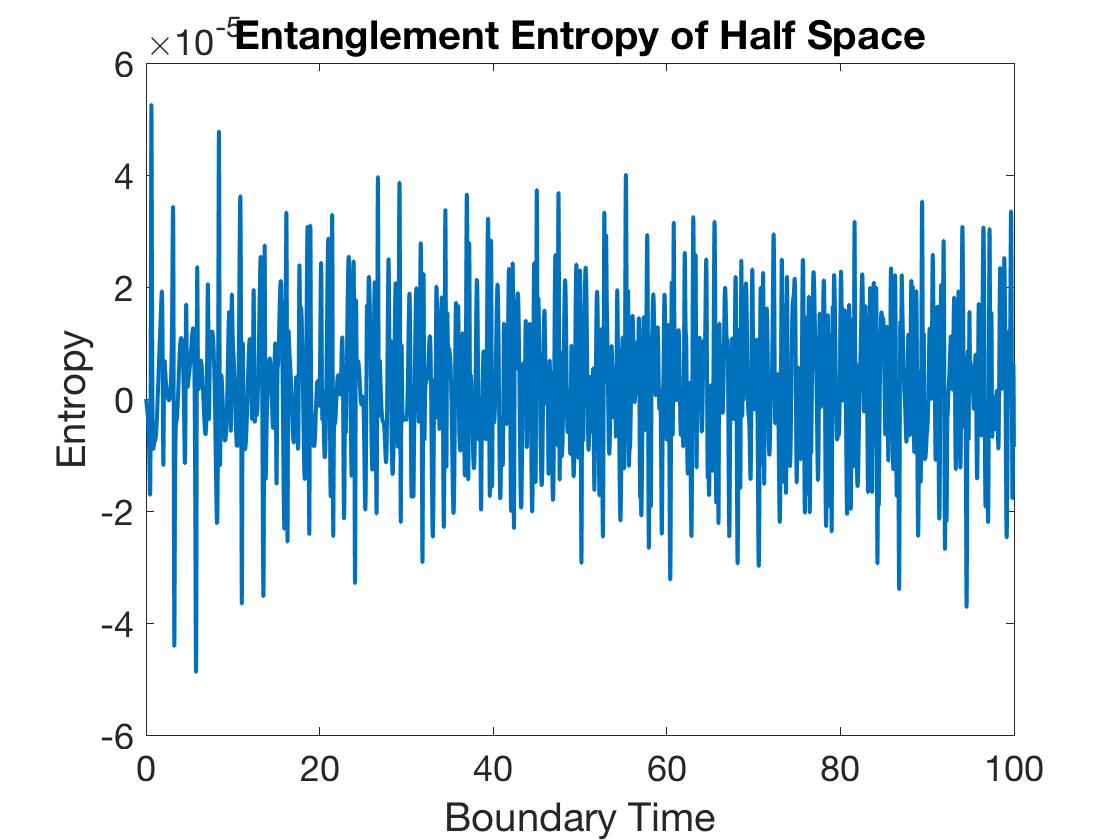}
\end{subfigure}
\begin{subfigure}{0.48\textwidth}
\includegraphics[width=1 \textwidth]{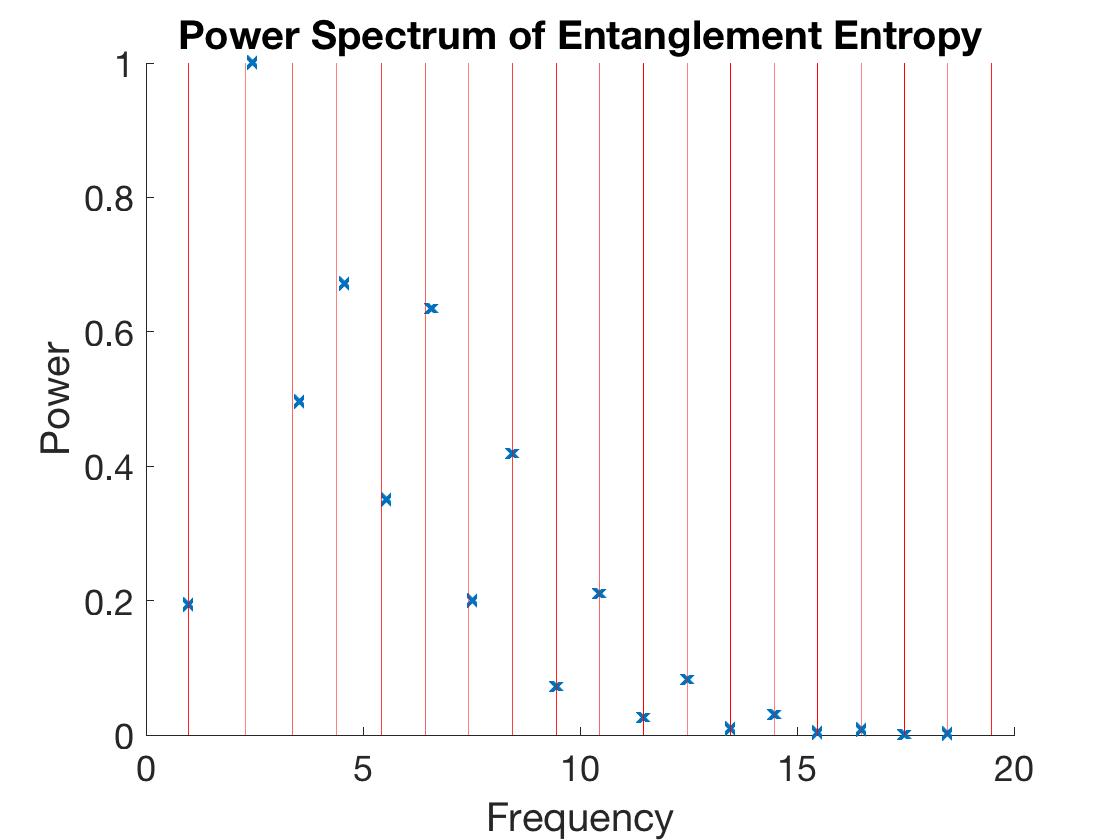}

\end{subfigure}
\caption{ The entanglement entropy of half-space for a weak gravitational quench with $a_b=a_\phi=10^{-5}$ and $\delta t=0.1$. In this case the filtering described in the text is ineffective, and the time-series of the entanglement entropy, shown on the left, shows a great deal of short-time structure. Indeed, the power spectrum, shown on the right, is focused on  the spectrum of normal modes of the metric excitation, similar to figure \ref{scalar1pt}. }
\label{EEcomp}
\end{figure}

So far we have shown examples where the entanglement entropy for weak quenches is smoothed out, as compared to the one-point functions. We return to describing different mechanisms for this behaviour below. However, we must add that in the weak gravitational quenches, the smoothing becomes ineffective for certain choices of quench parameters, \ie when the duration $\dt$ is sufficiently short, so as to excite mostly high frequency normal modes. In figure \ref{EEcomp}, we exhibit the time-series and power spectrum of the entanglement entropy for an example of such a quench. We see that the signal has complicated short-time features, and its power spectrum reveals a wide range of the normal modes of the background geometry. The latter is similar to what we found in studying the one-point functions of the weak quenches in section \ref{weakone}. It also is also an illustration of {\it entanglement spectroscopy} as suggested in \cite{pasquale}, and hence we can confirm that this spectroscopy works at least for certain classes of higher-dimensional quenches as well. We discuss the power spectra of the time-series exhibited in figure \ref{EEcomp} in more detail below.

\begin{figure}
\centering
\hspace{-30pt}
\begin{subfigure}{0.48\textwidth}
\includegraphics[width=1.0 \textwidth]{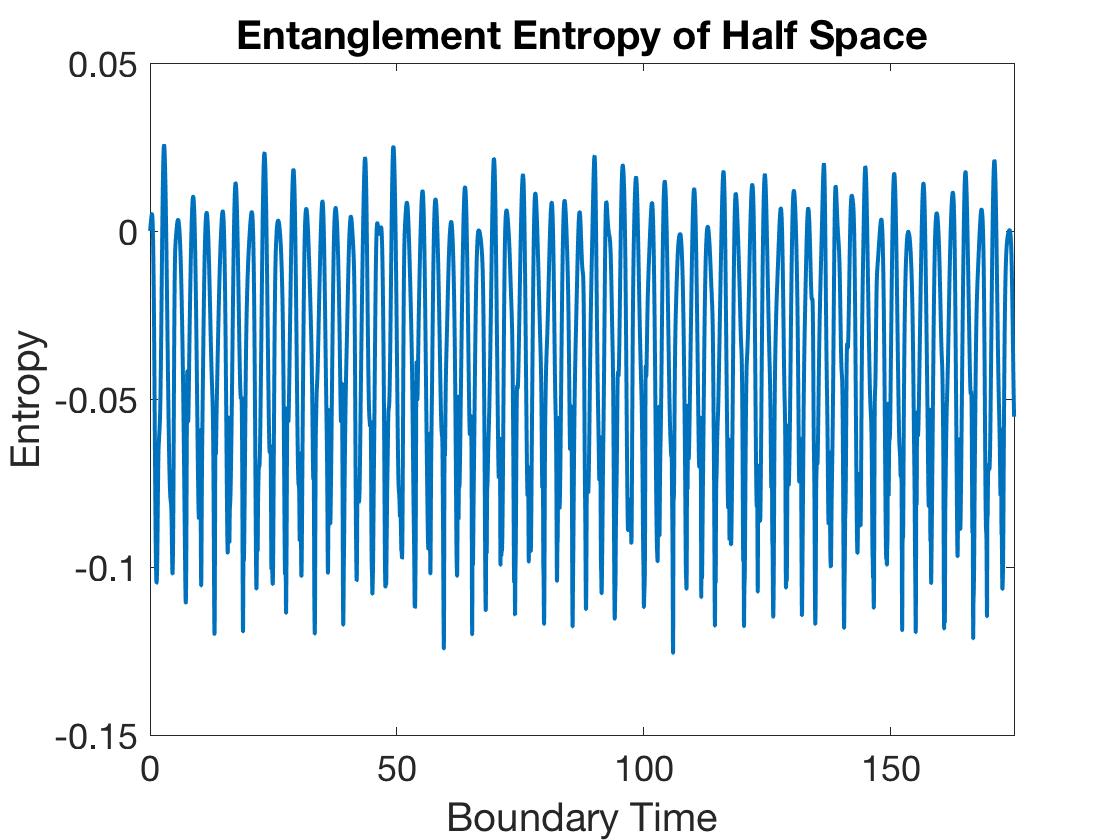}
\end{subfigure}
\begin{subfigure}{0.48\textwidth}
\hspace{-30pt}
\includegraphics[width=1.2 \textwidth]{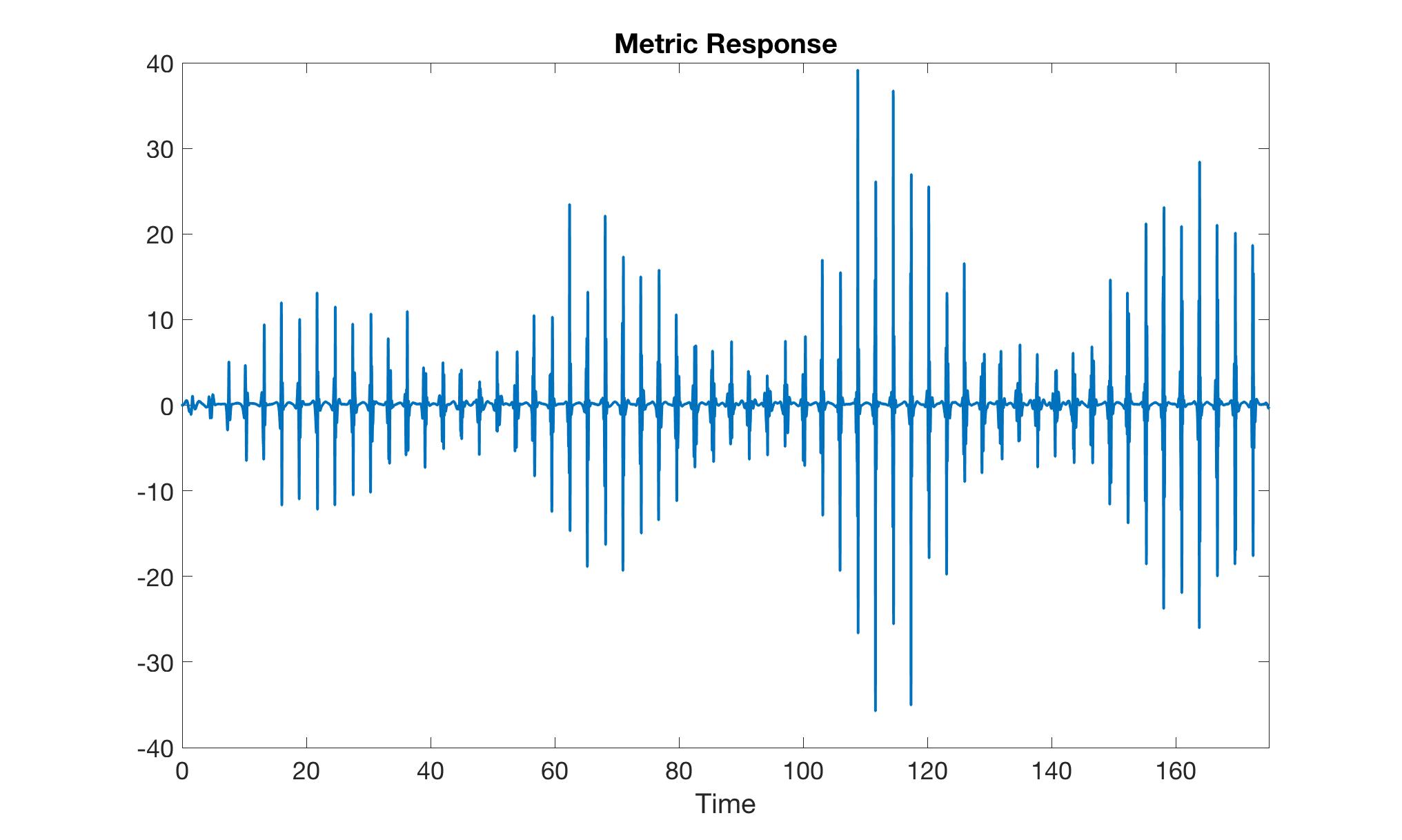}
\end{subfigure}
\caption{In the left panel, we show the entanglement entropy of half-space for a nonlinear gravitational quench with $a_b=0.05$, $a_\phi=0.05$ and $\delta t=0.5$. In the right panel we see a somewhat more erratic response in the metric one-point for the same quench. Again, we emphasize that the response in the right panel consists of a series of pulses, similar to those shown in left panel of figure \ref{scalarQ}. The scalar one-point response for this quench is similar to that of the metric in the right panel. }
\label{timeEE11}
\end{figure}

Finally, the behaviour of the entanglement entropy is distinct for nonlinear quenches, as shown in figure \ref{timeEE11}. The time variation of entanglement entropy is again dominated by smooth oscillations with $\omega\sim\wfun$, as for the weak scalar quenches, however, we see that the ``beat" modulation on longer time scales is again absent. Further, this response is again much smoother than the response in the one-point functions, which shows the usual pulse structure. The amplitude of the pulses in the one-point response also shows some modulation on longer time scales, at least in the initial period of time shown in the right panel. However,  we observe that the period of this modulation is quite different from that of the beats in figure \ref{timeEE}, with a scalar quench. On the other hand, this period does match (at least, roughly) the time required for the direct cascade and subsequent inverse cascade found in the power spectrum shown in figure \ref{drift}, for the same quench. Hence the long-time modulation seems to have a different origin here than in weak scalar quenches. We emphasize that  the amplitude of the oscillations in the entanglement entropy is remarkably stable here. Comparing to figure \reef{drift} for the scalar one-point function, we note that there is very little variation and, in fact, very little power in the vicinity of $\omega\sim\wfun$. 

These three examples are characteristic of the response of the entanglement entropy in the three distinct regimes. In particular, the beat phenomenon is a general feature of weak scalar quenches. As the amplitude of the scalar quenches is increased, the beats become less prominent and essentially disappear in the nonlinear regime. In the nonlinear regime for either scalar or gravitational quenches, we found that the response is again dominated by smooth oscillations, but with frequencies unrelated to $\wfun$. However we must say that our exploration of the nonlinear regime was necessarily incomplete, and other features may also emerge in different corners of the parameter space for nonlinear quenches.
In contrast, for weak gravitational quenches, the response of the entanglement entropy may be  smoother than the corresponding one-point functions but it still shows somewhat more structure than found in either the weak scalar or nonlinear quenches.
In particular, the precise character of the response depends on the details of the quench, \eg the duration $\delta t$. Further, when parameters are tuned in a weak gravitational quench to avoid an efficient smoothing of the entanglement entropy signal, one may recover the spectrum of the theory, \ie perform {\it entanglement spectroscopy}  along the lines of \cite{pasquale}.

\vspace{10pt}
\subsubsection{Smoothing Mechanisms} \label{smoother}

We have examined the response of the holographic entanglement entropy for a variety of different global quenches. A key feature in all of these was a smoothing of the time-series results in comparison to that for the one-point functions, studied in section \ref{onepo}. That is, short time features or high frequency contributions are filtered out in the response of the entanglement entropy. As shown by the different appearance of the responses, the details of this filtering differs in the three different types of quenches. As we explain below, there are three different mechanisms at play in the filtering and in each case, the interplay of different mechanisms produces the qualitatively different results seen above. The three filtering mechanisms are: {\it i}) radial integration; {\it ii}) near resonant driving; and {\it iii}) nonlinear cascade. 

\vspace{10pt}
\subsubsection*{{\it i}) Radial Integration:}

An important feature distinguishing the holographic entanglement entropy and other nonlocal boundary observables from the one-point functions is the nonlocal structure of the dual holographic construction in the bulk. That is, evaluating the holographic entanglement entropy requires performing a integral over all bulk radii from the IR to UV. 

To explain the role of this radial integration in the filtering of the response, we begin by adapting the calculation of the holographic entanglement entropy to the post-quench coordinates \reef{post9}, which yields
\beq
S_\mt{EE}=\frac1{4G}\int d^3\sigma \sqrt{h}
=\frac{L^3L_c\,L_{x_1}}{4G}\int_{-1}^1\! dx\ 
\frac1{(1+x)^2}\left(\frac{\hat B}{\hat A}\right)^{1/2}
\label{out1a}
\eeq
where $h_{ab}$ is the induced metric on the extremal bulk surface. Implicitly, we are examining the entanglement entropy for the half-space defined by $x_2>0$ (and $t=0$), and so $\sigma^a=\lbrace x,x_1,\theta\rbrace$. We have also regulated the $x_1$ integral by introducing the transverse length $L_{x_1}$ and we have used $\Delta\theta=2L_c$. Further recall that in these coordinates, the asymptotic boundary corresponds to $x=-1$ while the IR cap-off point is at $x=+1$.

Now for simplicity, let us examine the response of $S_\mt{EE}$ to a weak quench. The latter produces small metric perturbations in the bulk, which satisfy the linearized equations of motion. That is, the quench will deform $\hat B$ and $\hat A$ with normal mode excitations in $\beta$ and $\alpha$ --- see eq.~\reef{eq:redef}. We find that the leading order response of the entanglement entropy is then simply given by\footnote{We should also add that eq.~\reef{out1b} ignores the variation in the profile of the extremal surface. This is valid when calculating the leading contribution in a perturbative expansion of $\delta S_\mt{EE}$ \cite{Blanco:2013joa}.} 
\beq
\delta S_\mt{EE}(t)=\frac{\pi L^3L_{x_1}}{32G}\int_{-1}^1\! dx\,
\left(\beta(t,x) + \alpha(t,x)\right)\,.
\label{out1b}
\eeq

Now again, working to linear order in the bulk excitations, the integrand above can be expanded terms of the metric normal modes. The normal modes of $\beta$ are determined by solving the linearized versions of eqs.~\reef{evo1} and \reef{evo2}, which can be written as:
\begin{equation}\label{linbeta}
\frac2{(1+x)^2}\,\frac{d\ }{dx}\Big[(1+x)^3(1-x)(3+x)\beta^{\prime}\Big]-\frac{8(1-x^2)(3+x)(5-x)(7+x)}{\left(12-(1+x)^2\right)^2}\beta=-\omega^2\beta\;,
\end{equation}
where we have assumed a product form $\beta(x,t)=e^{-i\omega t}\,\beta(x)$.  
Therefore this defines a Sturm-Liouville system, whose inner product has weight $(1+x)^2$.\footnote{That is, the normal modes are orthogonal with the inner product: $\int_{-1}^1dx\,(1+x)^2\beta_m\beta_n \sim \delta_{mn}$. \label{footy789}} For each normal mode, this yields the radial profile $\beta_n(x)$ and the frequency $\omega_n$, given in table \ref{table33}. For simplicity, let us assume that each of these mode profiles is normalized to be one at the asymptotic boundary, \ie $\beta_n(x=-1)=+1$. Then if we expand this anisotropy field after a global quench as $\beta(t,x)=\sum b_n\, e^{-i\omega_n t}\,\beta_n(x) $, the corresponding one-point function is proportional to
\beq
\beta(t,x=-1)=\sum b_n\, e^{-i\omega_n t}\,.
\label{desk1}
\eeq

Now to evaluate eq.~\reef{out1b}, we must also consider the excitations of $\alpha$ but it turns out that these are not independent. Rather given a normal mode of $\beta$, the corresponding profile for $\alpha$ would be determined by the linearized versions of the constraint in eq.~\reef{eq:constraintalpha} and the evolution equation in eq.~\reef{eq:evoalpha}, as well as the boundary condition \reef{truck}. With this linearized analysis, the time dependence is given by the same normal mode frequency $\omega_n$ and the radial profile is given by
\beq
\alpha_n(x)=\frac{4(1+x)^2}{12-(1+x)^2}\,\beta_n(x)\,.
\label{aconstraint}
\eeq
The integrand in eq.~\reef{out1b} then has the normal mode expansion 
\beq
\beta(t,x) + \alpha(t,x)=\left[1+\frac{4(1+x)^2}{12-(1+x)^2}\right]\,\sum b_n\, e^{-i\omega_n t}\, \beta_n(x)\,.
\label{turk9}
\eeq
Therefore in the response $\delta S_\mt{EE}$, each of the normal modes $\beta_n(x)$ is integrated against a particular profile in the bulk. We can define the coefficients
\beq
s_n =\int_{-1}^1\! dx\ \left[1+\frac{4(1+x)^2}{12-(1+x)^2}\right]\,\beta_n(x)
\label{turk8}
\eeq
as a measure of the relative sensitivity of $\delta S_\mt{EE}$ to each of the normal modes. 
\begin{table}[htbp]
\centering
 \begin{tabular}{c|| c| c|c| c|c| c|c|c|c} 
 $n$ & 1  & 2& 3& 4& 5 & 6 & 7 & 8&9 \\ [0.5ex] 
 \hline
 $s_n\times 10^2$ &293.19& 52.56& 24.57& 14.40& 9.490& 6.736& 5.033& 3.904& 3.118\\ [0.6ex] 
 \hline
 ${s_n}/{s_1}\times 10^2$ & 100 & 17.93& 8.380 &  4.910& 3.237 &  2.298 & 1.717 &  1.332 & 1.063
\end{tabular}
\caption{Coefficients defined with the radial integration in eq.~\reef{turk8} and which measure the relative sensitivity of $\delta S_\mt{EE}$ to each of the normal modes.
We also show the ratio $s_n/s_1$, which demonstrates the effectiveness of the low frequency filtering provided by the radial integration.}
\label{table444}
\end{table}
In particular, with the expansion above eq.~\reef{desk1}, the response of the entanglement entropy becomes
\beq
\delta S_\mt{EE}(t)\sim\sum s_n\,b_n\, e^{-i\omega_n t}\,.
\label{desk2}
\eeq
Hence, the contribution of each mode to the power spectrum of the entanglement entropy is multiplied by a factor of $|s_n|^2$, as compared to its contribution in the power spectrum of the metric one-point function.

\begin{figure}
\centering
\includegraphics[width=.6\textwidth]{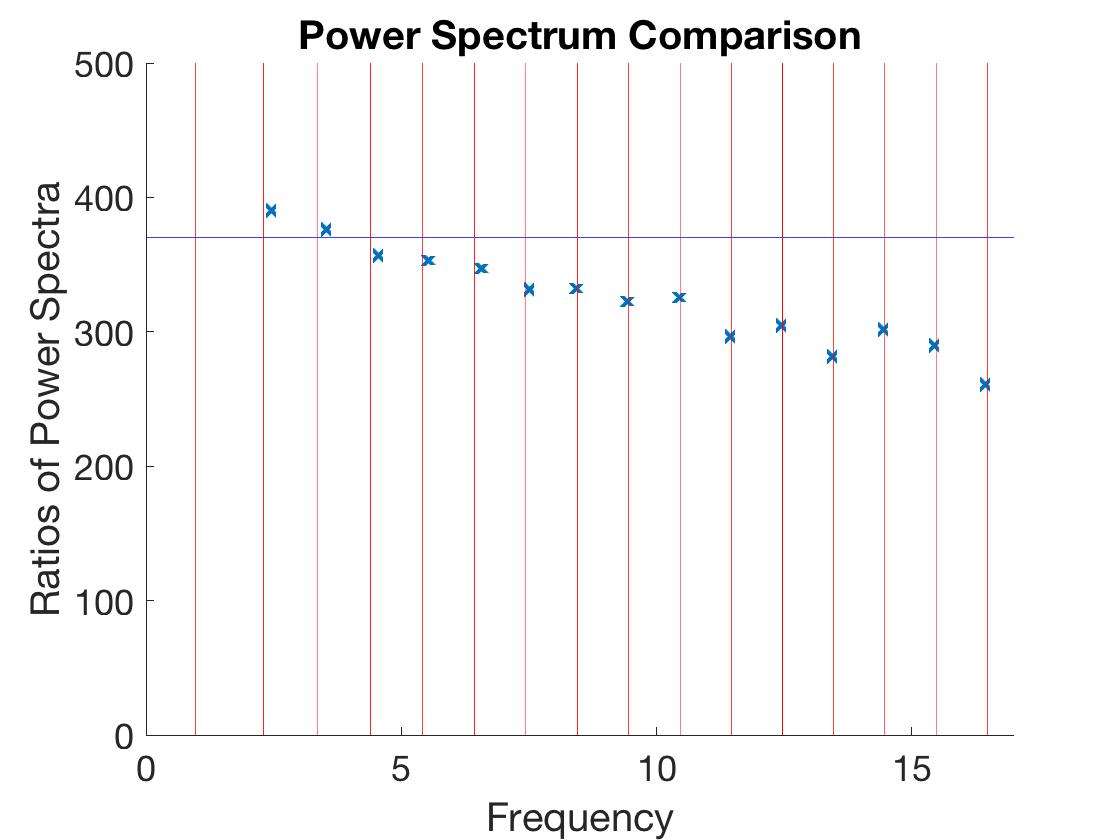}
\caption{Ratio of power spectra $R_n=P_n^\mt{EE}/(P_n^\mt{1pt}\,s_n^2)$ versus frequency, for a quench with $a_b=a_\phi=10^{-5}$ and $\delta t =0.1$. The individual power spectra for the metric one-point function and for the entanglement entropy are shown in figures \ref{scalar1pt} and \ref{EEcomp}, respectively, while the filtering coefficients $s_n$ are given in table \ref{table444}. The vertical lines indicate the normal mode frequencies. We expect these ratios to be constant and so the horizontal line is imply drawn as a guide to the eye. The latter is a best fit horizontal line for the four modes with the most power.}
\label{ratios}
\end{figure}

As a explicit illustration of this filtering mechanism, we analyze the power spectrum of the  entanglement entropy depicted in figure \ref{EEcomp}. We can then compare that spectrum to the power spectrum of the metric one-point function, shown in figure \ref{scalar1pt}, multiplied by the appropriate filtering factor discussed above. Namely, we expect $R_n=P_n^\mt{EE}/(P_n^\mt{1pt}\,s_n^2)$ to be a constant. In figure \ref{ratios}, we plot this ratio versus the frequency for the normal modes and find that the ratios remain fairly constant.\footnote{The ratios are not close to one because the two spectra in figures \ref{scalar1pt} and \ref{EEcomp} each have an arbitrary normalization.} However, the figure shows a definite trend that $R_n$ decreases with increasing $n$. We expect that primary source of this systematic discrepancy at high modes is due to our numerical imprecision in obtaining the power spectrum from the time-series. 

One can repeat a similar exercise for the quench illustrated in figure \ref{timeEEw2}. Since that quench is wider (\ie $\delta t=0.3$), fewer modes are excited and the power of the entanglement entropy (\ie the filtered response) is concentrated pre-dominantly in the lowest two modes. We find that in the power spectrum of the entanglement entropy ${P_2}/{P_1}= 2.23 \times 10^{-2}$, whereas our expectations based on the metric one-point function and the filtering coefficients in table \ref{table444} are ${P_2}/{P_1}= 2.67 \times 10^{-2}$. Hence we again find good agreement and the dominant source of discrepancy is in extracting the power spectrum from the time-series.

Given the coefficients and the ratios $s_n/s_1$ in table \ref{table444}, we see that the radial integration against the profile in eq.~\reef{turk9} naturally acts as a low frequency filter. That is, in comparison to the one-point function in eq.~\reef{desk1}, the response measured by $\delta S_\mt{EE}$ is amplified in the lowest lying mode of the metric perturbation and suppressed in all higher modes. Indeed, this radial integration appears in evaluating $\delta S_\mt{EE}$ irrespective of the quench protocol and we expect that it plays an important role for all of the different quenches in removing much of the short time scale structure, in comparison to the corresponding one-point functions. However, the radial integration is the only smoothing mechanism for the weak gravitational quench, whereas our mechanisms also come into play for the weak scalar and nonlinear quenches (as described below). Our two examples, shown in figures \ref{timeEEw2} and \ref{EEcomp}, illustrate that this low frequency filtering may or may not be particularly effective on its own depending on the details of the quench parameters for the weak gravitational quenches. In particular, as we found in section \ref{weakone}, when $\dt$ becomes smaller, the quench primarily excites modes with higher frequencies and there is little power in the lowest modes. As comparing figures \ref{scalar1pt} and \ref{EEcomp} illustrates, in these situations, the radial filtering certainly reshapes the power spectrum suppressing the relative power in the high frequencies, but it still leaves the response in $\delta S_\mt{EE}$ with significant power across a broad range of normal modes. 

\vspace{10pt}
\subsubsection*{{\it ii}) Near Resonant Driving:}

As well as the smoothing or filtering of the response, another distinctive feature of $\delta S_\mt{EE}$ was the appearance of ``beats" for weak scalar quenches, as is seen very clearly in the top panel of figure \ref{timeEE}. Near resonant driving is the mechanism responsible for this phenomenon and then becomes the main mechanism responsible for smoothing out the response for these quenches. Since the beat phenomenon arises for weak quenches, we can frame our explanation in terms of the scalar and metric normal modes, as in the previous discussion. 

The holographic entanglement entropy probes the metric perturbations, but the metric is not directly sourced in a scalar quench. Rather the metric is only excited at second order in a perturbative expansion, through the stress tensor in the Einstein equations \reef{volum}. Further as described above, $\delta S_\mt{EE}$ is most sensitive to the excitation in the lowest metric mode with frequency $\omega_1\simeq 0.9748\, \wfun$, as given in table \ref{table33}. While the metric modes in this table were determined by the source-free linearized Einstein equations, the response here comes from the same linearized equation but with a source, the stress tensor, which is quadratic in the scalar field perturbations. This means that in the Fourier decomposition of that source, we get not only the frequencies of the individual normal modes of the scalar field, but also the sums and differences of these frequencies. Of course, prominent among the differences of these normal mode frequencies is the asymptotic level spacing $\Delta\omega=\wfun$, especially for quenches sourcing the higher frequency modes of the scalar field\footnote{For the quench depicted in  figure \ref{timeEE}, the maximal power is in the twelfth scalar normal mode.} (\ie quenches with smaller $\delta t$). Since $\Delta\omega$ is the lowest frequency in the source and since it is very close the metric normal mode frequency $\omega_1$, this is defines the component of the source that is primarily responsible for exciting the lowest metric mode, and hence for producing the dominant response in  $\delta S_\mt{EE}$. The small splitting between the source and normal mode frequencies is also responsible for the beat phenomenon observed in figure \ref{timeEE}. In particular, we have $\omega_{modu} =\Delta\omega-\omega_1 \simeq 0.0251 \,\wfun$. 

The above explanation provides good quantitative agreement with the results shown in figure \ref{timeEE}.  For the scalar quench shown there, the Fourier decomposition of the time series for $\delta S_\mt{EE}$ over $0\le t\le500$ shows two dominant frequencies of approximately $\omega_{fast}\simeq2.40 \simeq \wfun$ and $\omega_{slow}\simeq2.335\simeq \omega_1$, with an error of roughly $10^{-2}$ and again, in good agreement with the previous discussion.

Let us consider the driving of the metric perturbations by the scalar field in more detail --- see also discussion in \cite{Craps:2015upq}. We can model any of the metric normal modes as a simple harmonic oscillator and the stress tensor as a sinusoidal driving force. That is, we consider the simpler (and well-known) problem $\ddot x +\omega_0^2\, x = f(t)\,\sin(\omega t)$, where we introduce a theta-function with $f(t)=f_0\, \theta(t)$ so that the source suddenly turns on as in our quenches. With the initial conditions $x(t=0)=0$ and $\dot x(t=0)=0$, it is straightforward to show that the response is given by
\beq
x(t) = \frac{f_0}{\omega_0^2-\omega^2}\,\left[\sin(\omega t) - \frac{\omega}{\omega_0}\,\sin(\omega_0 t)\right]\,.
\label{simp33}
\eeq
The most important feature of this solution is that the amplitude diverges as $\omega\to\omega_0$. That is, when the driving frequency approaches the resonant frequency of the oscillator, the amplitude of the response becomes arbitrarily large. In our holographic quenches, we effectively have many oscillator frequencies and many driving frequencies, however, $\Delta\omega-\omega_1$ gives by far the smallest splitting. Hence the corresponding response becomes the dominant contribution in $\delta S_\mt{EE}$ and we only observe the two overlapping oscillations in eq.~\reef{simp33} for this smallest splitting. Of course, as noted above, these two oscillations give rise to the distinctive beats found in the response for these weak scalar quenches, \eg see figure \ref{timeEE}. Another feature illustrated by this model  \reef{simp33} is that with $\omega/\omega_0\simeq1$, the amplitude of the two oscillations is nearly equal and as a result, the beat modulation almost completely extinguishes the signal at its minimum, just as is shown in figure \ref{timeEE}. 

The effectiveness of this near resonant driving of the metric excitations in the weak scalar quenches can also seen in the metric one-point function, as shown in the bottom right panel in figure \ref{timeEE}. Here, we see the short time structure of the pulses that are typical of the one-point functions but these pulses appear on the background of a smoothly modulated oscillation with $\omega\sim\wfun$.  The latter is again the contribution produced by the near resonant driving and as the figure shows, its amplitude is comparable to the amplitude of the pulses. As described above, the radial integration filters out the very high frequencies and so prevents a pulse-like response from appearing in the $\delta S_\mt{EE}$. Hence it is actually a combination of the amplification of the lowest mode produced by the near resonant driving and the filtering provided by the radial integration that leads to the very smooth response observed in the holographic entanglement entropy.

The beat phenomena is a common feature for weak scalar quenches, with varying amplitudes and durations. We have observed that as the amplitude of the scalar quenches increases, the beats become less prominent and essentially disappear in the nonlinear regime. This can be anticipated because the normal modes do not play a central role in defining the response to nonlinear quenches, \eg as illustrated in figure \ref{drift}. Hence we cannot expect a precise alignment of frequencies between the metric response and the scalar source in Einstein's equations. Given the above discussion, it is also clear why these beats do not appear in gravitational quenches, where either the metric or both the scalar and the metric are sourced --- see the example in figure \ref{timeEE11}. In such quenches, the metric is excited at first order and while the process described above is still active for weak quenches, it only produces beats at second order in the perturbative expansion. Hence the beats are no longer observable. 

The beats found here are essentially the same as those observed by \cite{Craps:2015upq} in the one-point functions after a scalar quench. However, the key difference is that there the beats were revealed by artificially evaluating the overlap of the metric response with the profile of the lowest lying mode. Here, as we explained above, this same filtering is naturally achieved by the radial integration involved in evaluating the holographic entanglement entropy. In the following, we find that a similar filtering occurs quite generally for nonlocal probes and therefore, the beat phenomena observed here for scalar quenches also appears in the response of other nonlocal observables. 

\vspace{10pt}
\subsubsection*{{\it iii}) Nonlinear Cascade:} 

The final case where we need to consider the smoothing is the nonlinear quenches. To begin, we expect that the filtering effect of the radial integration is still in effect for the nonlinear quenches, even though the description above involving expanding in terms of linearized modes no longer applies in this situation. However, in comparing responses in figures \ref{timeEEw2} and \ref{timeEE11}, we observe that the holographic entanglement entropy is smoother for the nonlinear quench than the weak gravitational quench. Hence another mechanism must also be in operation for the nonlinear quenches. In particular, we found in section \ref{nonQ} that at least on moderate time scales, the nonlinear dynamics of the bulk fields produced a quasi-periodic series of direct cascades, where much of the power is swept to higher frequencies, followed by inverse cascades, where the spectrum returns to lower frequencies, as illustrated in figures \ref{drift} and \ref{spectro}.  A close examination of either of these figures shows that the spectrum is relatively stable in the vicinity of $\omega\sim\wfun$ and that for the most part, the power transfer in these cascades is taking place at much higher frequencies. Therefore the nonlinear cascades sweep the power out of the low frequency regime but also keeps the most of power in the high frequency domain. Therefore we expect that in the nonlinear quenches, the smoothing is again produced by a combination of the two mechanisms: the cascades which carry most of the power to high frequencies and the low frequency filtering provided by the radial integration. 
However, we must reiterate that our explorations of the nonlinear regime have been necessarily incomplete, and we are not sure to what extent these results are sensitive to the precise choice of quench parameters.

\vspace{10pt}

Our holographic results in this section can also be compared with the results of \cite{pasquale}, which considered the entanglement entropy of half space for global quenches of a one-dimensional spin chain (in a confining phase). There the entanglement entropy was also found to exhibit oscillatory behaviour and the corresponding power spectra showed a rich structure with peaks at frequencies related to the energies of many of the bound states of the system. In our holographic model, the response of entanglement entropy is a smoother quantity and does not exhibit a very rich power spectrum in weak scalar quenches or in nonlinear quenches. However, in some weak gravitational quenches, even though the low frequency filtering of the radial integration smooths the signal somewhat, the response is still rich enough that we can extract the normal mode frequencies of a broad range of excitations, as shown in figure \ref{EEcomp}. Hence for this particular class of quenches, we can perform the same entanglement spectroscopy as introduced for the quenches of spin chains in \cite{pasquale}. To close, let us add that below we will find that the smoothing of the response is a general feature found for all of the nonlocal probes which we examine in this holographic model.

\vspace{10pt}
\subsection{Correlation Functions}

A second probe examined in \cite{pasquale} was the two-point correlator, which showed the most dramatic effect of confinement.  Recall that in the study of global quenches to gapless phases,  one finds ``light-cone" spreading of  correlations and entanglement. For example, for global quenches in two-dimensional CFTs, the entanglement entropy rises linearly, as if it was carried by entangled pairs of free massless particles created by the quench \cite{Cala05}. Similar behaviour has been found in higher dimensions as well, in both holographic and QFT models, \eg \cite{tsun1,tsun2,mark1,mark2,mark3}. For the confining phase studied in \cite{pasquale}, the two-point correlator revealed a similar spreading but it relied on the formation of mesonic bound states in the quench. Therefore the spreading was highly suppressed in these quenches, and further the ``light-cone" boundaries were defined by the meson velocity, which was distinct from the free particle (domain wall) velocity in the unconfined phase. Here we will examine our holographic model for similar features.

In the following, we consider the  two-point correlation function of two operators separated by a fixed distance $\ell$, as function of time following a global quench.
To simplify the gravitational computations, we will consider operators in the boundary theory with large conformal dimension, \ie $\Delta \sim\sqrt{N}$. The corresponding correlators can be calculated in the geometric optics limit as the length of a spacelike geodesic connecting the boundary insertions of the two operators \cite{geod1,geod2}. The boundary correlator is then $\langle\cO(\ell)\cO(0)\rangle_t \sim e^{-S(\ell,t)}$ where $S(\ell,t)$ denotes the regulated length\footnote{Since the geodesic connects two points on the asymptotic AdS geometry, the length naturally diverges. In figure \ref{pot}, which shows the two-point correlator for the AdS soliton geometry, we eliminate the UV divergence by subtracting the geodesic length for a fixed reference separation $\ell_0$, \ie we evaluate $-\log\left(\langle\cO(\ell)\cO(0)\rangle_{vac}/\langle\cO(\ell_0)\cO(0)\rangle_{vac}\right)$.
Figure \ref{2pt} makes a similar subtraction to produce a finite result --- see eq.~\reef{soup}.} of the corresponding geodesic. We will calculate the geodesics and their length numerically, using methods developed and explained in \cite{Rangamani:2015sha, Rangamani:2015agy, Rozali:2017bco}.

\vspace{10pt}
\subsubsection{Initial State} 
We start by discussing the results for the initial state, \ie the AdS soliton. 
To find the geodesic, we integrate out to the asymptotic boundary from the minimal radius reached by the geodesic. That is, we start at some radial position $r=r_0$ in the bulk with a horizontal boundary condition, \ie $r'=0$, and evolve the geodesic equation until the geodesic approaches close to the boundary at $r=1$. The spatial separation of the endpoints of the geodesic as it reaches the boundary determines $\ell$ for corresponding two-point correlator.  

\begin{figure}
\centering
\begin{subfigure}{0.49\textwidth}
\includegraphics[width=\textwidth]{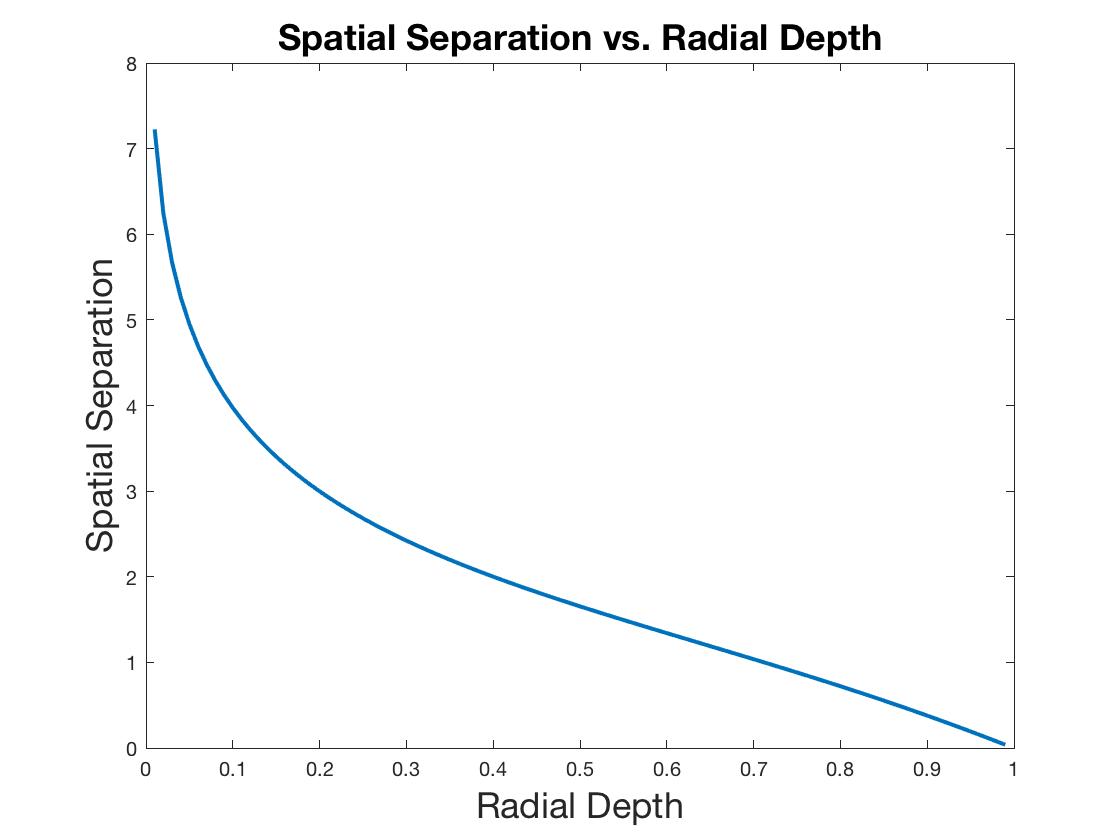}
\end{subfigure}
\begin{subfigure}{0.49\textwidth}
\includegraphics[width=\textwidth]{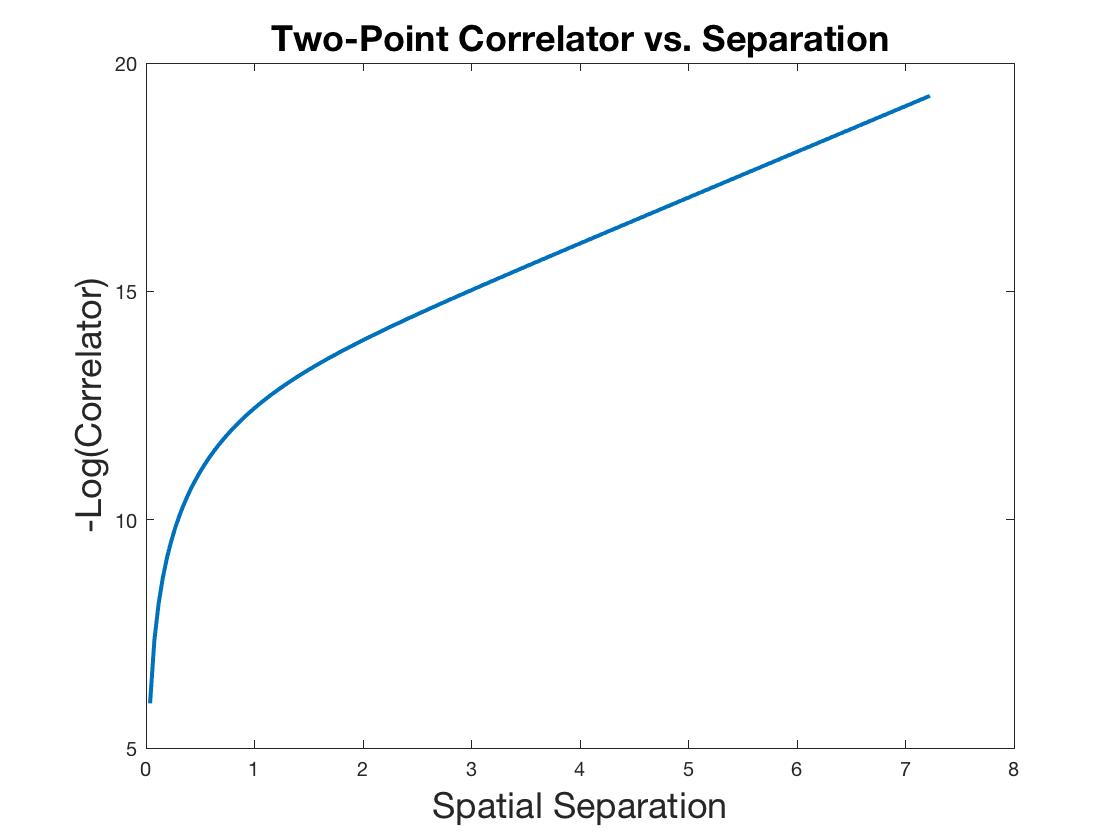}
\end{subfigure}
\caption{Two-point correlation function for the AdS soliton geometry. In the left panel, we display the boundary separation as function of radial depth, we see that
the relation is monotonic. In particular asymptotically wide separations
are obtained from geodesics grazing the cap-off point $r=0$. In the right
panel, we show the logarithm of the correlator $-\log \langle\cO(\ell)\cO(0)\rangle$ as function of the separation $\ell$.
We see that for large separations this quantity becomes linear, exhibiting the exponential decay expected for a gapped or confined phase.}
\label{pot}
\end{figure}

In the left panel of figure \ref{pot}, we plot the relation between the initial radial position $r_0$ and the boundary separation $\ell$. We see that the deeper the geodesics goes into the bulk, the larger the separation of the boundary operators grows. In particular, that relation is monotonic,\footnote{As a result, there is a difference here from the holographic entanglement entropy. Recall that for the entanglement entropy, the connected surfaces, which reached deep into the IR, simply played no role, \ie these surfaces were not the minimal area surfaces for either the narrow or wide strips.} and covers all possible values of $\ell$ as we take $r_0 \rightarrow 0$.

In the right panel of figure \ref{pot}, we plot $-\log\langle\cO(\ell)\cO(0)\rangle$ as function of the separation $\ell$. In particular, we see that for sufficiently large $\ell$, this quantity grows linearly, as expected for the exponentially decaying correlators in a gapped phase. From the figure, we may observe that the correlator exhibits this exponential decay for $\ell\gtrsim \tfun\simeq 2.622$, \ie when the separation exceeds the correlation length associated with confinement. In that regime, we are probing the IR regions of the bulk geometry, where the geodesic stays close to $r=0$ for much of its evolution. 

These holographic results are a variant of the well-known results for the Wilson line in confining theories --- see discussion in the next section. As noted above, the exponential decay of the correlator is expected from the interpretation of the theory as being in a gapped or confined phase. We now explore the time dependence of the correlator following our quench.

\vspace{10pt}
\subsubsection{Time Dependence}

\begin{figure}
\centering
\begin{subfigure}{0.48\textwidth}
\includegraphics[width=1.0\textwidth]{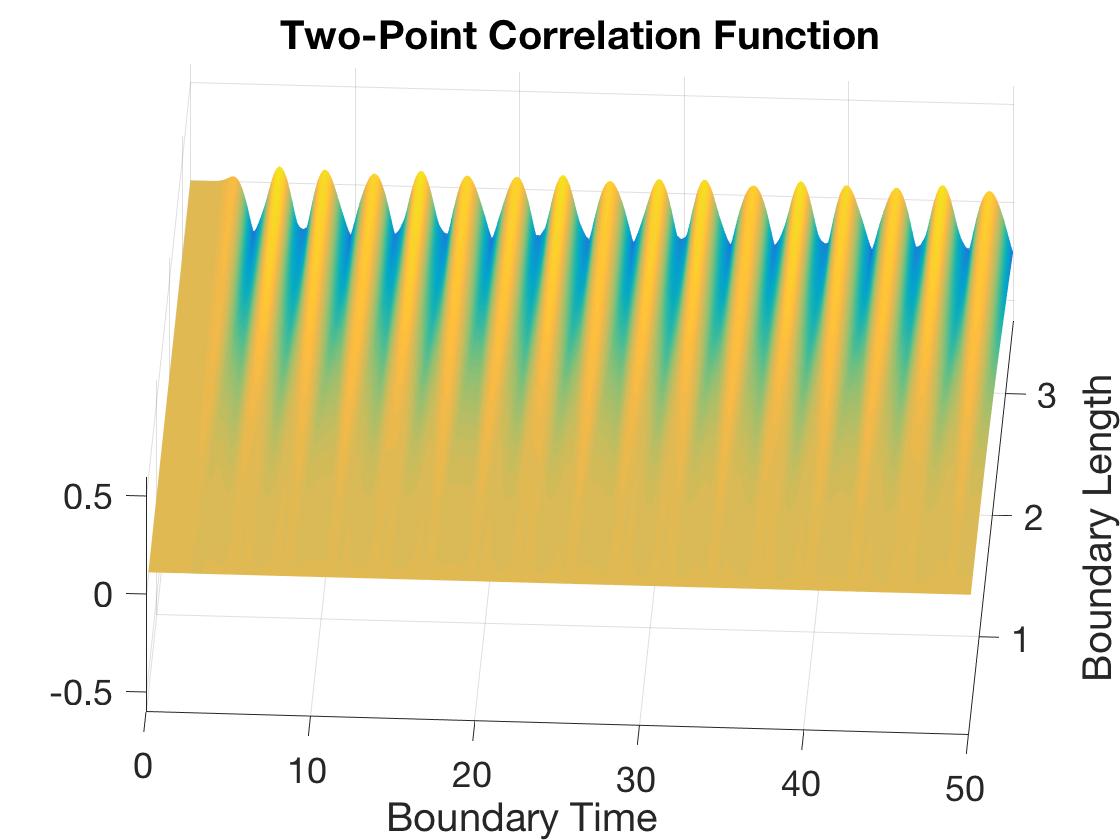}
\end{subfigure}
\ \ 
\begin{subfigure}{0.48\textwidth}
\includegraphics[width=1.0\textwidth]{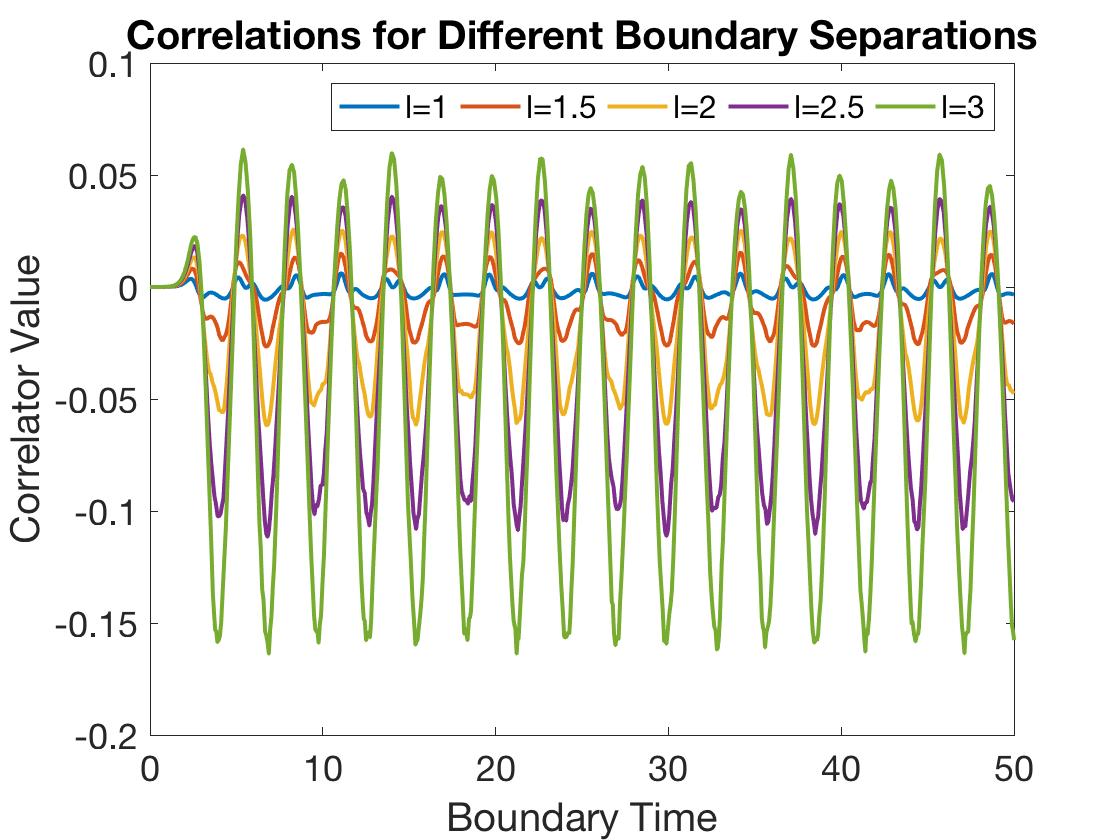}
\end{subfigure}
\caption{Time-dependent two-point correlation function for
a quench with $a_b=0.05$, $a_\phi=0.05$ and $\delta t=0.5$. To enhance
the time-dependent features, we divide by the static correlator
(which decays exponentially for large separations) and display the logarithm of
that ratio in the left panel. The result exhibits oscillatory behaviour in the tails of the correlator, \ie at large separations. The right panel shows several slices at fixed separation (see inset) running through the 3D plot on the left. We see that the overall amplitude of our observable grows with larger separation, but also that the structure on short-time scales is more and more smoothed out.}
\label{2pt}
\end{figure}

In our holographic calculations, we see that in contrast that the light-cone spreading of correlations described above is replaced by an oscillatory behaviour. However, we observe that in general, the correlator is not very sensitive to the time dependence for small $\ell$. This simply reflects that the correlator becomes very large as $\ell\to0$ and so the time-dependent excitations become a tiny fraction of the total. On the other hand, for large separations where the correlator is naturally small, and our computations show that the oscillatory component comprises a larger relative contribution to the total correlator, \ie the (regulated) geodesic length becomes more sensitive to the time-dependence in the geometry at large separations. The time-dependence of the correlator is thus more pronounced in the tails of the correlator (which actually decays exponentially with distance in the confined phase). Hence to enhance
the time-dependent features of the correlator, we normalized the time-dependent correlator by the correlator in the static vacuum (with the same $\ell$) in figure \ref{2pt}, \ie we are plotting 
\beq
-\log\left[\frac{\langle\cO(\ell)\cO(0)\rangle(t)}{ \langle\cO(\ell)\cO(0)\rangle_{vac}}\right] \sim S(\ell,t)-S_{vac}(\ell)
\label{soup}
\eeq
where $S(\ell,t)$ denotes the length of the corresponding geodesics in the bulk. The left panel shows a 3D plot of the normalized correlator at a function of both $\ell$ and $t$, while the right panel shows time sequences at fixed separation running through the 3D plot on the left. We see that the overall amplitude of the time-dependence in the normalized correlator grows with larger separation, but also that the structure on short-time scales is more and more smoothed out as $\ell$ increases.

We expect that this ``tuneable filtering," exhibited in the right panel of figure \ref{2pt}, has an explanation that is more or less the same as discussed for the entanglement entropy. The holographic two-point correlator again has a 
nonlocal structure in the bulk so that this observable samples the metric excitations by integrating them against a profile which is determined by the geodesic. As the separation in the correlator is increased, the geodesic probes deeper into the bulk geometry and the corresponding profile becomes less sensitive to the higher modes in the metric perturbation. As a result, at large $\ell$, the short time structure is almost completely filtered out and the response in the two-point correlator is completely dominated by the oscillations of the lowest lying metric mode. As in the previous section,  it is straightforward to set up a framework where the filtering can be quantitatively evaluated for weak gravitational quenches. However, we do not pursue this direction here because we did not think that it adds very much to our investigation.

However, we can add that the above interpretation is further supported by the appearance of the same ``beat" phenomenon in the two-point correlator for weak scalar quenches, in the same pattern as found for the holographic entanglement entropy. That is, for large separations, weak scalar quenches exhibit prominent beats, analogous to those shown in figure \reef{timeEE}. However, they become washed out as the amplitude is increased and the quench goes into the nonlinear regime. In contrast, the gravitational quenches (where the metric is also sourced directly) show stable oscillations with $\omega\simeq\wfun$, but don't exhibit beats --- again for gravitational quenches in both the weak and nonlinear regimes. 

\vspace{10pt}

We can compare our holographic results for the two-point correlator with those of \cite{pasquale} for global quenches in a one-dimensional spin chain (in a confining phase). There, the two-point correlator showed ``light-cone" spreading which was propagated by entangled pairs of mesonic bound states produced in the quench.
In contrast, we found no evidence of such ``light-cone" spreading in our holographic quenches, which would have appeared as a ridge extending diagonally across the ($t$,$\ell$)-plane in the left panel of figure \ref{2pt}.\footnote{Note that in this figure, the quench profile \reef{proto} is centred at $t=t_0=5$ with $\dt=0.5$ and hence we might expect to see a fairly wide ridge around this time.} However, if we recall the spectra for weak quenches as shown in figure  \ref{scalar1pt}, we observe that although our holographic quenches excite many individual glueball states with zero momentum, there are no pairs (with equal and opposite momenta) being created. If the latter were created, the power spectra would include a continuum of excitations beyond twice the lowest normal mode frequency, \ie for $\omega>2\omega_1$. Of course, the spectra reveals the excitation of individual glueball states in this energy range, but our holographic quench protocol simply does not excite pairs of such bound states. From the bulk perspective, this failure to excite pairs is a reflection of the consistency of solving the equations of motion within the translation invariant ansatz of eq.~\reef{metric} or \reef{post9}. In other words, the translation invariant backgrounds form a super-selection sector in our theory and a perfectly homogeneous source can not excite modes (even pairs of modes) carrying nonvanishing momentum. Hence we can expect that this suppression extends to the nonlinear regime, such as in the quench for which we show the two-point correlator in figure \ref{2pt}. 

Clearly, the oscillatory behaviour shown there indicates that the glueball excitations effect the two-point correlator in a nontrivial way. However, our holographic quenches do not exhibit the physical propagation of correlations by pairs of bound-state particles through the confining phase, as was found in \cite{pasquale}. We expect that this feature could be changed by modifying the quench protocol so that the source was not completely uniform in our holographic model. For example, in holographic {\it local} quenches, one can see that the influence of the quench at length scale $\ell$ is felt mainly at time $t$, such that $\ell \propto t$ with the constant being the analogue of a Lieb-Robinson velocity \cite{Rangamani:2015agy}.

Another feature of the ``light-cone" spreading found in \cite{pasquale} was that the boundaries of the ``light-cone" were defined by the velocity of the mesons, which was distinct from the free particle velocity in the unconfined phase. However, in our holographic model, we expect that all of the glueball states will still propagate at the usual speed of light, \eg the bulk metric \reef{soliton} maintains Lorentz invariance in the ($t$,$x_1$,$x_2$) subspace for all values of $z$. However, there are holographic models where quenches may show this kind of behaviour, \eg in certain models the velocity of meson excitations lies below the speed of light when propagating through a thermal plasma \cite{fast1,fast2}.


\vspace{10pt}
\subsection{String tension}

Wilson lines are another interesting family of observables in our holographic model \cite{wilson1,wilson2}. The holographic calculation of these nonlocal observables is again similar to that of the entanglement entropy or the two-point correlation functions.  The latter involve evaluating the ``area" of extremal three- and one-dimensional surfaces in the bulk, while the Wilson lines require evaluating the area of two-dimensional extremal surfaces. 

In particular, the holographic calculation of the correlator of two parallel Wilson lines   is very similar to that for the two-point correlator, in the previous section. Technically, the corresponding integrand differs from that for the geodesic calculation by a factor of $\sqrt{g_{x_2x_2}}$, for two parallel Wilson lines extending along the $x_2$-direction. Since that metric component is bounded and nonvanishing in the soliton geometry, qualitative features of our previous results will extend to the Wilson line calculation.

Another interesting Wilson line would be that surrounding the compact direction, which acquires a nonvanishing expectation value in this confining phase \cite{edward}. This calculation would be a close analog of the half-space entanglement entropy, which was carried out above. In this case, the technical difference between the two calculations is that the integrand for the entanglement entropy has an extra factor of $\sqrt{g_{x_1x_1}}$, for the half-space defined by $x_2>0$ (and $t=0$) --- see eq.~\reef{out1a}. However, we again expect that the qualitative features of our previous results for the entanglement entropy given in section \ref{HEEr} will extend to this Wilson line calculation.

Given these similarities, we focus the following discussion of Wilson lines to examining the behaviour of string tension after a holographic quench. The string tension gives the strength of the linear potential between two static quarks at large separations in the confining phase.
Alternatively, we can describe the string tension as characterizing the linear growth of the expectation value of two parallel (time-like) Wilson lines at large separations. In our holographic framework, this tension becomes the tension of a fundamental string stretched out along the IR cap-off radius in the bulk geometry. For example, in the unperturbed AdS soliton \reef{soliton}, this tension is given by 
\beq
T_\mt{string}=\frac{\sqrt{-g_{tt}\,g_{x_1x_1}}|_{z=z_0}}{2\pi\,\alpha'}=\frac{2\pi L^2}{\ell_s^2}\,\frac{1}{\Delta\theta^2}=8\pi\,\frac{\sqrt{2\lambda}}{L_c^2}
\label{tense}
\eeq
where $\lambda$ denotes the 't Hooft coupling in the boundary theory\footnote{The standard AdS/CFT dictionary allows us to translate $L^4 = 2\lambda\,\ell_s^4$.} .

\begin{figure}
\centering
\begin{subfigure}{0.48\textwidth}
\centering
\includegraphics[width=.9\textwidth]{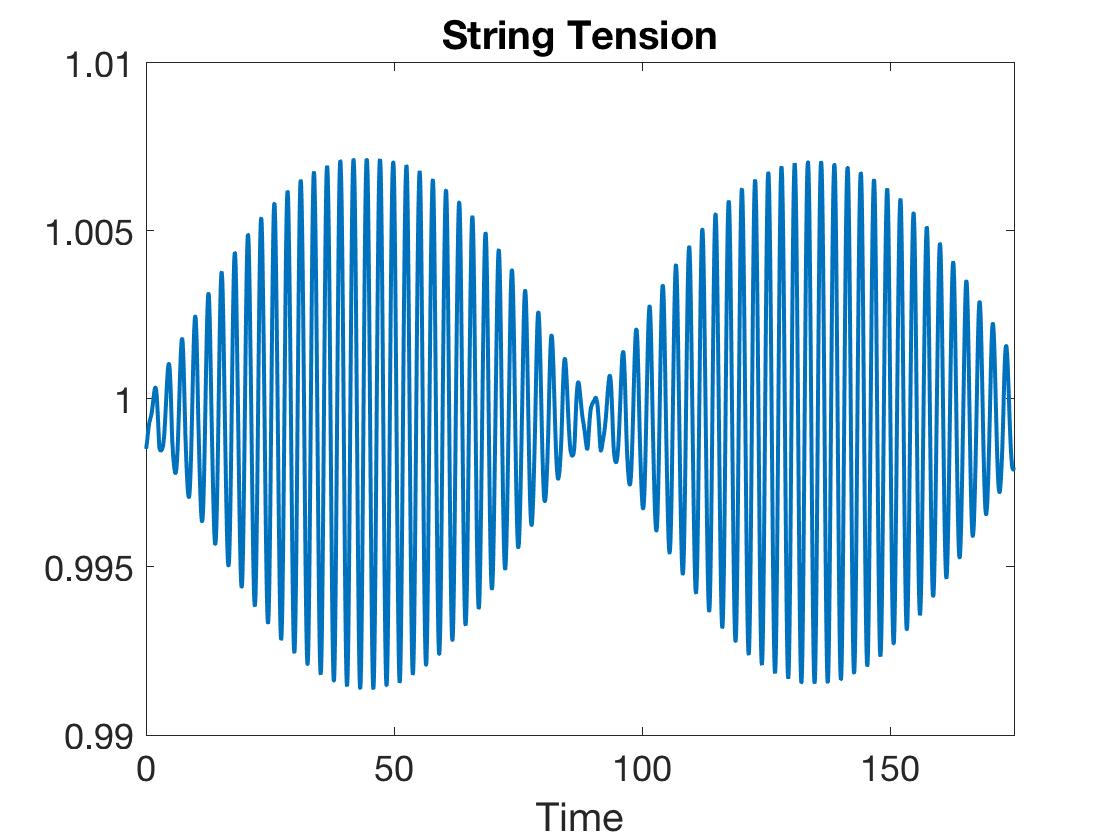}
\end{subfigure}
\begin{subfigure}{0.48\textwidth}
\centering
\includegraphics[width=.9\textwidth]{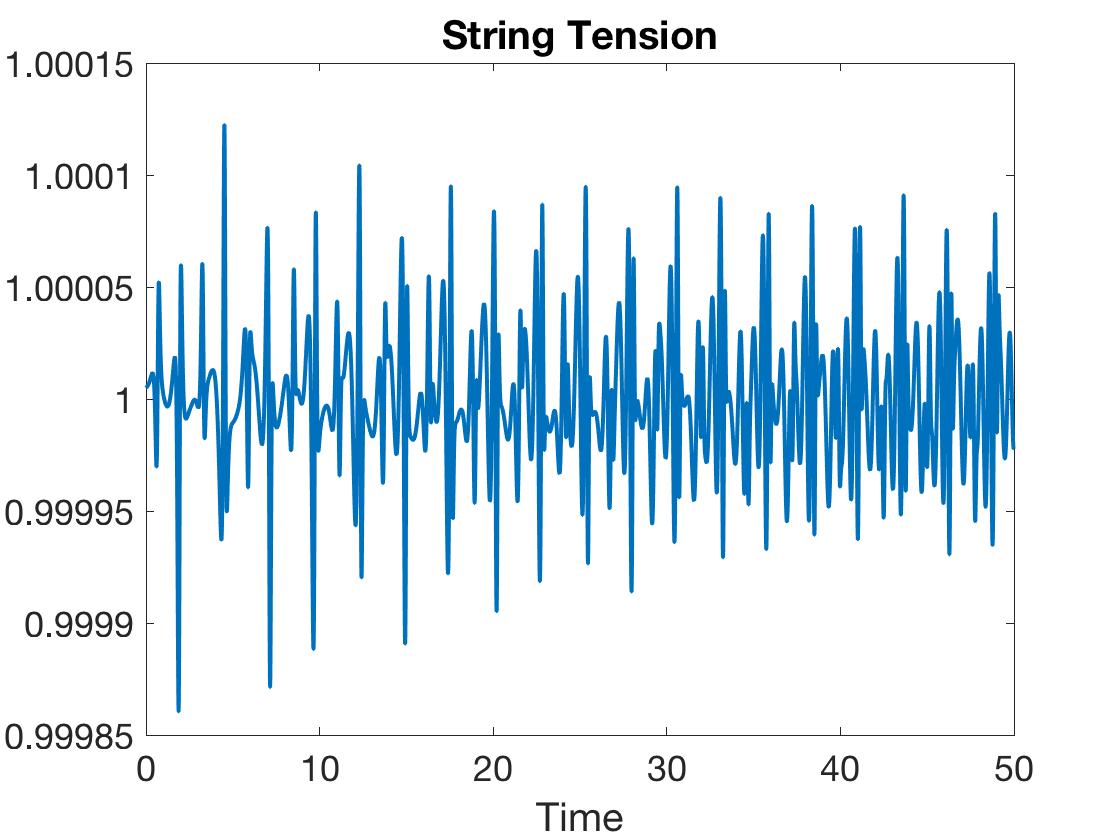}
\end{subfigure}
\begin{subfigure}{0.48\textwidth}
\centering
\includegraphics[width=1.0\textwidth]{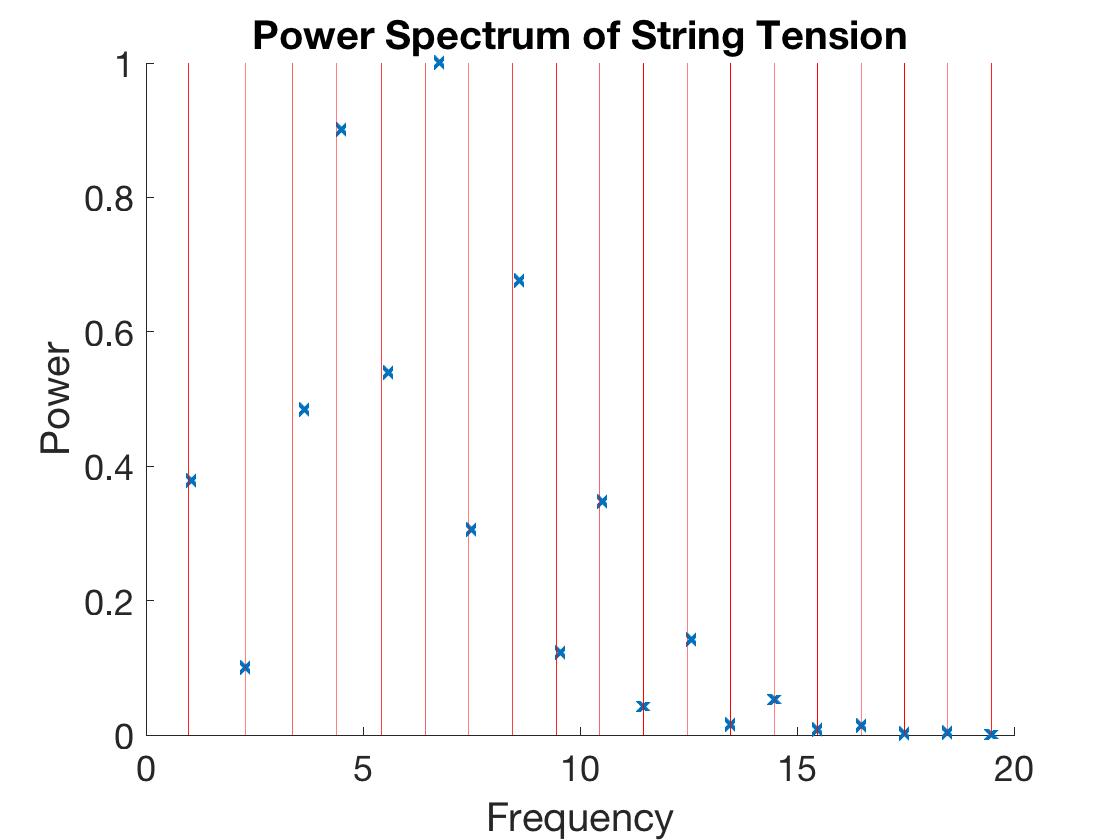}
\end{subfigure}
\caption{In the left panel, we exhibit the time dependence of the string tension for a weak scalar quench with $a_\phi=10^{-4}$, $a_b=0$ and $\delta
t=0.05$. We see beats in the time series, where the oscillatory behaviour is modulated on a long time scale. This is, of course, similar to the behaviour seen for weak scalar quenches with the entanglement entropy and the two-point correlator --- note $\delta S_\mt{EE}$ for the same quench appears in figure \ref{timeEE}. In the right panel, we exhibit the time-dependence for the string tension for a weak gravitational quench with  $a_b=a_\phi=10^{-5}$ and $\delta t=0.1$. The ``beats" are notably absent and there is significantly more structure on short-time scales --- note the change in scale on the time axis. In the bottom figure we performs spectroscopy on that signal, showing the excitations overlap chiefly with the normal modes of the geometry. Again, this behaviour is similar to that seen other nonlocal probes --- note $\delta S_\mt{EE}$ (and the corresponding power spectrum for the same quench appears in figure \ref{EEcomp}. }
\label{tension}
\end{figure}

In figure \ref{tension}, we show the string tension as function of time after two different weak quenches. Since these are weak quenches, we can evaluate the change in the string tension perturbatively and to linear order, eq.~\reef{tense} yields\footnote{Evaluating the string tension for nonlinear quenches is somewhat more challenging because of the time shift between the turning point at the maximal bulk radius of the extremal surface and where it reaches the asymptotic boundary. Of course, this effect arises for all of the nonlocal observables which were considered in this section.}
\beq
\delta T_\mt{string}=\frac{1}{2\pi\,\alpha'}\,\left(
\delta g_{x_1x_1}- \delta g_{tt}\right)_{x=+1}=-\frac{L^2}{2\pi\,\alpha'}\,\left(\beta+\alpha+2\delta\right)_{x=+1}
\label{tense2}
\eeq
using the post-quench coordinates \reef{post9}. 

The left panel of figure \ref{tension} shows the string tension after a weak scalar quench. In this case, we observe a smooth response with the same ``beat'' phenomenon that was observed for the entanglement entropy and the two-point correlator. This feature again appears quite generally for weak scalar quenches but we expect that it is washed out when the amplitude is increased to produce quenches in the nonlinear regime.

As noted above, the nonlocal quantities are much smoother than the local ones, as is clearly the case here for the string tension for the scalar quench. The same is true for weak gravitational quenches (and for nonlinear quenches, we expect), following the discussion in section \ref{smoother} for the entanglement entropy. However, similar to that case as well, one can tune the duration $\delta t$ for weak gravitational quenches as to make the smoothing of nonlocal quantities less effective.  As an example, the right panel of figure  \ref{tension} shows the string tension after a weak gravitational quench. While the response is smoothed somewhat in comparison to the corresponding one-point function, it still shows significant structure on short time scales.  Indeed, one can examine the power spectrum of the time-series in the right panel of figure \ref{tension}, and once again recover the normal mode frequencies for the AdS soliton geometry. This is shown in the bottom panel of figure \ref{tension}.

\begin{figure}
\centering
\includegraphics[width=.6\textwidth]{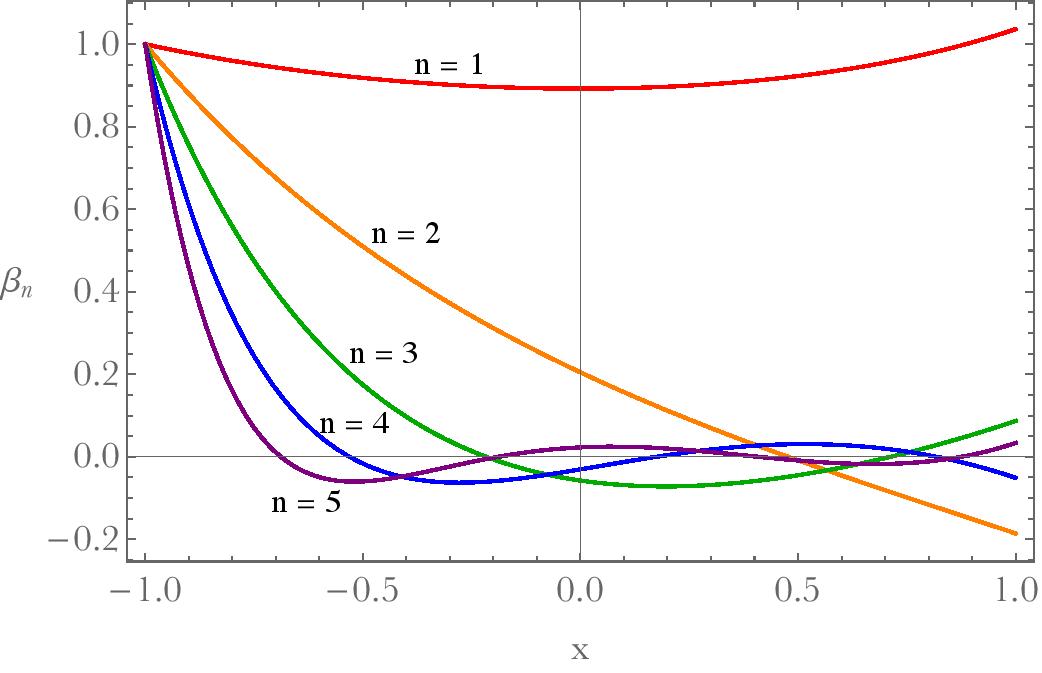}
\caption{The radial profiles $\beta_n(x)$ for the first five normal modes of the anisotropy field. Note that these profiles are normalized such that $\beta_n(x=-1)=1$ at the asymptotic boundary. }
\label{mad}
\end{figure}

We can examine the smoothing mechanism for the string tension in greater detail by applying a linear mode analysis, similar to what was done in section \ref{smoother}.  From eq.~\reef{tense2}, we see that we must consider excitations of $\alpha$ and $\delta$, as well as $\beta$. However, neither of these gives an independent variation of the metric \reef{post9}. In particular, we already found that in the linear regime that 
$\alpha(t,x)$ is directly proportional to $\beta(t,x)$, as given in eq.~\reef{aconstraint}. 
Evaluating the latter at the IR cap-off point, as required in eq.~\reef{tense}, we find the radial profiles $\delta_n(x)$ satisfy
\beq
\alpha_n(x=1)=2\beta_n(x=1)\,. \label{endc1}
\eeq
Similarly, $\delta(t,x)$ can be evaluated in terms of $\beta(t,x)$ by solving the linearized version of the constraint in eq.~\reef{dconstraint}. Focusing on a particular normal mode $\beta=e^{-i\omega_nt}\beta_n(x)$ and the linearized solution has the same time dependence and the radial profile takes the form
\beq
\delta_n(x)=-\frac{16\beta_n(x)}{12-(1+ x)^2}+\frac{384}{(1+x)^4}\int_{-1}^x\mathrm d\tilde x\frac{(1+\tilde x)^3\,\beta_n(\tilde x)}{[12-(1+\tilde x)^2]^2}\;,
\label{space0}
\eeq
For the string tension, we evaluate this radial profile at the IR cap-off point to find
\beq
\delta_n(x=1)=-2\beta_n(x=1)+24\int_{-1}^1\mathrm d x\frac{(1+ x)^3\,\beta_n( x)}{[12-(1+x)^2]^2}\;.
\label{endc2}
\eeq
Now with an expansion of the metric excitation as $\beta(t,x)=\sum b_n\, e^{-i\omega_n t}\,\beta_n(x)$,  eq.~\reef{tense} yields the response of the string tension as
\beq
\delta T\sim\sum r_n\,b_n\, e^{-i\omega_n t}\,,
\label{desk3}
\eeq
where we have combine eqs.~\reef{endc1} and \reef{endc2} to define
\beq
r_n\equiv-\beta_n(x=1)+24\int_{-1}^1\mathrm dx\frac{(1+x)^3\beta_n(x)}{[12-(1+x)^2]^2}\,.
\label{alfonso}
\eeq
Recall that the corresponding one-point function is given by eq.~\reef{desk1}. Hence the above coefficients \reef{alfonso} measure the relative sensitivity of $\delta T$ to the normal modes.

The values of $r_n$ for the first few normal modes are given in table \ref{table456}  and we see that the nonlocal geometric structure of $\delta T$ again acts like a low frequency filter. That is, in comparison to the one-point function in eq.~\reef{desk1}, the response measured by $\delta T$ is slightly amplified in the lowest lying mode and suppressed in all higher modes. This behaviour can be understood as follows. The integral on the left-hand side of eq.~\reef{alfonso} is a Sturm-Liouville inner product of a smooth function with $\beta_n$ --- see footnote \ref{footy789}.  This integral therefore rapidly decreases exponentially with increasing mode number for the normal modes $\beta_n$. The contribution for higher modes is therefore dominated by the first term $-\beta_n(x=1)$. Recall that we normalized the radial profiles of the normal modes at the asymptotic boundary such that $\beta_n(x=-1)=1$. With this normalization, figure \ref{mad} illustrates that the normal modes generally have a much smaller amplitude at the IR cap-off point, \ie at $x=+1$. The only exception to this statement is the lowest lying mode at $n=1$, which is roughly constant across the entire radial direction. We also show the corresponding values of the radial profiles for the first few normal modes in table \ref{table456}.\footnote{Cmparing with the coefficients in table \ref{table444}, we found that $\beta_n(x=1)/\beta_1(x=1)=(-)^{n+1} s_n/s_1$. While we have no explanation for this surprising match, we found that the agreement extends to the machine precision of our numerical calculations. An amusing consequence is as follows: Consider the string tension for two spacelike Wilson lines at a fixed time and, \eg extended along the $x_2$ direction while separated in the $x_1$ direction. In this case, eq.~\reef{tense2} is replaced by $\delta T_\mt{space} = - \frac{L^2}{\pi \alpha'}\, \beta(t,x=1)$. Hence the power spectrum for the string tension would be identical to that for $\delta S_\mt{EE}$.}

\begin{figure}
\centering
\includegraphics[width=.6\textwidth]{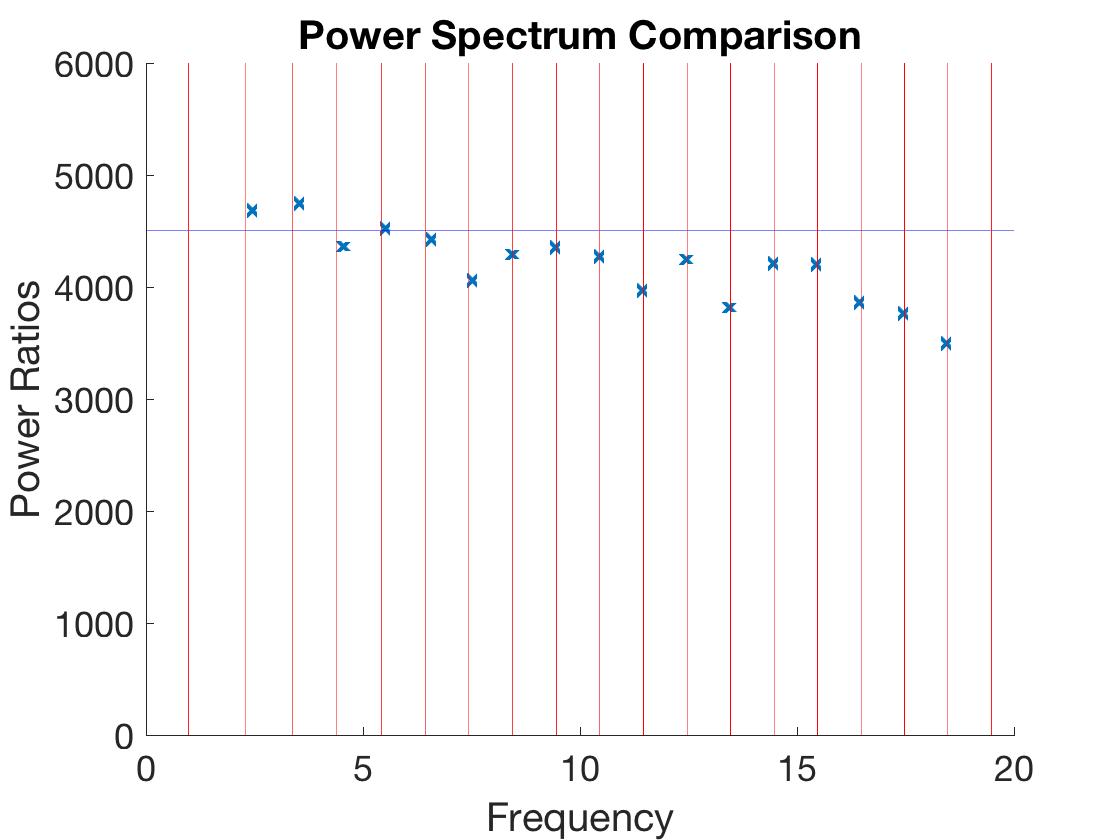}
\caption{Ratio of power spectra $\tilde R_n=P_n^\mt{T}/(P_n^\mt{1pt}\,r_n^2)$ versus frequency, for a quench with $a_b=a_\phi=10^{-5}$ and $\delta t =0.1$. The individual power spectra for the metric one-point function and for the string tension are shown in figures \ref{scalar1pt} and \ref{tension}, respectively, while the filtering coefficients $r_n$ are given in table \ref{table456}. The vertical lines indicate the normal mode frequencies. We expect these ratios to be constant and so the horizontal line is imply drawn as a guide to the eye. The latter is a best fit horizontal line for the four  modes with the most power. }
\label{str_ratios}
\end{figure}

\begin{table}[htbp]
\hspace{-25pt}
 \begin{tabular}{c|| c| c|c| c|c| c|c|c|c} 
 $n$ & 1  & 2& 3& 4& 5 & 6 & 7 & 8&9 \\ [0.6ex] 
 \hline
 $r_n\times 10^2$ & 114.88 & 4.814 & --6.844 & 4.765 & --3.290 & 2.368 & --1.776 & 1.380 & --1.102 \\ [0.6ex] 
 \hline
 ${r_n}/{r_1}\times 10^2$ & 100 &4.190& --5.957 & 4.148& --2.864 & 2.061 & --1.546 & 1.291 & --0.959\\ [0.6ex] 
 \hline
$\beta_n(x=1)\times 10^2$ & 103.66 & --18.58 & 8.686 & --5.090 & 3.355 & --2.382 & 1.779 & --1.380 & 1.102 
\end{tabular}
\caption{The coefficients $r_n$ defined in eq.~\reef{alfonso} for the first few normal modes $\beta_n$. These coefficients measure the relative sensitivity of $\delta T$ to each of these normal modes. We also show the ratio $r_n/r_1$, which demonstrates the effectiveness of the low frequency filtering provided by the present mechanism, and the values of the normal mode profiles at the IR cap-off point. }
\label{table456}
\end{table}
Next we present a explicit illustration of the present filtering mechanism similar to the discussion of filtering in the context of the entanglement entropy. In particular, the power spectrum of the string tension is given in (the bottom panel of) figure \ref{tension} for the quench depicted in the top right panel. Our expectation is that this spectrum should match the power spectrum of the metric one-point function, shown in figure \ref{scalar1pt}, multiplied by the filtering factor  $r_n^2$. That is, we expect $\tilde R_n=P_n^\mt{T}/(P_n^\mt{1pt}\,r_n^2)$ to be a constant. In figure \ref{str_ratios}, we plot these ratios versus the frequency for each of the normal modes and find that the ratios remain fairly constant.\footnote{The ratios are not close to one because the two spectra in figures \ref{scalar1pt} and \ref{EEcomp} each have a different normalization.} There is again a trend for $\tilde R_n$ to decrease with increasing $n$ but this is less pronounced than in figure \ref{ratios} for the entanglement entropy. We again expect that the deviation for the high modes is likely due to numerical imprecision in obtaining the spectrum.

As noted above, the nonlocal geometric structure of $\delta T$ again acts like a low frequency filter. To compare the present filtering mechanism with that from the radial integration for $\delta S_\mt{EE}$, table \ref{table456} also shows the ratios $r_n/r_1$, which can be compared to the analogous ratios $s_n/s_1$ in table \ref{table444}. While $\delta T$ is somewhat more effective at suppressing the $n=2$ and 3 modes, we see that both mechanisms have more or less the same effect on the higher modes.

\vspace{20pt}
\section{Discussion} \label{discuss}
We now summarize our main results. We have studied holographic quenches in the confining phase defined by the AdS soliton geometry. In particular, we focused on quenches where the energy injected was insufficient to reach the deconfined phase. 

We began by examining the time-series representing the response in the expectation value of local operators in section \ref{onepo}. For weak quenches, the spectrum of these one-point functions was stable and revealed the normal mode frequencies of glueball excitations in the confining phase. The response was more interesting in strong quenches where the dynamics of the bulk fields became nonlinear. In this case, the spectra of the one-point functions initially exhibit a direct cascade, where the energy shifts towards higher frequencies, but this was followed by a shift of energies back to lower frequencies in an inverse cascade process. The quasi-periodic sequence of direct and inverse cascades repeats, seemingly indefinitely. We discussed various notions of thermalization.  Up to the timescale that we have studied, we find no evidence of thermalization in our holographic model, even in the weakest sense. 

In section \ref{nonloco}, we also investigated the behaviour of nonlocal boundary observables:  the entanglement entropy of half-space, two-point correlations functions of heavy operators and the string tension. A common feature in all of these was that the time series was smoothed out in comparison to the one-point functions. A general feature that played a role in this smoothing of the response was the nonlocal structure of the corresponding holographic construction in the bulk. For example, the holographic entanglement entropy is given by the area of an extremal surface in the bulk which extends from the asymptotic boundary to the IR cap-off point. We demonstrated in section \ref{smoother} that the radial integral effectively acts as a low frequency filter suppressing the contribution of the high frequency excitations in the bulk. 

Two other interesting effects were also observed in particular classes of quenches. In weak scalar quenches, the metric normal modes are driven by the excitations injected into the scalar field by the quench. There is a near degeneracy between the frequency of the lowest metric mode and the driving frequency $\Delta\omega$, corresponding to the splitting of the higher normal frequencies of the scalar. Hence there is a near resonant driving of this lowest mode, which greatly amplifies its amplitude compared to the other modes. Combined with the low frequency filtering, this mechanism produces a distinctive response exhibiting beats for the nonlocal observables in these quenches, as shown in figures \ref{timeEE} and \ref{tension}.

In the nonlinear quenches, the cascades mentioned above sweep most of the energy to high frequencies and while there is an interesting nonlinear dynamics at these high frequencies, the excitations at low frequencies remain relatively stable. Combined again with the low frequency filtering, this mechanism produces a generally smooth
response in the nonlocal observables. Another mechanism which may play a role in smoothing the response for nonlinear quenches is as follows: Generally the dual holographic construction began by determining an extremal surface in the bulk (of dimension three, two and one for entanglement entropy, Wilson lines and two-point correlators, respectively). In a strongly dynamic background, these extremal surfaces will extend over some finite time interval. That is, in the coordinates of eq.~\reef{metric}, if the surface starts at the minimum radius at $t=t_0$, it will generally reach the asymptotic boundary at some later time $t=t_b$. Of course, this effect is a negligible second order effect for weak quenches. However, for nonlinear quenches, the feature that the bulk surfaces extend over a finite time may play a role in smoothing of the time-series for the corresponding nonlocal boundary observables.

To a large extent, our present study was motivated by \cite{pasquale}, which investigated global quenches in the confining phase of a one-dimensional spin chain. With our confining holographic model, we are investigating a higher dimensional version of these quenches and we were able to study the response of a similar set of observables. We found that {\it quench spectroscopy} was a feature of both systems, at least in the appropriate regime. That is, observables such as the one-point functions or half-space entanglement entropy exhibited oscillatory behaviour after a quench and the dominant frequencies in the corresponding power spectra matched the masses of various bound states in the confining phase. In the holographic framework, this spectroscopy applied to the weak quenches where the ``glueball" states in the boundary theory could be associated with the normal mode excitations of the bulk fields. In the nonlinear regime, these normal modes are not useful in describing the gravitational dynamics in the bulk but of course, the power spectra were still very useful to characterize the post-quench behaviour.

The main effect of confinement observed in \cite{pasquale} was a dramatic change of the ``light-cone" structure in the spreading of correlations. In our holographic studies, we can certainly say that this spreading was suppressed, in agreement with the results for the spin chain. However, the two-point correlator in \cite{pasquale} showed ``light-cone" spreading which was propagated by entangled pairs of mesonic bound states produced in the quench. Unfortunately, it is an feature of our model that global quenches do not produce such pairs of bound states and so we did not observe this effect --- see further discussion below. Note that we expect this feature to arise in any quantum field theory model of quenches where the source is perfectly homogeneous, \ie it is not simply an idiosyncrasy of our holographic model.

We would add that the smoothing of the response of nonlocal observables in comparison to the one-point functions observed in our holographic model does not seem to have a counterpart in the spin chain quenches of \cite{pasquale}. A key factor for producing this low frequency filtering effect seemed to be nonlocal radial character of the dual holographic constructions of the nonlocal observables. Roughly, we can translate this feature into a statement that the nonlocal observables are sensitive to many different energy scales in the boundary theory, but in particular, they seem to be most sensitive to IR excitations near the gap of the confining phase. 
This structure may be an effect of the strong coupling or large central charge limits, which are implicit in our holographic framework. 

\ 

We now turn to discussing some directions for future research:  It is interesting to explain the long time scale emerging in our analysis of time dependence of the geometry, as some aspects of the nonlinear long time dynamics. On the one hand, it looks like the initial signal, after each cycle of direct and inverse cascades, reconstructs itself momentarily. This is reminiscent of the story of ``revivals" \cite{2014PhRvL.112v0401C, 2016JPhA...49O5401C}; if so, it would be interesting to make the connection more explicit. On the other hand, similar emergence of long time scale in the context of AdS instability was observed in the two-timescale formalism of \cite{Balasubramanian:2014cja}. It would be interesting to apply that formalism to the present context, though that may be more difficult here since the normal mode spectrum is not known analytically.

It would be interesting to understand in what sense late time features of the system become stationary, at least for some initial states and some observables. Such features will also characterize the final state of a quantum quench past a quantum critical point, and into a gapped phase. Indeed, probing criticality at zero temperature, with various protocols of quenching or driving, shows interesting universal results in the the post-quench late time physics, at least in some instances (see for example \cite{2009arXiv0910.3692G,2010PhRvE} for the case of quenching, and \cite{Berdanier:2017kmd} for the case of a periodic drive). We expect holography to be a useful tool for such investigations.

Finally, one other future direction would be the study of inhomogeneous quenches in our holographic model. The present investigation was enormously simplified because translation invariant solutions form a superselection sector for the bulk equations of motion. This feature was highlighted in our comparison of two-point correlation functions in our holographic model with analogous correlators in the spin chain in \cite{pasquale}. In particular, we did not observe the spreading of correlations found there because our homogeneous holographic quenches did not produce (pairs of) bound state excitations carrying nonvanishing momentum. It is also very likely that the homogeneity plays an important role in the failure of the holographic system to thermalize on the time scales which we were able to simulate. In particular, with an inhomogeneous quench, we might expect to see the formation of local black holes. Of course, a ``partially thermalized" phase of local plasma balls \cite{fireball,fire2} would be interesting in its own right. Hence it would be very interesting to consider inhomogeneous quenches to see which of the behaviours observed in our present study of homogeneous solutions survive for more generic holographic quenches.

\vspace{20pt}
\section*{Acknowledgements}
We thank Sumit Das and Luis Lehner for useful conversations, as well as Ben Craps and Jonathan Lingren for a useful correspondence. Research at Perimeter Institute is supported by the Government of Canada through the Department of Innovation, Science and Economic Development and by the Province of Ontario
through the Ministry of Research \& Innovation. RCM, MR, and BW were also supported
by NSERC Discovery grants. RCM is also supported by research funding
from the Canadian Institute for Advanced Research and from the Simons Foundation through ``It from Qubit" Collaboration.

\vspace{20pt}
\bibliographystyle{JHEP}
\bibliography{bibl}

\providecommand{\href}[2]{#2}\begingroup\raggedright\begin{thebibliography}{10}

\bibitem{more1}
S.~Mondal, D.~Sen, and K.~Sengupta, {\it Non-equilibrium dynamics of quantum
  systems: order parameter evolution, defect generation, and qubit transfer},
  in {\em Quantum Quenching, Annealing and Computation}, pp.~21--56.
\newblock Springer, 2010.

\bibitem{more2}
J.~Dziarmaga, {\it Dynamics of a quantum phase transition and relaxation to a
  steady state},  {\em Advances in Physics} {\bf 59} (2010), no.~6 1063--1189.

\bibitem{more3}
A.~Polkovnikov, K.~Sengupta, A.~Silva, and M.~Vengalattore, {\it Colloquium:
  Nonequilibrium dynamics of closed interacting quantum systems},  {\em Reviews
  of Modern Physics} {\bf 83} (2011), no.~3 863.

\bibitem{more4}
A.~Lamacraft, J.~Moore, K.~Levin, and D.~Stamper-Kurn, {\it Potential insights
  into non-equilibrium behavior from atomic physics},  {\em Ultracold bosonic
  and fermionic gases} {\bf 5} (2012).

\bibitem{LL}
L.~D. Landau and E.~M. Lifshitz, {\em Quantum mechanics: non-relativistic
  theory}, vol.~3.
\newblock Elsevier, 2013.

\bibitem{Cala06}
P.~Calabrese and J.~L. Cardy, {\it {Time-dependence of correlation functions
  following a quantum quench}},  {\em Phys. Rev. Lett.} {\bf 96} (2006) 136801,
  [\href{http://arxiv.org/abs/cond-mat/0601225}{{\tt cond-mat/0601225}}].

\bibitem{Cala07}
P.~Calabrese and J.~Cardy, {\it {Quantum Quenches in Extended Systems}},  {\em
  J. Stat. Mech.} {\bf 0706} (2007) P06008,
  [\href{http://arxiv.org/abs/0704.1880}{{\tt arXiv:0704.1880}}].

\bibitem{cala08}
P.~Calabrese, C.~Hagendorf, and P.~Le~Doussal, {\it Time evolution of
  one-dimensional gapless models from a domain wall initial state: stochastic
  loewner evolution continued?},  {\em Journal of Statistical Mechanics: Theory
  and Experiment} {\bf 2008} (2008), no.~07 P07013.

\bibitem{sotir08}
S.~Sotiriadis and J.~Cardy, {\it Inhomogeneous quantum quenches},  {\em Journal
  of Statistical Mechanics: Theory and Experiment} {\bf 2008} (2008), no.~11
  P11003.

\bibitem{Cala05}
P.~Calabrese and J.~L. Cardy, {\it {Evolution of entanglement entropy in
  one-dimensional systems}},  {\em J. Stat. Mech.} {\bf 0504} (2005) P04010,
  [\href{http://arxiv.org/abs/cond-mat/0503393}{{\tt cond-mat/0503393}}].

\bibitem{kollath07}
C.~Kollath, A.~M. L{\"a}uchli, and E.~Altman, {\it Quench dynamics and
  nonequilibrium phase diagram of the bose-hubbard model},  {\em Physical
  review letters} {\bf 98} (2007), no.~18 180601.

\bibitem{cramer08}
M.~Cramer, C.~M. Dawson, J.~Eisert, and T.~J. Osborne, {\it Exact relaxation in
  a class of nonequilibrium quantum lattice systems},  {\em Physical Review
  Letters} {\bf 100} (2008), no.~3 030602.

\bibitem{roux2009}
G.~Roux, {\it Quenches in quantum many-body systems: One-dimensional
  bose-hubbard model reexamined},  {\em Physical Review A} {\bf 79} (2009),
  no.~2 021608.

\bibitem{sotir09q}
S.~Sotiriadis, P.~Calabrese, and J.~Cardy, {\it Quantum quench from a thermal
  initial state},  {\em EPL (Europhysics Letters)} {\bf 87} (2009), no.~2
  20002.

\bibitem{rigol2007}
M.~Rigol, V.~Dunjko, V.~Yurovsky, and M.~Olshanii, {\it Relaxation in a
  completely integrable many-body quantum system: an ab initio study of the
  dynamics of the highly excited states of 1d lattice hard-core bosons},  {\em
  Physical review letters} {\bf 98} (2007), no.~5 050405.

\bibitem{man07}
S.~R. Manmana, S.~Wessel, R.~M. Noack, and A.~Muramatsu, {\it Strongly
  correlated fermions after a quantum quench},  {\em Physical review letters}
  {\bf 98} (2007), no.~21 210405.

\bibitem{rigol2008}
M.~Rigol, V.~Dunjko, and M.~Olshanii, {\it Thermalization and its mechanism for
  generic isolated quantum systems},  {\em Nature} {\bf 452} (2008), no.~7189
  854--858.

\bibitem{cala11q}
P.~Calabrese, F.~H. Essler, and M.~Fagotti, {\it Quantum quench in the
  transverse-field ising chain},  {\em Physical review letters} {\bf 106}
  (2011), no.~22 227203.

\bibitem{joh}
M.~Ammon and J.~Erdmenger, {\em {Gauge/gravity duality}}.
\newblock Cambridge Univ. Pr., Cambridge, UK, 2015.

\bibitem{Das:2010yw}
S.~R. Das, T.~Nishioka, and T.~Takayanagi, {\it {Probe Branes, Time-dependent
  Couplings and Thermalization in AdS/CFT}},  {\em JHEP} {\bf 07} (2010) 071,
  [\href{http://arxiv.org/abs/1005.3348}{{\tt arXiv:1005.3348}}].

\bibitem{Rangamani:2015agy}
M.~Rangamani, M.~Rozali, and A.~Vincart-Emard, {\it {Dynamics of Holographic
  Entanglement Entropy Following a Local Quench}},  {\em JHEP} {\bf 04} (2016)
  069, [\href{http://arxiv.org/abs/1512.03478}{{\tt arXiv:1512.03478}}].

\bibitem{Caputa:2017ixa}
P.~Caputa, S.~R. Das, M.~Nozaki, and A.~Tomiya, {\it {Quantum Quench and
  Scaling of Entanglement Entropy}},
  \href{http://arxiv.org/abs/1702.04359}{{\tt arXiv:1702.04359}}.

\bibitem{Basu:2013soa}
P.~Basu, D.~Das, S.~R. Das, and K.~Sengupta, {\it {Quantum Quench and Double
  Trace Couplings}},  {\em JHEP} {\bf 12} (2013) 070,
  [\href{http://arxiv.org/abs/1308.4061}{{\tt arXiv:1308.4061}}].

\bibitem{Basu:2012gg}
P.~Basu, D.~Das, S.~R. Das, and T.~Nishioka, {\it {Quantum Quench Across a Zero
  Temperature Holographic Superfluid Transition}},  {\em JHEP} {\bf 03} (2013)
  146, [\href{http://arxiv.org/abs/1211.7076}{{\tt arXiv:1211.7076}}].

\bibitem{Basu:2011ft}
P.~Basu and S.~R. Das, {\it {Quantum Quench across a Holographic Critical
  Point}},  {\em JHEP} {\bf 01} (2012) 103,
  [\href{http://arxiv.org/abs/1109.3909}{{\tt arXiv:1109.3909}}].

\bibitem{Buchel:2012gw}
A.~Buchel, L.~Lehner, and R.~C. Myers, {\it {Thermal quenches in N=2*
  plasmas}},  {\em JHEP} {\bf 08} (2012) 049,
  [\href{http://arxiv.org/abs/1206.6785}{{\tt arXiv:1206.6785}}].

\bibitem{Buchel:2013lla}
A.~Buchel, L.~Lehner, R.~C. Myers, and A.~van Niekerk, {\it {Quantum quenches
  of holographic plasmas}},  {\em JHEP} {\bf 05} (2013) 067,
  [\href{http://arxiv.org/abs/1302.2924}{{\tt arXiv:1302.2924}}].

\bibitem{Buchel:2013gba}
A.~Buchel, R.~C. Myers, and A.~van Niekerk, {\it {Universality of Abrupt
  Holographic Quenches}},  {\em Phys. Rev. Lett.} {\bf 111} (2013) 201602,
  [\href{http://arxiv.org/abs/1307.4740}{{\tt arXiv:1307.4740}}].

\bibitem{Craps:2015upq}
B.~Craps, E.~J. Lindgren, and A.~Taliotis, {\it {Holographic thermalization in
  a top-down confining model}},  {\em JHEP} {\bf 12} (2015) 116,
  [\href{http://arxiv.org/abs/1511.00859}{{\tt arXiv:1511.00859}}].

\bibitem{Balasubramanian:2014cja}
V.~Balasubramanian, A.~Buchel, S.~R. Green, L.~Lehner, and S.~L. Liebling, {\it
  {Holographic Thermalization, Stability of Anti-de-Sitter Space and the
  Fermi-Pasta-Ulam Paradox}},  {\em Phys. Rev. Lett.} {\bf 113} (2014), no.~7
  071601, [\href{http://arxiv.org/abs/1403.6471}{{\tt arXiv:1403.6471}}].

\bibitem{Buchel:2014xwa}
A.~Buchel, S.~R. Green, L.~Lehner, and S.~L. Liebling, {\it {Conserved
  quantities and dual turbulent cascades in anti-de-Sitter spacetime}},  {\em
  Phys. Rev.} {\bf D91} (2015), no.~6 064026,
  [\href{http://arxiv.org/abs/1412.4761}{{\tt arXiv:1412.4761}}].

\bibitem{AbajoArrastia:2010yt}
J.~Abajo-Arrastia, J.~Aparicio, and E.~Lopez, {\it {Holographic Evolution of
  Entanglement Entropy}},  {\em JHEP} {\bf 11} (2010) 149,
  [\href{http://arxiv.org/abs/1006.4090}{{\tt arXiv:1006.4090}}].

\bibitem{Albash:2010mv}
T.~Albash and C.~V. Johnson, {\it {Evolution of Holographic Entanglement
  Entropy after Thermal and Electromagnetic Quenches}},  {\em New J. Phys.}
  {\bf 13} (2011) 045017, [\href{http://arxiv.org/abs/1008.3027}{{\tt
  arXiv:1008.3027}}].

\bibitem{Balasubramanian:2010ce}
V.~Balasubramanian, A.~Bernamonti, J.~de~Boer, N.~Copland, B.~Craps,
  E.~Keski-Vakkuri, B.~Muller, A.~Schafer, M.~Shigemori, and W.~Staessens, {\it
  {Thermalization of Strongly Coupled Field Theories}},  {\em Phys. Rev. Lett.}
  {\bf 106} (2011) 191601, [\href{http://arxiv.org/abs/1012.4753}{{\tt
  arXiv:1012.4753}}].

\bibitem{Balasubramanian:2011ur}
V.~Balasubramanian, A.~Bernamonti, J.~de~Boer, N.~Copland, B.~Craps,
  E.~Keski-Vakkuri, B.~Muller, A.~Schafer, M.~Shigemori, and W.~Staessens, {\it
  {Holographic Thermalization}},  {\em Phys. Rev.} {\bf D84} (2011) 026010,
  [\href{http://arxiv.org/abs/1103.2683}{{\tt arXiv:1103.2683}}].

\bibitem{Aparicio:2011zy}
J.~Aparicio and E.~Lopez, {\it {Evolution of Two-Point Functions from
  Holography}},  {\em JHEP} {\bf 12} (2011) 082,
  [\href{http://arxiv.org/abs/1109.3571}{{\tt arXiv:1109.3571}}].

\bibitem{Allais:2011ys}
A.~Allais and E.~Tonni, {\it {Holographic evolution of the mutual
  information}},  {\em JHEP} {\bf 01} (2012) 102,
  [\href{http://arxiv.org/abs/1110.1607}{{\tt arXiv:1110.1607}}].

\bibitem{Galante:2012pv}
D.~Galante and M.~Schvellinger, {\it {Thermalization with a chemical potential
  from AdS spaces}},  {\em JHEP} {\bf 07} (2012) 096,
  [\href{http://arxiv.org/abs/1205.1548}{{\tt arXiv:1205.1548}}].

\bibitem{Baron:2012fv}
W.~Baron, D.~Galante, and M.~Schvellinger, {\it {Dynamics of holographic
  thermalization}},  {\em JHEP} {\bf 03} (2013) 070,
  [\href{http://arxiv.org/abs/1212.5234}{{\tt arXiv:1212.5234}}].

\bibitem{Balasubramanian:2012tu}
V.~Balasubramanian, A.~Bernamonti, B.~Craps, V.~Keranen, E.~Keski-Vakkuri,
  B.~Muller, L.~Thorlacius, and J.~Vanhoof, {\it {Thermalization of the
  spectral function in strongly coupled two dimensional conformal field
  theories}},  {\em JHEP} {\bf 04} (2013) 069,
  [\href{http://arxiv.org/abs/1212.6066}{{\tt arXiv:1212.6066}}].

\bibitem{Keranen:2011xs}
V.~Keranen, E.~Keski-Vakkuri, and L.~Thorlacius, {\it {Thermalization and
  entanglement following a non-relativistic holographic quench}},  {\em Phys.
  Rev.} {\bf D85} (2012) 026005, [\href{http://arxiv.org/abs/1110.5035}{{\tt
  arXiv:1110.5035}}].

\bibitem{Caceres:2012em}
E.~Caceres and A.~Kundu, {\it {Holographic Thermalization with Chemical
  Potential}},  {\em JHEP} {\bf 09} (2012) 055,
  [\href{http://arxiv.org/abs/1205.2354}{{\tt arXiv:1205.2354}}].

\bibitem{Das:2014jna}
S.~R. Das, D.~A. Galante, and R.~C. Myers, {\it {Universal scaling in fast
  quantum quenches in conformal field theories}},  {\em Phys. Rev. Lett.} {\bf
  112} (2014) 171601, [\href{http://arxiv.org/abs/1401.0560}{{\tt
  arXiv:1401.0560}}].

\bibitem{Das:2014hqa}
S.~R. Das, D.~A. Galante, and R.~C. Myers, {\it {Universality in fast quantum
  quenches}},  {\em JHEP} {\bf 02} (2015) 167,
  [\href{http://arxiv.org/abs/1411.7710}{{\tt arXiv:1411.7710}}].

\bibitem{Das:2015jka}
S.~R. Das, D.~A. Galante, and R.~C. Myers, {\it {Smooth and fast versus
  instantaneous quenches in quantum field theory}},  {\em JHEP} {\bf 08} (2015)
  073, [\href{http://arxiv.org/abs/1505.05224}{{\tt arXiv:1505.05224}}].

\bibitem{Das:2016lla}
S.~R. Das, D.~A. Galante, and R.~C. Myers, {\it {Quantum Quenches in Free Field
  Theory: Universal Scaling at Any Rate}},  {\em JHEP} {\bf 05} (2016) 164,
  [\href{http://arxiv.org/abs/1602.08547}{{\tt arXiv:1602.08547}}].

\bibitem{pasquale}
M.~{Kormos}, M.~{Collura}, G.~{Tak{\'a}cs}, and P.~{Calabrese}, {\it {Real time
  confinement following a quantum quench to a non-integrable model}},  {\em
  ArXiv e-prints} (Apr., 2016) [\href{http://arxiv.org/abs/1604.03571}{{\tt
  arXiv:1604.03571}}].

\bibitem{edward}
E.~Witten, {\it {Anti-de Sitter space, thermal phase transition, and
  confinement in gauge theories}},  {\em Adv. Theor. Math. Phys.} {\bf 2}
  (1998) 505--532, [\href{http://arxiv.org/abs/hep-th/9803131}{{\tt
  hep-th/9803131}}].

\bibitem{Horowitz:1998ha}
G.~T. Horowitz and R.~C. Myers, {\it {The AdS / CFT correspondence and a new
  positive energy conjecture for general relativity}},  {\em Phys. Rev.} {\bf
  D59} (1998) 026005, [\href{http://arxiv.org/abs/hep-th/9808079}{{\tt
  hep-th/9808079}}].

\bibitem{Craps:2013iaa}
B.~Craps, E.~Kiritsis, C.~Rosen, A.~Taliotis, J.~Vanhoof, and H.-b. Zhang, {\it
  {Gravitational collapse and thermalization in the hard wall model}},  {\em
  JHEP} {\bf 02} (2014) 120, [\href{http://arxiv.org/abs/1311.7560}{{\tt
  arXiv:1311.7560}}].

\bibitem{daSilva:2016nah}
E.~da~Silva, E.~Lopez, J.~Mas, and A.~Serantes, {\it {Holographic Quenches with
  a Gap}},  {\em JHEP} {\bf 06} (2016) 172,
  [\href{http://arxiv.org/abs/1604.08765}{{\tt arXiv:1604.08765}}].

\bibitem{Lopez:2017hjg}
E.~Lopez and G.~M. del Bosch, {\it {A simple holographic scenario for gapped
  quenches}},  {\em JHEP} {\bf 02} (2017) 130,
  [\href{http://arxiv.org/abs/1701.02671}{{\tt arXiv:1701.02671}}].

\bibitem{stress}
R.~C. Myers, {\it {Stress tensors and Casimir energies in the AdS / CFT
  correspondence}},  {\em Phys. Rev.} {\bf D60} (1999) 046002,
  [\href{http://arxiv.org/abs/hep-th/9903203}{{\tt hep-th/9903203}}].

\bibitem{Maliborski:2013via}
M.~Maliborski and A.~Rostworowski, {\it {Lecture Notes on Turbulent Instability
  of Anti-de Sitter Spacetime}},  {\em Int. J. Mod. Phys.} {\bf A28} (2013)
  1340020, [\href{http://arxiv.org/abs/1308.1235}{{\tt arXiv:1308.1235}}].

\bibitem{daSilva:2014zva}
E.~da~Silva, E.~Lopez, J.~Mas, and A.~Serantes, {\it {Collapse and Revival in
  Holographic Quenches}},  {\em JHEP} {\bf 04} (2015) 038,
  [\href{http://arxiv.org/abs/1412.6002}{{\tt arXiv:1412.6002}}].

\bibitem{2014PhRvL.112v0401C}
J.~{Cardy}, {\it {Thermalization and Revivals after a Quantum Quench in
  Conformal Field Theory}},  {\em Physical Review Letters} {\bf 112} (June,
  2014) 220401, [\href{http://arxiv.org/abs/1403.3040}{{\tt arXiv:1403.3040}}].

\bibitem{2016JPhA...49O5401C}
J.~{Cardy}, {\it {Quantum revivals in conformal field theories in higher
  dimensions}},  {\em Journal of Physics A Mathematical General} {\bf 49}
  (Oct., 2016) 415401, [\href{http://arxiv.org/abs/1603.08267}{{\tt
  arXiv:1603.08267}}].

\bibitem{Minahan98}
J.~A. Minahan, {\it {Glueball mass spectra and other issues for supergravity
  duals of QCD models}},  {\em JHEP} {\bf 01} (1999) 020,
  [\href{http://arxiv.org/abs/hep-th/9811156}{{\tt hep-th/9811156}}].

\bibitem{Constable99}
N.~R. Constable and R.~C. Myers, {\it {Spin two glueballs, positive energy
  theorems and the AdS / CFT correspondence}},  {\em JHEP} {\bf 10} (1999) 037,
  [\href{http://arxiv.org/abs/hep-th/9908175}{{\tt hep-th/9908175}}].

\bibitem{Brower99}
R.~C. Brower, S.~D. Mathur, and C.-I. Tan, {\it {Discrete spectrum of the
  graviton in the AdS(5) black hole background}},  {\em Nucl. Phys.} {\bf B574}
  (2000) 219--244, [\href{http://arxiv.org/abs/hep-th/9908196}{{\tt
  hep-th/9908196}}].

\bibitem{Brow00}
R.~C. Brower, S.~D. Mathur, and C.-I. Tan, {\it {Glueball spectrum for QCD from
  AdS supergravity duality}},  {\em Nucl. Phys.} {\bf B587} (2000) 249--276,
  [\href{http://arxiv.org/abs/hep-th/0003115}{{\tt hep-th/0003115}}].

\bibitem{Auzzi:2013pca}
R.~Auzzi, S.~Elitzur, S.~B. Gudnason, and E.~Rabinovici, {\it {On periodically
  driven AdS/CFT}},  {\em JHEP} {\bf 11} (2013) 016,
  [\href{http://arxiv.org/abs/1308.2132}{{\tt arXiv:1308.2132}}].

\bibitem{Rangamani:2015sha}
M.~Rangamani, M.~Rozali, and A.~Wong, {\it {Driven Holographic CFTs}},  {\em
  JHEP} {\bf 04} (2015) 093, [\href{http://arxiv.org/abs/1502.05726}{{\tt
  arXiv:1502.05726}}].

\bibitem{hamming}
\url{https://en.wikipedia.org/wiki/Window_function#Hamming_window}.

\bibitem{2011PhRvL.106e0405B}
M.~C. {Ba{\~n}uls}, J.~I. {Cirac}, and M.~B. {Hastings}, {\it {Strong and Weak
  Thermalization of Infinite Nonintegrable Quantum Systems}},  {\em Physical
  Review Letters} {\bf 106} (Feb., 2011) 050405,
  [\href{http://arxiv.org/abs/1007.3957}{{\tt arXiv:1007.3957}}].

\bibitem{gge1}
M.~Rigol, V.~Dunjko, V.~Yurovsky, and M.~Olshanii, {\it Relaxation in a
  completely integrable many-body quantum system: an ab initio study of the
  dynamics of the highly excited states of 1d lattice hard-core bosons},  {\em
  Physical review letters} {\bf 98} (2007), no.~5 050405.

\bibitem{gge2}
M.~Rigol, A.~Muramatsu, and M.~Olshanii, {\it Hard-core bosons on optical
  superlattices: Dynamics and relaxation in the superfluid and insulating
  regimes},  {\em Physical Review A} {\bf 74} (2006), no.~5 053616.

\bibitem{Buchel:2012uh}
A.~Buchel, L.~Lehner, and S.~L. Liebling, {\it {Scalar Collapse in AdS}},  {\em
  Phys. Rev.} {\bf D86} (2012) 123011,
  [\href{http://arxiv.org/abs/1210.0890}{{\tt arXiv:1210.0890}}].

\bibitem{RT1}
S.~Ryu and T.~Takayanagi, {\it {Holographic derivation of entanglement entropy
  from AdS/CFT}},  {\em Phys. Rev. Lett.} {\bf 96} (2006) 181602,
  [\href{http://arxiv.org/abs/hep-th/0603001}{{\tt hep-th/0603001}}].

\bibitem{RT2}
S.~Ryu and T.~Takayanagi, {\it {Aspects of Holographic Entanglement Entropy}},
  {\em JHEP} {\bf 08} (2006) 045,
  [\href{http://arxiv.org/abs/hep-th/0605073}{{\tt hep-th/0605073}}].

\bibitem{HRT}
V.~E. Hubeny, M.~Rangamani, and T.~Takayanagi, {\it {A Covariant holographic
  entanglement entropy proposal}},  {\em JHEP} {\bf 07} (2007) 062,
  [\href{http://arxiv.org/abs/0705.0016}{{\tt arXiv:0705.0016}}].

\bibitem{Klebanov:2007ws}
I.~R. Klebanov, D.~Kutasov, and A.~Murugan, {\it {Entanglement as a probe of
  confinement}},  {\em Nucl. Phys.} {\bf B796} (2008) 274--293,
  [\href{http://arxiv.org/abs/0709.2140}{{\tt arXiv:0709.2140}}].

\bibitem{Blanco:2013joa}
D.~D. Blanco, H.~Casini, L.-Y. Hung, and R.~C. Myers, {\it {Relative Entropy
  and Holography}},  {\em JHEP} {\bf 08} (2013) 060,
  [\href{http://arxiv.org/abs/1305.3182}{{\tt arXiv:1305.3182}}].

\bibitem{tsun1}
H.~Liu and S.~J. Suh, {\it {Entanglement Tsunami: Universal Scaling in
  Holographic Thermalization}},  {\em Phys. Rev. Lett.} {\bf 112} (2014)
  011601, [\href{http://arxiv.org/abs/1305.7244}{{\tt arXiv:1305.7244}}].

\bibitem{tsun2}
H.~Liu and S.~J. Suh, {\it {Entanglement growth during thermalization in
  holographic systems}},  {\em Phys. Rev.} {\bf D89} (2014), no.~6 066012,
  [\href{http://arxiv.org/abs/1311.1200}{{\tt arXiv:1311.1200}}].

\bibitem{mark1}
H.~Casini, H.~Liu, and M.~Mezei, {\it {Spread of entanglement and causality}},
  {\em JHEP} {\bf 07} (2016) 077, [\href{http://arxiv.org/abs/1509.05044}{{\tt
  arXiv:1509.05044}}].

\bibitem{mark2}
M.~Mezei, {\it {On entanglement spreading from holography}},
  \href{http://arxiv.org/abs/1612.00082}{{\tt arXiv:1612.00082}}.

\bibitem{mark3}
J.~S. Cotler, M.~P. Hertzberg, M.~Mezei, and M.~T. Mueller, {\it {Entanglement
  Growth after a Global Quench in Free Scalar Field Theory}},  {\em JHEP} {\bf
  11} (2016) 166, [\href{http://arxiv.org/abs/1609.00872}{{\tt
  arXiv:1609.00872}}].

\bibitem{geod1}
V.~Balasubramanian and S.~F. Ross, {\it {Holographic particle detection}},
  {\em Phys. Rev.} {\bf D61} (2000) 044007,
  [\href{http://arxiv.org/abs/hep-th/9906226}{{\tt hep-th/9906226}}].

\bibitem{geod2}
J.~Louko, D.~Marolf, and S.~F. Ross, {\it {On geodesic propagators and black
  hole holography}},  {\em Phys. Rev.} {\bf D62} (2000) 044041,
  [\href{http://arxiv.org/abs/hep-th/0002111}{{\tt hep-th/0002111}}].

\bibitem{Rozali:2017bco}
M.~Rozali and A.~Vincart-Emard, {\it {Comments on Entanglement Propagation}},
  \href{http://arxiv.org/abs/1702.05869}{{\tt arXiv:1702.05869}}.

\bibitem{fast1}
D.~Mateos, R.~C. Myers, and R.~M. Thomson, {\it {Thermodynamics of the brane}},
   {\em JHEP} {\bf 05} (2007) 067,
  [\href{http://arxiv.org/abs/hep-th/0701132}{{\tt hep-th/0701132}}].

\bibitem{fast2}
Q.~J. Ejaz, T.~Faulkner, H.~Liu, K.~Rajagopal, and U.~A. Wiedemann, {\it {A
  Limiting velocity for quarkonium propagation in a strongly coupled plasma via
  AdS/CFT}},  {\em JHEP} {\bf 04} (2008) 089,
  [\href{http://arxiv.org/abs/0712.0590}{{\tt arXiv:0712.0590}}].

\bibitem{wilson1}
J.~M. Maldacena, {\it {Wilson loops in large N field theories}},  {\em Phys.
  Rev. Lett.} {\bf 80} (1998) 4859--4862,
  [\href{http://arxiv.org/abs/hep-th/9803002}{{\tt hep-th/9803002}}].

\bibitem{wilson2}
S.-J. Rey and J.-T. Yee, {\it {Macroscopic strings as heavy quarks in large N
  gauge theory and anti-de Sitter supergravity}},  {\em Eur. Phys. J.} {\bf
  C22} (2001) 379--394, [\href{http://arxiv.org/abs/hep-th/9803001}{{\tt
  hep-th/9803001}}].

\bibitem{2009arXiv0910.3692G}
V.~{Gritsev} and A.~{Polkovnikov}, {\it {Universal Dynamics Near Quantum
  Critical Points}},  {\em ArXiv e-prints} (Oct., 2009)
  [\href{http://arxiv.org/abs/0910.3692}{{\tt arXiv:0910.3692}}].

\bibitem{2010PhRvE}
F.~{Pollmann}, S.~{Mukerjee}, A.~{Green}, and J.~{Moore}, {\it {Dynamics after
  a sweep through a quantum critical point}},
  \href{http://arxiv.org/abs/0907.3206}{{\tt arXiv:0907.3206}}.

\bibitem{Berdanier:2017kmd}
W.~Berdanier, M.~Kolodrubetz, R.~Vasseur, and J.~E. Moore, {\it {Floquet
  Dynamics of Boundary-Driven Systems at Criticality}},
  \href{http://arxiv.org/abs/1701.05899}{{\tt arXiv:1701.05899}}.

\bibitem{fireball}
O.~Aharony, S.~Minwalla, and T.~Wiseman, {\it {Plasma-balls in large N gauge
  theories and localized black holes}},  {\em Class. Quant. Grav.} {\bf 23}
  (2006) 2171--2210, [\href{http://arxiv.org/abs/hep-th/0507219}{{\tt
  hep-th/0507219}}].

\bibitem{fire2}
P.~Figueras and S.~Tunyasuvunakool, {\it {Localized Plasma Balls}},  {\em JHEP}
  {\bf 06} (2014) 025, [\href{http://arxiv.org/abs/1404.0018}{{\tt
  arXiv:1404.0018}}].

\end{thebibliography}\endgroup

\end{document}